\documentclass[11pt,a4paper]{article}
\usepackage{eurosym}
\usepackage{array}
\usepackage{float}
\usepackage{amsmath}
\usepackage{amsfonts}
\usepackage{pdflscape}
\usepackage{amssymb}
\usepackage{latexsym}
\usepackage{epsfig}
\usepackage{graphicx}
\usepackage{booktabs}
\usepackage{sw20elba}
\usepackage{mathrsfs}
\usepackage{fancyhdr}
\usepackage{sectsty}
\usepackage[longnamesfirst]{natbib}
\usepackage[labelsep=colon,font=it,labelfont=sc]{caption}
\usepackage{theorem}
\usepackage{url}
\usepackage{rotating}
\usepackage[truetex]{color}

\newtheorem{theorem}{Theorem}

{
\theorembodyfont{\upshape} 
\theoremheaderfont{\scshape} 

}

\newtheorem{corollary}{Corollary}

{\theorembodyfont{\upshape} 

}
\newtheorem{exercise}{Assumption}

\newenvironment{proof}[1][Proof]{\bigskip \noindent \textbf{#1:} }{\  \rule{0.5em}{0.5em}}

\theoremheaderfont{\bfseries}
\theorembodyfont{\upshape}
\sectionfont{\centering \sc}
\subsectionfont{\centering \sc}
\subsubsectionfont{\centering \sc}
\allsectionsfont{\centering \sc}
\allowdisplaybreaks

\bibpunct[:~]{(}{)}{;}{a}{,}{;}
\renewcommand{\cite}{\citet}
\graphicspath{{C:/Work/forthcoming/CNT-fractional-BS/}{C:/Users/gcava/Documents/ARTICOLI/FRB2016/}{C:/Users/gcava/OneDrive - Alma Mater Studiorum Universit  di Bologna/Documents/Articoli/FRB2016/revision 1}}
\input{tcilatex}

\begin{document}

\begin{center}
\bigskip

{\Large {}PARAMETERS ON THE BOUNDARY IN PREDICTIVE REGRESSION}\textsc{%
{\LARGE {}{* }}}

\mbox{}

\ \vspace{-0.15cm}

\textsc{Giuseppe Cavaliere}$^{a,b}$\textsc{, Iliyan Georgiev}$^{a}$ \textsc{%
and Edoardo Zanelli}$^{a}$

\textsc{\vspace{-0.15in} }

\vspace{0.45cm}

{\small {}First draft: May 3, 2022; Revised: February 1, 2024 and July 1,
2024; \\This version: September 19, 2024}

\renewcommand{\thefootnote}{}
\footnote{
\hspace{-7.2mm}
$^{*}$ We thank Rasmus S\o ndergaard Pedersen and Mervyn Silvapulle for comments. 
Financial support from the Italian Ministry of University and Research (PRIN 2020 Grant 2020B2AKFW) is gratefully acknowledged. 
Address correspondence to: Giuseppe Cavaliere, Department of Economics, University of Bologna, 
Piazza Scaravilli 2, 40126 Bologna, Italy; email: giuseppe.cavaliere@unibo.it.
\\
$^{a}$ Department of Economics, University of Bologna, Italy.
\\
$^{b}$ Department of Economics, University of Exeter, UK.}
\addtocounter{footnote}{-2}
\renewcommand{\thefootnote}{\arabic{footnote}}%
\end{center}

\par\begingroup\leftskip=1 cm
\rightskip=1 cm
\small

\begin{center}
\textsc{Abstract\vspace{-0.15cm}}
\end{center}

{\small We consider bootstrap inference in predictive (or Granger-causality)
regressions when the parameter of interest may lie on the boundary of the
parameter space, here defined by means of a smooth inequality constraint.
For instance, this situation occurs when the definition of the parameter
space allows for the cases of either no predictability or sign-restricted
predictability. We show that in this context constrained estimation gives
rise to bootstrap statistics whose limit distribution is, in general,
random, and thus distinct from the limit null distribution of the original
statistics of interest. This is due to both (i) the possible location of the
true parameter vector on the boundary of the parameter space, and (ii) the
possible non-stationarity of the posited predicting (resp. Granger-causing)
variable. We discuss a modification of the standard fixed-regressor wild
bootstrap scheme where the bootstrap parameter space is shifted by a
data-dependent function in order to eliminate the portion of limiting
bootstrap randomness attributable to the boundary, and prove validity of the
associated bootstrap inference under non-stationarity of the predicting
variable as the only remaining source of limiting bootstrap randomness. Our
approach, which is initially presented in a simple location model, has
bearing on inference in parameter-on-the-boundary situations beyond the
predictive regression problem. \bigskip {} }

\noindent \textsc{{\small {}Keywords: }}{\small {}}Parameter on the
boundary, random measures, weak convergence in distribution, asymptotic
inference, uniform inference.

\par\endgroup\normalsize

\setcounter{footnote}{0}

\numberwithin{equation}{section}%
\numberwithin{theorem}{section}
\numberwithin{corollary}{section}
\numberwithin{lemma}{section}%
\newpage

\section{Introduction}

In this paper we revisit the well-known problem of bootstrap inference in
regressions with parameter space defined by means of smooth inequality
constraints. For instance, consider the setup of a regression $y_{t}=\text{$%
\alpha+$}\beta x_{t-1}+\varepsilon_{t}$ where the parameter space for $%
(\alpha,\beta)$ is defined by the constraint $\beta\geq0$. This framework
arises when only the possibilities $\beta=0$ of no predictability (or no
first-order Granger causality, generalizable to higher orders), and $\beta>0$
of sign-restricted predictability, are entertained, and the model is
estimated under the constraint $\beta\in\lbrack0,\infty)$. In applications,
economic theory is often informative about the direction of predictability,
and such information could be used to improve the efficiency of estimators
and increase the power of hypotheses tests. A prominent example is provided
by predictive regressions for financial returns; see, e.g., Phillips (2014)
and the references therein. Interest can then be in testing the very
hypothesis of no predictability (i.e., $\beta=0$) by means of a one-sided
test, or a special case of this hypothesis (e.g., $\alpha=\beta=0$), or a
hypothesis where the parameter vector may but need not lie on the boundary
of the parameter space (e.g., $\alpha+\beta=0$).

While in this context the bootstrap is potentially useful, its application
is not straightforward if the parameter vector may lie on the boundary of
the parameter space; see Andrews (2000). In particular, as we discuss in the
following, even in a simple location model where the parameter space is a
closed half-line, the cumulative distribution function {[}cdf{]} of the
parametric bootstrap $t$-statistic, conditional on the original data,
converges weakly to a random cdf, rather than to the target asymptotic
distribution of the $t$-statistic computed from the original data.

Our first contribution is to show that in predictive regressions with
parameter values on the boundary, the distribution of fixed regressor%
\footnote{%
We focus on `fixed regressor' bootstrap schemes as they do not require
knowledge on the regressor generating process. For instance, and in contrast
to recursive-based schemes, they can be applied to both I(0) and I(1)
settings.} bootstrap statistics, like the $t$-statistic for $\beta =0$ in
the regression above, may be random in the limit. Limiting randomness may
arise in two ways. A first possible source of randomness in the limit
bootstrap measure is in the non-stationarity of the regressor, which
operates through the random limits of sample product moments. This is hardly
surprising, see e.g. Georgiev et al. (2019). A second potential source of
randomness is the location of the parameter vector on the boundary of the
parameter space. Invalidity of standard bootstrap schemes when a parameter
is on the boundary was initially discussed in Andrews (2000), where a simple
location-model example was given; see also Chatterjee and Lahiri (2011). In
the context of hypotheses tests in predictive regressions, we revisit
Andrews' result and show that, for a general bootstrap scheme, the
occurrence or non-occurrence of limiting bootstrap randomness due to the
possible location of a parameter on the boundary of the parameter space
depends on how well the bootstrap scheme approximates the mutual position of
three objects: (i)\ the boundary, (ii)\ the parameter set identified by the
null hypothesis, and (iii)\thinspace\ the true parameter value. Standard
bootstrap approximations of this mutual position may not be sufficiently
precise, giving rise to complex conditioning in the limit bootstrap
distribution, with ensuing bootstrap validity only for special types of
statistics.

Our second contribution is to show that certain non-standard bootstrap
schemes, designed to provide a better match with the geometric configuration
in the original parameter space, give rise to limit bootstrap distributions
where randomness, if present, is not attributable to the boundary value of
the parameter vector. This fact allows us to establish bootstrap validity in
an `unconditional' sense; see Cavaliere and Georgiev (2020). That is,
although randomness of the limiting bootstrap cdf prevents the possibility
that the bootstrap could mimic the asymptotic distribution of the original
statistic, we can show that in large samples bootstrap tests and asymptotic
tests are correctly sized for essentially the same set of nominal sizes.

Formally, we make use of the following definition, which generalizes the
definition of unconditional bootstrap validity given in Cavaliere and
Georgiev (2020, p.2555). Let $p_{n}$ and $p_{n}^{\ast }$ be respectively the 
\emph{p}-value of an asymptotic test and of its bootstrap analogue. Let also 
\begin{equation*}
C:=\{q\in (0,1):\lim_{n\rightarrow \infty }P(p_{n}\leq q)=q|\mathsf{H}_{0}\},
\end{equation*}%
such that a test rejecting for $p_{n}\leq q$ (or for $p_{n}>q$) is correctly
sized for nominal significance levels $q$ (resp. $1-q$) with $q\in C$, as $%
n\rightarrow \infty $.\footnote{{\small {}{}For $q\in \mathrm{int}C$ it
holds that $P(p_{n}=q)\rightarrow 0$ and rejections for $p_{n}\leq q$ (or $%
p_{n}>q$) are asymptotically equivalent to rejections for $p_{n}<q$ (or $%
p_{n}\geq q$).}} If, under the null hypothesis $\mathsf{H}_{0}$, 
\begin{equation}
P(p_{n}^{\ast }\leq q)\rightarrow q\text{ for all $q\in \limfunc{int}C$,}
\label{eq unconditional validity}
\end{equation}%
where $\limfunc{int}C$ denotes the interior of the set $C$, we say that the
bootstrap test based on $p_{n}^{\ast }$ is valid for $\mathsf{H}_{0}$.%
\footnote{%
Bootstrap unconditional validity as in Cavaliere and Georgiev (2020) is
obtained as the special case $\limfunc{int}$$C=(0,1)$.} The meaning is that
the bootstrap test and the asymptotic test are first-order asymptotically
equivalent in terms of correct size control. In particular, bootstrap
validity for simple hypotheses $\mathsf{H}_{0}$ characterizes pointwise size
control.

Notice that bootstrap validity as in (1.1) is implied by the classic
definition of bootstrap consistency, namely that $\sup_{x\in\mathbb{R}%
}|F_{n}^{\ast}(x)-F(x)|\rightarrow_{p}0$ for a bootstrap statistic with cdf $%
F_{n}^{\ast}$ conditionally on the data and an original test statistic with
continuous asymptotic cdf $F$. The converse does not hold; that is, (\ref{eq
unconditional validity}) does not imply classic bootstrap consistency, see
the discussion in Cavaliere and Georgiev (2020).

For test statistics whose asymptotic distribution is continuous, it holds
that $\limfunc{int}$$C=(0,1)$ and hence condition (\ref{eq unconditional
validity}) should hold for all $q\in (0,1)$ for the bootstrap to be valid.
Unfortunately, parameter values on the boundary of the parameter space may
induce discontinuities in the limiting cdf's, such that not even the exact 
\emph{p}-values of the associated tests are asymptotically standard uniform
on $[0,1]$. This makes the above weaker version of the validity definition
unavoidable.

Finally, we turn to the special case of one-sided tests for the null
hypothesis that the parameter vector lies on the boundary of the parameter
space, such that the boundary coincides with the parameter set identified by
the null hypothesis. This case provides a transparent example of a limit
bootstrap cdf which is random only on a subset of its domain. Then, if
bootstrap validity is defined as in (\ref{eq unconditional validity}), in
this case also some standard bootstrap schemes can be proved to be valid.

This paper is related to recent work by Fang and Santos (2019) and Hong and
Li (2020). The latter two papers propose nonstandard bootstrap schemes --
involving a tuning tool -- which correct the inconsistency of `classic'
bootstrap methods in settings that cover parameters on the boundary as a
special case. The main difference from the present contribution is that our
theory applies to random limit bootstrap measures. Thus, Fang and Santos
(2019) consider bootstrap inference in settings where the target asymptotic
distribution, say that of a random element $\tau $, can be thought of as a
transformation $\varphi $ of another random element $\tau ^{\prime }$, and
both the distribution of $\tau ^{\prime }$ and the transformation $\varphi $
need to be estimated; see also the related works by D\"{u}mbgen (1993),
Hirano and Porter (2012), Fang (2014) and Chen and Fang (2019). Although
Fang and Santos (2019) consider deterministic $\varphi $ and the
unconditional distribution of $\tau ^{\prime }$, such that their results are
not directly applicable here, their way of conceptualizing the problem
remains fruitful also in the case of random $\varphi $ and random
conditional distributions $\text{$\tau ^{\prime }|\tau ^{\prime \prime }$
(for some random element }\tau ^{\prime \prime }$)$\text{. }$We discuss this
in Section \ref{subsec:Fang-and-Santos}.

Our contribution is also related to Hong and Li (2020), who propose a
`numerical bootstrap' which is valid in settings where a parameter space can
be approximated locally by a cone with vertex at the true value of the
parameters; see Geyer (1994) for a detailed discussion of the approximation.
Both the approaches in this paper and that by Hong and Li (2020) are
connected to the large body of literature considering estimation and
inference for constrained M-estimators; see, among others, Geyer (1994),
Andrews (1999, 2000), and the references therein. In Section \ref%
{subsec:Fang-and-Santos} we argue that, when applied to a restricted
predictive regression, the `numerical bootstrap' of Hong and Li (2020)
performs a geometric approximation of the kind we propose, though at the
cost of a slower-than-standard convergence rate for the resulting bootstrap
estimator.

We present our main idea using first a simple location model for i.i.d.
scalar data whose location parameter is constrained to be positive. This is
done in Section \ref{sec example}. The predictive regression framework is
presented in Section \ref{sec par on the boundary setup}; in this section we
also show that the bootstrap limit measure associated with standard fixed
regressor wild bootstrap schemes is random. A new family of bootstrap
algorithms and their validity are discussed in Section \ref{sec par on the
boundary main}. Results on the validity of one-sided tests, connections to
the previous literature, and uniform size control for the bootstrap tests
are discussed in Section \ref{sec:Discussion}. Section \ref{sec:Numerical}
provides simulation evidence, whereas Section \ref{sec conclusion}
concludes. Proofs are collected in the Appendix.

\subsection*{Notation and definitions}

We use the following notation throughout. The spaces of c\`{a}dl\`{a}g
functions $[0,1]\rightarrow \mathbb{R}^{n}$, $[0,1]\rightarrow \mathbb{R}%
^{m\times n}$ and $\mathbb{R}\rightarrow \mathbb{R}$, all equipped with the
respective Skorokhod $J_{1}$-topologies, are denoted by $\mathscr{D}{}_{n}$, 
$\mathscr{D}{}_{m\times n}$ and $\mathscr{D}(\mathbb{R})$, respectively; see
Kallenberg (1997, Appendix A2). For $n=1$, the subscript in $\mathscr{D}{}%
_{n}$ is suppressed. $\mathscr{C}_{n}(\mathbb{R}^{n})$ is the space of
continuous functions from $\mathbb{R}^{n}$ to $\mathbb{R}^{n}$ equipped with
the topology of uniform convergence on compacts. Integrals are over $[0,1]$
unless otherwise stated, $\Phi $ is the standard Gaussian cdf, $U_{[0,1]}$
is the uniform distribution on $[0,1]$ and $\mathbb{I}_{\{\cdot \}}$ is the
indicator function. If $F$ is a cdf, possibly random, $F^{-1}$ stands for
the right-continuous generalized inverse, i.e., $F^{-1}(u):=\sup \{v\in 
\mathbb{R}:F\left( v\right) \leq u\}$, $u\in \mathbb{R}$. Unless differently
specified, limits are for $n\rightarrow \infty $.

With $(Z_{n},Y_{n})$ and $(Z,Y)$ being random elements of the metric spaces $%
S_{Z}\times S_{Y_{n}}$ and $S_{Z}\times S_{Y}$ ($n\in \mathbb{N}$), and
defined on a common probability space, we denote by `$Z_{n}|Y_{n}\overset{w}{%
\rightarrow }_{p}Z|Y$' (resp. `$Z_{n}|Y_{n}\overset{w}{\rightarrow }%
_{a.s.}Z|Y$') the fact that $E\left\{ g\left( Z_{n}\right) |Y_{n}\right\} {%
\rightarrow }E\left\{ g\left( Z\right) |Y\right\} $ in probability (resp.
a.s.) for all bounded continuous functions $g:S_{Z}\rightarrow \mathbb{R}$.
When $Z_{n}$ is a bootstrap statistic and $Y_{n}$ denotes the original data,
we write `$Z_{n}\overset{w^{\ast }}{\rightarrow }_{p}Z|Y$' (resp. `$Z_{n}%
\overset{w^{\ast }}{\rightarrow }_{a.s.}Z|Y$'). Finally, with $(Z_{n},Y_{n})$
and $(Z,Y)$ possibly defined on different probability spaces, `$Z_{n}|Y_{n}%
\overset{w}{\rightarrow }_{w}Z|Y$' means that $E(g(Z_{n})|Y_{n})\overset{w}{%
\rightarrow }E(g(Z)|Y)$ for all bounded continuous functions $%
g:S_{Z}\rightarrow \mathbb{R}$, see Kallenberg (1997, 2017); we label this
fact `weak convergence in distribution'. For the special case of scalar
random variables $Z_{n}$ and $Z$, if the conditional distribution $Z|Y$ is
diffuse (non-atomic), weak convergence in distribution is equivalent to the
following weak convergence in $\mathscr{D}(\mathbb{R})$: 
\begin{equation}
F_{n}(\cdot |Y_{n}):=P(Z_{n}\leq \cdot |Y_{n})\overset{w}{\rightarrow }%
P(Z\leq \cdot |Y)=:F(\cdot |Y).
\label{eq weak convergence of the conditional cdf}
\end{equation}%
When $Z_{n}$ is a bootstrap statistic and conditioning is on the original
data, we use the notation `$\overset{w^{\ast }}{\rightarrow }_{w}$'. For
multivariate generalizations we refer to Cavaliere and Georgiev (2020,
Appendix A).

\section{Preview of the results in a location model}

\label{sec example}

To illustrate the main arguments that will be proposed in the predictive
regression framework later, consider as in Andrews (2000) and Cavaliere et
al. (2017) the location model 
\begin{equation*}
y_{t}=\theta+\varepsilon_{t}\text{ (}t=1,...,n\text{)}
\end{equation*}
where the $\varepsilon_{t}$'s are i.i.d.$\left(0,1\right)$ and the parameter
space is $\Theta:=\{\theta\in\mathbb{R}:\theta\geq0\}$. Interest is in
inference on the true value $\theta_{0}$ of $\theta$ by using the Gaussian
QMLE, $\hat{\theta}$. With $l_{n}\left(\theta\right):=-\frac{1}{2}%
\sum_{t=1}^{n}(y_{t}-\theta)^{2}$, we find $\hat{\theta}:=\arg\max_{\theta%
\in\Theta}l_{n}\left(\theta\right)=\max\{0,\bar{y}_{n}\}$, $\bar{y}%
_{n}:=n^{-1}\sum_{t=1}^{n}y_{t}$. If $\theta_{0}$ is an interior point of $%
\Theta$, i.e. $\theta_{0}>0$, then $n^{1/2}(\hat{\theta}-\theta_{0})\overset{%
w}{\rightarrow}\xi$, $\xi\sim N\left(0,1\right)$. In contrast, if $%
\theta_{0} $ is on the boundary of $\Theta$, i.e. $\theta_{0}=0$, the
asymptotic distribution of $\hat{\theta}$ is 
\begin{equation}
n^{1/2}(\hat{\theta}-\theta_{0})=n^{1/2}\hat{\theta}\overset{w}{\rightarrow}%
\ell:=\max\{0,\xi\}  \label{eq loc model asy distr}
\end{equation}
again with $\xi\sim N\left(0,1\right)$.

The first takeaway of this section is the fact that the location of a
parameter on the boundary of the parameter space may induce limiting
bootstrap randomness of a kind that invalidates bootstrap inference. To see
this, consider in the context of the location model a standard Gaussian
parametric bootstrap based on the bootstrap sample 
\begin{equation*}
y_{t}^{\ast }=\hat{\theta}+\varepsilon _{t}^{\ast },
\end{equation*}%
where the $\varepsilon _{t}^{\ast }$'s are i.i.d.$N\left( 0,1\right) $
independent of the original data. The bootstrap counterpart of $l_{n}\left(
\theta \right) $ is $l_{n}^{\ast }\left( \theta \right) :=-\frac{1}{2}%
\sum_{t=1}^{n}(y_{t}^{\ast }-\theta )^{2}$, and the usual bootstrap QMLE is $%
\hat{\theta}^{\ast }:=\arg \max_{\theta \in \Theta }l_{n}^{\ast }\left(
\theta \right) =\max \{0,\bar{y}_{n}^{\ast }\}$, $\bar{y}_{n}^{\ast }:=\hat{%
\theta}+\bar{\varepsilon}_{n}^{\ast }$, $\bar{\varepsilon}_{n}^{\ast
}:=n^{-1}\sum_{t=1}^{n}\varepsilon _{t}^{\ast }$. Conditionally on the
original sample, $\hat{\theta}^{\ast }$'s exact distribution is 
\begin{equation}
n^{1/2}(\hat{\theta}^{\ast }-\hat{\theta})=n^{1/2}\max \{-\hat{\theta},\bar{%
\varepsilon}_{n}^{\ast }\}\sim \max \{-n^{1/2}\hat{\theta},\xi ^{\ast }\}|%
\hat{\theta}\text{, }\xi ^{\ast }|\hat{\theta}\sim \xi \sim N\left(
0,1\right) ,  \label{eq loc boot}
\end{equation}%
with associated conditional cdf given by 
\begin{equation}
P^{\ast }(n^{1/2}(\hat{\theta}^{\ast }-\hat{\theta})\leq x)=\Phi \left(
x\right) \mathbb{I}_{\{x\geq -n^{1/2}\hat{\theta}\}}\text{, }x\in \mathbb{R}%
\text{.}  \label{eq bootstrap cdf loc mod}
\end{equation}%
Now, when $\theta _{0}$ is an interior point of $\Theta $, $-n^{1/2}\hat{%
\theta}$ diverges to $-\infty $ in probability and the distribution of $%
n^{1/2}(\hat{\theta}^{\ast }-\hat{\theta})$ given the data converges weakly
in probability to the non-random distribution of $\xi ^{\ast }$; the
bootstrap therefore mimics the $N\left( 0,1\right) $ asymptotic distribution
of the original statistic, the bootstrap distributional approximation is
consistent and bootstrap inference is valid in the sense of (\ref{eq
unconditional validity}), with $\limfunc{int}$$C=(0,1)$. Conversely, when $%
\theta _{0}$ is on the boundary of the parameter space, the cdf in (\ref{eq
bootstrap cdf loc mod}) converges weakly in $\mathscr{D}(\mathbb{R})$ to the
random cdf $\Phi \left( x\right) \mathbb{I}_{\{x\geq -\ell \}}$. In terms of
weak convergence in distribution, 
\begin{equation}
n^{1/2}(\hat{\theta}^{\ast }-\hat{\theta})\overset{w^{\ast }}{\rightarrow }%
_{w}\ell ^{\ast }|\ell \text{, }\ell ^{\ast }:=\max \{-\ell ,\xi ^{\ast }\}%
\text{,}  \label{eq:sy loc}
\end{equation}%
where $\ell $ is distributed as in (\ref{eq loc model asy distr}) and is
independent of $\xi ^{\ast }$. The limit distribution in (\ref{eq:sy loc})
is random, since its cdf is a stochastic process depending on the
conditioning random variable $\ell $. Thus, it is distinct from the limit
distribution in (\ref{eq loc model asy distr}), which is unconditional and
hence characterized by a non-random cdf. Because the bootstrap limit
distribution is random, the bootstrap approximation is not consistent for
the limit in (\ref{eq loc model asy distr}).

As we shall see in Section \ref{sec par on the boundary main}, limiting
bootstrap randomness could be of two kinds: `benign', thus not compromising
the validity of bootstrap inference in the sense of (\ref{eq unconditional
validity}), or `malignant', thus invalidating bootstrap inference. In this
example, a bootstrap test employing a bootstrap statistic $\text{$\tau
_{n}^{\ast }:=$}\phi (n^{1/2}(\hat{\theta}^{\ast }-\hat{\theta}))$ as the
analogue of a statistic $\tau _{n}:=$$\phi (n^{1/2}\hat{\theta})$, where $%
\phi $ is a real function, may not be valid in the sense of (\ref{eq
unconditional validity}) under the null hypothesis $\mathsf{H}_{0}:\theta
_{0}=0$ even if the function $\phi $ is continuous, thus implying
`malignant' randomness.

To get some further insight into the source of limiting bootstrap
randomness, which will be exploited in the next sections, it is useful to
notice that the asymptotic distributions in (\ref{eq loc model asy distr})
and (\ref{eq loc boot}) can be written as 
\begin{eqnarray*}
\ell &=&\max \{0,\xi \}=\arg \min\nolimits_{\lambda \in \Lambda }|\lambda
-\xi |\text{, }\Lambda :=\{\lambda \in \mathbb{R}:\lambda \geq 0\} \\
\ell ^{\ast }|\ell &=&\max \{-\ell ,\xi ^{\ast }\}|\ell =\left( \arg
\min\nolimits_{\lambda \in \Lambda (\ell )}|\lambda -\xi ^{\ast }|\right) %
\big|\ell \text{, }\Lambda (\ell ):=\{\lambda \in \mathbb{R}:\lambda \geq
-\ell \}=\Lambda -\ell \text{,}
\end{eqnarray*}%
respectively. Hence, bootstrap randomness, and the implied bootstrap
invalidity, can be attributed to the fact that in the bootstrap world the
limit constraint set for the objective function $|\lambda -\xi ^{\ast }|$ is
the \emph{random} half line $\Lambda (\ell )$ rather than the original fixed
half line $\Lambda =\Lambda (0)$. That is, the chosen bootstrap scheme
shifts the constraint set by the random variable $-\ell $, which is non-zero
with probability $1/2$.

The second takeaway of this section is the fact that bootstrap validity
could be restored by offsetting properly the previous shift of the limit
constraint set. Specifically, this requires an ad hoc construction of a
bootstrap parameter space intended to approximate well the mutual position
of the true parameter value and the boundary of the original parameter space.

Consider a bootstrap scheme where the boundary of the bootstrap parameter
space $\Theta ^{\ast }$ is chosen in a data-driven way such that the mutual
position of $\theta _{0}$ and the boundary of $\Theta $ is well approximated
irrespective of whether $\theta _{0}$ belongs to $\partial \Theta $ or not.
To this aim, introduce the half line $\Theta ^{\ast }:=\{\theta :\theta \geq
g^{\ast }(\hat{\theta})\}$, where $g^{\ast }(\theta ):=\theta -|\theta
|^{1+\kappa }$, $\kappa >0$, and the associated $\hat{\theta}^{\ast }:=\arg
\max_{\theta \in \Theta ^{\ast }}l_{n}^{\ast }\left( \theta \right) =\max
\{g^{\ast }(\hat{\theta}),\bar{y}_{n}^{\ast }\}$. The bootstrap QMLE
statistic is then given by $n^{1/2}(\hat{\theta}^{\ast }-\hat{\theta}%
)=n^{1/2}\max \{g^{\ast }(\hat{\theta})-\hat{\theta},\bar{\varepsilon}%
_{n}^{\ast }\}$. Conditionally on the data, it is distributed as $\max
\{n^{1/2}(g^{\ast }(\hat{\theta})-\hat{\theta}),\xi ^{\ast }\}|\hat{\theta}$%
, with $\xi ^{\ast }|\hat{\theta}\sim N(0,1)$. If $\theta _{0}=0$, it then
follows that $n^{1/2}(g^{\ast }(\hat{\theta})-\hat{\theta})=-n^{1/2}\hat{%
\theta}^{1+\kappa }\overset{p}{\rightarrow }0$, and the bootstrap statistic
conditionally on the data converges weakly in probability to $\ell$ of (\ref%
{eq loc model asy distr}). Conversely, if $\theta _{0}>0$ then $%
n^{1/2}(g^{\ast }(\hat{\theta})-\hat{\theta})=-n^{1/2}\hat{\theta}^{1+\kappa
}\overset{p}{\rightarrow }-\infty $ and the bootstrap statistic
conditionally on the data converges weakly in probability to the $N\left(
0,1\right) $ distribution. In both cases, the bootstrap mimics the
asymptotic distribution of $n^{1/2}(\hat{\theta}-\theta _{0})$ and bootstrap
validity in the sense of (\ref{eq unconditional validity}) can be seen to be
successfully restored.\medskip {}

\noindent \textsc{Remark}. In the location model, an appropriate choice of $%
\Theta ^{\ast }$ simultaneously restores bootstrap validity and removes all
the randomness from the limit bootstrap distribution. In the predictive
regression framework we shall conclude that, in order to achieve bootstrap
validity, it is essential to remove only the portion of limiting bootstrap
randomness that is due to the location of the parameter vector on the
boundary of the parameter space. As no other sources of limiting bootstrap
randomness exist in the context of the location model, in this section the
previous conclusion simplifies to eliminating all the limiting bootstrap
randomness.$\hfill \square $\medskip {}

Before moving on to predictive regressions, we notice that when a test of $%
\mathsf{H}_{0}:\theta _{0}=0$ against $\mathsf{H}_{1}:\theta _{0}>0$ is
performed, employing $\tau _{n}^{\ast }:=n^{1/2}(\hat{\theta}^{\ast }-\hat{%
\theta})$ as the bootstrap analogue of $\tau _{n}:=n^{1/2}\hat{\theta}$, the
standard parametric bootstrap with $\Theta ^{\ast }=\Theta $ is valid in the
sense of (\ref{eq unconditional validity}); see also Andrews (2000).
Specifically, the bootstrap test rejects $\mathsf{H}_{0}$ when the bootstrap 
\emph{p}-value $\tilde{p}_{n}^{\ast }=1-p_{n}^{\ast }$ is small, with the
following convergence satisfied under the null hypothesis: 
\begin{equation*}
p_{n}^{\ast }=P^{\ast }(\tau _{n}^{\ast }\leq \tau _{n})=P^{\ast }(\xi
^{\ast }\leq \tau _{n})\overset{w}{\rightarrow }\Phi (\ell ).
\end{equation*}%
A similar convergence is satisfied by the \emph{p}-value $\tilde{p}%
_{n}=1-p_{n}$ of the asymptotic test, with

\begin{equation*}
p_{n}=P(\ell\leq u)|_{u=\tau_n}=\tfrac{1}{2}\mathbb{I}_{\{\tau_n=0\}}+\Phi(%
\tau_n)\mathbb{I}_{\{\tau_n>0\}}=\Phi(\tau_n)\overset{w}{\rightarrow}%
\Phi(\ell).
\end{equation*}
As $\ell $ is distributed like $\Phi ^{-1}(U)\mathbb{I}_{\{U>1/2\}}$, $U\sim
U_{[0,1]}$, it follows that $\Phi(\ell)$ is distributed like $\Phi (\Phi
^{-1}(U)\mathbb{I}_{\{U>1/2\}})$. As a result, both the bootstrap and the
asymptotic test are correctly sized for nominal levels below $1/2$. This
phenomenon, whose extensions to predictive regression are discussed in
Section \ref{sec one sided tests}, does not generalize to hypotheses where
one-sided tests are not appropriate or straightforward. Therefore, a remedy
is necessary for the inference-invalidating limiting bootstrap randomness
induced by the location of a parameter on the boundary.

\section{The predictive regression setup}

\label{sec par on the boundary setup}

Consider the following predictive regression in a triangular array setup: 
\begin{equation}
y_{t}=\theta _{1}+\theta _{2}x_{n,t-1}+\varepsilon _{t},\quad \text{ (}%
t=1,...,n;\text{ }n=1,2,...\text{)},  \label{eq:prr}
\end{equation}%
where $\varepsilon _{t}$ is a martingale difference sequence {[}mds{]} and $%
x_{n,t}$ is a non-stationary posited predicting variable satisfying the
following assumption; see, e.g. M\"{u}ller and Watson (2008) for references
to primitive conditions.

\begin{exercise}
Let $z_{n,t}:=n^{-1/2}\sum_{s=1}^t \varepsilon_s$. Then:

\begin{itemize}
\item[(a)] $\{\varepsilon _{t}\}$ is an mds w.r.t. some filtration to which $%
(x_{n,t},z_{n,t})$ is adapted, with $E\varepsilon _{t}^{2}=\omega _{zz}\in
(0,\infty ).$

\item[(b)] a law of large numbers holds as $n \rightarrow \infty$:%
\begin{equation*}
\sum_{t=1}^{n}\left( 
\begin{array}{c}
\Delta x_{n,t} \\ 
\Delta z_{n,t}%
\end{array}%
\right) \left( 
\begin{array}{cc}
\Delta x_{n,t} & \Delta z _{n,t}%
\end{array}%
\right) \overset{p}{\rightarrow }\Omega: =\left( 
\begin{array}{cc}
\omega _{x x } & \omega _{x z } \\ 
\omega _{x z } & \omega _{z z }%
\end{array}%
\right)>0.
\end{equation*}

\item[(c)] an invariance principle holds in $\mathscr{D}_2$ as $n
\rightarrow \infty$: 
\begin{equation*}
(x_{n,\left\lfloor n\cdot\right\rfloor }, z_{n,\left\lfloor
n\cdot\right\rfloor })^{\prime}\overset{w}{\rightarrow}(X,Z)^{\prime}\sim
BM(0,\Omega),
\end{equation*}
a bivariate Brownian motion on $[0,1]$.
\end{itemize}
\end{exercise}

Assumption 1 covers the specification $x_{n,t}=n^{-1/2}x_{t}$ for an $I(1)$
process $x_{t}$ driven by an mds that could be contemporaneously correlated
with $\varepsilon _{t}$.\footnote{%
As the bootstrap p-values discussed in the paper are invariant to rescaling
of the regressor, the normalization of $x_{t}$ by $n^{-1/2}$ has no
practical implication. It is equivalent to specifying a local-to-zero
regression coefficient, as is frequent in applications where $y_{t}$ is a
financial return and $x_{t}$ is non-stationary.}$^{,}$\footnote{%
Results under two alternative stochastic specifications of $x_{n,t}$, as a
near-unit root and as a stationary process, are given in the accompanying
supplement, Section \ref{sec more DGPs}.{}}

Assumption 1 implies that $\sum_{t=1}^{n}x_{n,t-1}\Delta z_{n,t}\overset{w}{%
\rightarrow }\int XdZ$, which need not have a mixed Gaussian distribution
because $X$ and $Z$ need not be independent. Nevertheless, it holds that $%
\sum_{t=1}^{n}x_{n,t-1}(\Delta z_{n,t}-\omega _{xz}\omega _{xx}^{-1}\Delta
x_{n,t})\overset{w}{\rightarrow }\int Xd(Z-\omega _{xz}\omega _{xx}^{-1}X)$,
which is zero-mean mixed Gaussian with conditional variance $\sigma
_{e}^{2}\int X^{2}$, where $\sigma _{e}^{2}:=\omega _{zz}-\omega
_{xz}^{2}\omega _{xx}^{-1}$ is the variance of $\varepsilon _{t}$ corrected
for $\Delta x_{n,t}$. The bootstrap schemes discussed below all rely on the
independence of the processes $X$ and $Z-\omega _{xz}\omega _{xx}^{-1}X$.

Further, Assumption 1 imposes unconditional homoskedasticity for simplicity.
As all the bootstrap schemes below are based on `wild' bootstrap schemes,
unconditional heteroskedasticity can be accommodated at only a notational
cost.

The next assumption specifies the parameter space, say $\Theta$, by means of
a smooth inequality constraint.

\begin{exercise}
The parameter space is $\Theta:=\{\theta=(\theta_{1},\theta_{2})^{\prime}\in%
\mathbb{R}^{2}:g(\theta)\geq0\}$, with non-empty boundary $%
\partial\Theta:=\{\theta\in\mathbb{R}^{2}:g(\theta)=0\}$, where $g:$ $%
\mathbb{R}^{2}\rightarrow\mathbb{R}$ is continuously differentiable in some
neighborhood of the true parameter value $\theta_{0}:=(\theta_{1,0},%
\theta_{2,0})^{\prime}$ with gradient $\tfrac{\partial}{\partial\theta^{%
\prime}}g(\theta)\neq0$ in that neighborhood.
\end{exercise}

In the following, $\dot{g}$ will denote the gradient of the function $g$
evaluated at $\theta_0$.

Assumption 2 generalizes the leading example of the parameter space $\Theta=%
\mathbb{R}\times\lbrack0,\infty)$ obtained by setting $g(\theta)=(0,1)%
\theta=\theta_{2}$. The boundary of $\Theta$ then corresponds to the case $%
\theta_{2}=0$ of no predictability of $y_{t}$ by $x_{n,t-1}$ whereas the
interior of $\Theta$ corresponds to the case of sign-restricted
predictability.

Interest is in bootstrap inference on a null hypothesis $\mathsf{H}_{0}$
identifying a set of parameter values that has a non-empty intersection with
the boundary of the parameter space. In particular, we consider the
following mutual positions of the boundary, the parameter set identified by $%
\mathsf{H}_{0}$ and the true value $\theta_{0}$:

\begin{description}
\item[$\mathscr{G}{}_{1}$.] $\mathsf{H}_{0}$ is the hypothesis that $%
\theta_{0}$ belongs to the boundary: $\mathsf{H}_{0}:g(\theta_{0})=0$;

\item[$\mathscr{G}{}_{2}$.] $\mathsf{H}_{0}$ is a simple null hypothesis on
the boundary: $\mathsf{H}_{0}:\theta_{0}=\bar{\theta}$, $g(\bar{\theta})=0$;

\item[$\mathscr{G}{}_{3}$.] $\mathsf{H}_{0}:h(\theta_{0})=0$, where $%
\{\theta\in\mathbb{R}^{2}:h\left(\theta\right)=0\}$ is not a subset of the
boundary $\partial\Theta$, but meets $\partial\Theta$ at a singleton set.
\end{description}

\noindent For example, let again $g(\theta)=\theta_{2}$, such that the
parameter space is $\mathbb{R}\times\lbrack0,\infty)$ with boundary $%
\partial\Theta=\mathbb{R}\times\{0\}$. Then the hypothesis of no
predictability $\mathsf{H}_{0}:\theta_{2,0}=0$ falls under $\mathscr{G}{}%
_{1} $. The hypothesis $\mathsf{H}_{0}:\theta_{0}=\bar{\theta}%
=(0,0)^{\prime} $ that $y_{t}$ is unpredictable with zero mean falls under $%
\mathscr{G}{}_{2} $. Finally, the hypothesis $\mathsf{H}_{0}:(1,1)%
\theta_{0}=\theta_{1,0}+\theta_{2,0}=0$ falls under $\mathscr{G}{}_{3}$ by
setting $h\left(\theta\right):=(1,1)\theta$; in this case, the intersection
point of the boundary and the parameter set identified by $\mathsf{H}_{0}$
is $(0,0)^{\prime}$ which might, but need not, be the true value under $%
\mathsf{H}_{0}$.

\subsection{Asymptotic distributions}

\label{subsec:Asymptotic-distributions}

Let $\hat{\theta}$ be the OLS estimator of $(\theta _{1},\theta
_{2})^{\prime }$ in the equation 
\begin{equation}
y_{t}=\theta _{1}+\theta _{2}x_{n,t-1}+\delta \Delta x_{n,t}+e_{t}
\label{eq:pr}
\end{equation}%
subject to the constraint $\hat{\theta}\in \Theta $, i.e. $g(\hat{\theta}%
)\geq 0$, and where the role of the regressor $\Delta x_{n,t}$ is to ensure
that the residuals are asymptotically uncorrelated with the innovations
driving $x_{n,t}$, a convenient prerequisite for the bootstrap
implementations. The existence, with probability approaching one, of a
measurable minimizer of the residual sum of squares (\ref{eq:pr}) over the
set $\Theta $ can be established in a similar but simpler way than that of
its bootstrap counterpart in our detailed proof of Theorem \ref{Lemma
bootstrap with boundary}. Moreover, any two such minimizers are first-order
asymptotically equivalent, explaining our usage of `the' associated with the
constrained OLS estimator. Specifically, any such minimizer $\hat{\theta}$
satisfies $n^{1/2}(\hat{\theta}-\theta _{0})\overset{w}{\rightarrow }\ell
(\theta _{0})$, with $\ell (\theta _{0})$ depending on the position of $%
\theta _{0}$ relative to the boundary $\partial \Theta $. Thus, $\ell
(\theta _{0})=\tilde{\ell}:=M^{-1/2}\xi $ if $\theta _{0}\in $$\limfunc{int}$%
$\Theta :=\Theta \setminus \partial \Theta $, where $M:=\int \tilde{X}\tilde{%
X}^{\prime }$, $\tilde{X}:=(1,X)^{\prime }$, $\xi \sim N\left( 0,\sigma
_{e}^{2}I_{2}\right) $ is independent of $X$, and $\sigma _{e}^{2}>0$ is the
variance of $\varepsilon _{t}$ corrected for $\Delta x_{n,t}$, whereas 
\begin{equation}
n^{1/2}(\hat{\theta}-\theta _{0})\overset{w}{\rightarrow }\ell (\theta
_{0})=\ell :=\underset{\lambda \in \Lambda }{\arg \min }||\lambda
-M^{-1/2}\xi ||_{M}\text{, }\Lambda :=\{\lambda \in \mathbb{R}^{2}:\dot{g}%
^{\prime }\lambda \geq 0\}  \label{eq asy distribution}
\end{equation}%
if $g(\theta _{0})=0$, with $||x||_{M}:=(x^{\prime }Mx)^{1/2}$ for $x\in 
\mathbb{R}^{2}$; see Section 12 in the working paper version of Andrews
(1999) or the proof of Theorem \ref{Lemma bootstrap with boundary} for the
bootstrap counterpart.

The previous asymptotic result is sufficient in order to see that the
possibility of having $\theta_{0}$ at the boundary of the parameter space $%
\Theta$ induces a dichotomy in the limit distribution of $n^{1/2}(\hat{\theta%
}-\theta_{0})$ similar to the dichotomy established in the introductory
location-model example. Replicating the constraint set in the limit
distribution by means of a bootstrap scheme will be our main concern in what
follows.

\subsection{Standard bootstrap invalidity}

\label{subsec:Standard-bootstrap-invalidity}

Consider first a fixed-regressor wild bootstrap sample generated as 
\begin{equation}
y_{t}^{\ast }=\hat{\theta}_{1}+\hat{\theta}_{2}x_{n,t-1}+\varepsilon
_{t}^{\ast }\text{,}  \label{eq BS for PR}
\end{equation}%
where $\varepsilon _{t}^{\ast }=\hat{e}_{t}w_{t}^{\ast }$, $t=1,...,n$, with 
$\hat{e}_{t}$ the residuals of (\ref{eq:pr}) and $w_{t}^{\ast }$ i.i.d. $%
N(0,1)$, independent of the original data.\footnote{%
{}{}The conclusions do not change if another zero-mean unit-variance
distribution with a finite fourth moment is used instead of the standard
Gaussian distribution.} Then the distribution of $n^{1/2}(\hat{\theta}%
-\theta _{0})$ could be tentatively approximated by the distribution of $%
n^{1/2}(\hat{\theta}^{\ast }-\hat{\theta})$ conditional on the original
data, where $\hat{\theta}^{\ast }$ is obtained by regressing $y_{t}^{\ast }$
on $(1,x_{n,t-1})^{\prime }$ under the constraint $\hat{\theta}^{\ast }\in
\Theta ^{\ast }=\Theta $, i.e., $g(\hat{\theta}^{\ast })\geq 0$ as for the
original estimator; see Andrews (2000)\footnote{%
Note that the term $\Delta x_{n,t}$ is no longer necessary because $%
x_{n,t-1} $ and $\varepsilon _{t}^{\ast }$ are independent conditionally on
the data.}.

To motivate the analysis in the next section, it is useful to anticipate
some asymptotic properties of $\hat{\theta}^{\ast}$ which obtain by
specializing Theorem \ref{Lemma bootstrap with boundary} below to the
fixed-regressor wild bootstrap scheme. For $\theta_{0}\in$$\limfunc{int}$$%
\Theta$, it turns out that the bootstrap distribution converges to a
conditional version of the limit distribution of $n^{1/2}(\hat{\theta}%
-\theta_{0})$ found earlier: 
\begin{equation}
n^{1/2}(\hat{\theta}^{\ast}-\hat{\theta})=n^{1/2}(\tilde{\theta}^{\ast}-\hat{%
\theta})+o_{p}(1)\overset{w^{\ast}}{\rightarrow}_{w}\tilde{\ell}|M\text{, }
\label{eq random limit depending on M only}
\end{equation}
where $\tilde{\theta}^{\ast}$ denotes the unconstrained OLS estimator from
the bootstrap sample. The limit bootstrap distribution is, therefore,
random. The vehicle of limiting bootstrap randomness is the random matrix $M$%
, such that limiting bootstrap randomness is fully attributable to the
stochastic properties of the regressor. Due to the fact that the bootstrap
replicates a conditional version of the limit distribution of the original
estimator $\hat{\theta}$, bootstrap inference is not invalidated. Rigorous
statements in this sense will be provided in Corollary \ref{corollary
bootstrap with boundary}.

On the other hand, if $\theta _{0}\in \partial \Theta $ the bootstrap
statistic converges as follows: 
\begin{eqnarray}
&&n^{1/2}(\hat{\theta}^{\ast }-\hat{\theta})\overset{w^{\ast }}{\rightarrow }%
_{w}\ell ^{\ast }|(M,\ell )
\label{eq asy for standard BS with param on the boundary} \\
&&\ell ^{\ast }\overset{}{:=}\underset{\lambda \in \Lambda _{\ell }^{\ast }}{%
\arg \min }||\lambda -M^{-1/2}\xi ^{\ast }||_{M}\text{, }\Lambda _{\ell
}^{\ast }\overset{}{:=}\left. \{\lambda \in \mathbb{R}^{2}:\dot{g}^{\prime
}\lambda \overset{}{\geq }-\dot{g}^{\prime }\ell \}\right.  \notag
\end{eqnarray}%
where $\xi ^{\ast }\sim N(0,\sigma _{e}^{2}I_{2})$ is independent of $%
(M,\ell )$. In contrast with the case $\theta _{0}\in $$\limfunc{int}$$%
\Theta $ and additionally to the random matrix $M$, in (\ref{eq asy for
standard BS with param on the boundary}) also the random vector $\ell $
appears as a vehicle of limiting bootstrap randomness. Moreover, the limit
in (\ref{eq asy for standard BS with param on the boundary}) is not a
conditional version of the limit of $n^{1/2}(\hat{\theta}-\theta _{0})$,
inasmuch as $\Lambda _{\ell }^{\ast }$ in (\ref{eq asy for standard BS with
param on the boundary}) is a random half-plane, rather than the original
admissible set $\Lambda $ of (\ref{eq asy distribution}). The kind of
limiting bootstrap randomness introduced by $\ell $ is similar to the one
established in the introductory location model and, in general, it
invalidates bootstrap inference. The reason for the discrepancy between $%
\Lambda $ and $\Lambda _{\ell }^{\ast }$ is that the parameter space of the
standard fixed-regressor wild bootstrap does not approximate well the
original mutual position of the true value $\theta _{0}$ and the boundary,
unless $g(\hat{\theta})=0$. Other, non-standard bootstrap schemes may be
designed in order to provide better approximations, at least under the null
hypothesis. Under these schemes the possible boundary position of $\theta
_{0}$ is no longer a vehicle of limiting bootstrap randomness, while the
role of the random matrix $M$ in the limit bootstrap distribution is
maintained. This topic is analyzed in the next section.

\section{Asymptotically valid bootstrap schemes}

\label{sec par on the boundary main}

In order to unify the discussion of several bootstrap schemes for inference
on $\mathsf{H}_{0}$ under the three cases $\mathscr{G}{}_{1}$, $\mathscr{G}{}%
_{2}$ and $\mathscr{G}{}_{3}$, consider a bootstrap sample generated as in (%
\ref{eq BS for PR}) and, more generally than before, a bootstrap OLS
estimator $\hat{\theta}^{\ast }$ constrained to belong to a bootstrap
parameter space $\Theta^{\ast}$ satisfying the following assumption.

\begin{exercise}
\label{asstheta} The bootstrap parameter space is $\Theta ^{\ast }:=\{\theta
\in \mathbb{R}^{2}:g(\theta )\geq g^{\ast } (\hat{\theta})\}$ for some
function $g^{\ast }:\mathbb{R}^{2}\rightarrow \mathbb{R}$ which is
continuously differentiable in a neighborhood of $\theta _{0}$ and satisfies 
$g^{\ast }(\theta )\leq g(\theta )$ for $\theta \in \Theta $.
\end{exercise}

The standard bootstrap considered in Section \ref{sec par on the boundary
setup} obtains by setting $g^{\ast }=0$, such that $\Theta ^{\ast }=\Theta $%
, the original parameter space. Alternatively, setting $g^{\ast }=g$
restricts the bootstrap true value $\hat{\theta}$ to lie on the boundary of
the bootstrap parameter space $\Theta ^{\ast }$.\footnote{{\small {}}{}As $%
\hat{\theta}\overset{p}{\rightarrow }\theta _{0}$ and $\dot{g}(\theta
_{0})\neq 0$, it follows by continuity that $P(\dot{g}(\hat{\theta})\neq
0)\rightarrow 1$, such that, with probability approaching one, $\hat{\theta}$
is not a stationary point of $g$. In particular, with probability
approaching one, $\hat{\theta}$ is not a local minimizer of $g$, implying
that $\hat{\theta}\in \partial $$\Theta ^{\ast }$ under $\Theta ^{\ast
}=\{\theta \in R^{2}:g(\theta )\geq g(\hat{\theta})\}$.} Finally, setting $%
g^{\ast }=g-|g|^{1+\kappa }$ for some $\kappa >0$ introduces a correction,
in the spirit of an alternative to the standard bootstrap mentioned in
Andrews (2000, p.403, Method two), Fang and Santos (2019, Example 2.1) and
Cavaliere et al. (2022), where the bootstrap true value either shrinks to
the boundary of the bootstrap parameter space at a proper rate or remains
bounded away from this boundary, according to whether $\theta _{0}$ belongs
to the original boundary $\partial \Theta $ or not. Other choices of $%
g^{\ast }$ with the same implication are discussed in Sections \ref%
{subsec:Fang-and-Santos} and \ref{sec tuning par}.

To formulate the next theorem, recall $M$ and $\ell(\theta_{0})$ introduced
in Section \ref{subsec:Asymptotic-distributions}, and let $%
\xi^{\ast}|(M,\ell(\theta_{0}))\sim N(0,\sigma_e^2 I_2)$ as in Section \ref%
{subsec:Standard-bootstrap-invalidity}. Let also $D_{n}=\{y_{t},x_{n,t-1}%
\}_{t=1}^{n}$ denote the original data. Finally, call a convergence in
distribution $Z_{n}\overset{w}{\rightarrow}Z$ and a weak convergence of
random distributions $Z_{n}^{*}|D_{n}\overset{w}{\rightarrow}_{w}Z^{*}|Y$
joint, denoted as $(Z_{n},(Z_{n}^{*}|D_{n}))$$\overset{w}{\rightarrow}%
_{w}(Z,(Z^{*}|Y))$, if $(Z_{n},E\{g(Z_{n}^{*})|D_{n}\})\overset{w}{%
\rightarrow}(Z,E\{g(Z)|Y\})$ for all continuous and bounded real functions $%
g $ with matching domain.

\begin{theorem}
\label{Lemma bootstrap with boundary}Under a null hypothesis $\mathsf{H}_{0}$
as in $\mathscr{G}{}_{1}$--$\,\mathscr{G}{}_{3}$ and under Assumptions 1-3,
the bootstrap estimator $\hat{\theta}^{\ast}$ obtained by regressing $%
y_{t}^{\ast}$ of (\ref{eq BS for PR}) on $(1,x_{n,t-1})^{\prime}$ under the
constraint $\hat{\theta}^{\ast}\in\Theta^{\ast}$, satisfies 
\begin{equation*}
(n^{1/2}(\hat{\theta}-\theta_{0}),(n^{1/2}(\hat{\theta}^{\ast}-\hat{\theta}%
)|D_{n}))\overset{w}{\rightarrow}_{w}\left(\ell(\theta_{0}),(\ell^{\ast}(%
\theta_{0})|(M,\ell(\theta_{0})))\right)\text{, }
\label{eq asy distrib for general bootstraps}
\end{equation*}
where in the case $g^{\ast}(\theta_{0})<g(\theta_{0})$, 
\begin{equation}
\ell^{\ast}(\theta_{0})=\tilde{\ell}^{\ast}:=M^{-1/2}\xi^{\ast}\text{ with }%
\tilde{\ell}^{\ast}|(M,\ell(\theta_{0}))=\tilde{\ell}|M  \label{eq semno}
\end{equation}
in the sense of a.s. equality of conditional distributions, whereas in the
case $g^{\ast}(\theta_{0})=g(\theta_{0})$, 
\begin{equation}
\ell^{\ast}(\theta_{0})=\ell^{\ast}:=\underset{\lambda\in\Lambda_{\ell}^{%
\ast}}{\arg\min}||\lambda-M^{-1/2}\xi^{\ast}||_{M}\text{, }%
\Lambda_{\ell}^{\ast}:=\{\lambda\in\mathbb{R}^{2}:\dot{g}^{\prime}\lambda%
\geq(\dot{g}^{\ast}-\dot{g})^{\prime}\ell(\theta_{0})\}\text{. }
\label{eq amans}
\end{equation}
\end{theorem}

\noindent The following conclusions could be drawn.

\begin{itemize}
\item[(i)] Consider first configurations $\mathscr{G}{}_{1}$ and $%
\mathscr{G}{}_{2}$ under $\mathsf{H}_{0}$, such that $g(\theta _{0})=0$.
Consider the magnitude order, in probability, of the distance between the
bootstrap `true' value $\hat{\theta}$ and the bootstrap boundary $\partial
\Theta ^{\ast }$ as a measure of how precisely the bootstrap approximates
the geometry of $\mathscr{G}_{1}$ and $\mathscr{G}_{2}$. As seen previously,
the standard bootstrap corresponds to $g^{\ast }=0$ and approximates the
geometry up to an exact magnitude order of $n^{-1/2}$, resulting in a
situation where the belonging of $\theta _{0}$ to the boundary contributes
to the randomness of the limit bootstrap distribution given by (\ref{eq asy
for standard BS with param on the boundary}) and (\ref{eq amans}) via
conditioning on the r.v. $\ell (\theta _{0})=\ell $. Conversely, bootstrap
schemes employing $g^{\ast }(\theta _{0})=g(\theta _{0})$ and $\dot{g}^{\ast
}=\dot{g}$, such that the bootstrap boundary is tangent to the original
boundary at $\theta _{0}$, give rise to approximations of order $%
o_{p}(n^{-1/2})$ and all the randomness in the bootstrap limit is due to the
properties of the stochastic regressor via the random variable $M$, as now $%
\ell ^{\ast }|(M,\ell )\overset{}{=}\ell |M$ in the sense of a.s. equality
of random distributions; see (\ref{eq asy distribution}) and (\ref{eq amans}%
). Moreover, for such schemes the bootstrap mimics a conditional version of
the asymptotic distribution of the original estimator: $n^{1/2}(\hat{\theta}%
^{\ast }-\hat{\theta})\overset{w^{\ast }}{\rightarrow }_{w}\ell |M$.
Examples are the `restricted' bootstrap based on $g^{\ast }=g$, which
replicates the geometry of the original data under $\mathsf{H}_{0}$ by
putting $\hat{\theta}$ on the bootstrap boundary, and the choices $g^{\ast
}=g-|g|^{1+\kappa }$ for some $\kappa >0$. In general, the limit
distribution of the resulting bootstrap estimator is random, with randomness
depending on both the stochastic regressor and the position of $\theta _{0}$
relative to the boundary.

\item[(ii)] Consider now the case in $\mathscr{G}{}_{3}$, such that $%
g(\theta _{0})=0$ need not, but may hold under $\mathsf{H}_{0}$. Among the
bootstraps considered in (i), the standard one would fail to mimic a
conditional version of the original limit distribution if $g(\theta _{0})=0$%
, while the `restricted' one would fail if $g(\theta _{0})>0$. As an
alternative, consider the bootstrap based on $g^{\ast }=g-|g|^{1+\kappa }$
for some $\kappa >0$. If $\theta _{0}\in \partial \Theta $, then this choice
puts the bootstrap true value $\hat{\theta}$ at the asymptotically
negligible distance of $o_{p}(n^{-1/2})$ from the bootstrap boundary,
whereas if $\theta _{0}\in $$\limfunc{int}$$\Theta $, then $\hat{\theta}$ is
bounded away from the bootstrap boundary, in probability. This guarantees
bootstrap validity under some regularity conditions, see (ii)\ in Corollary %
\ref{corollary bootstrap with boundary} below.
\end{itemize}

\medskip{}

In general, bootstrap validity in the sense of (\ref{eq unconditional
validity}) can be evaluated through the following corollary of Theorem \ref%
{Lemma bootstrap with boundary} above.

\begin{corollary}
\label{corollary bootstrap with boundary}Under the assumptions of Theorem %
\ref{Lemma bootstrap with boundary}, a necessary and sufficient condition
for the convergence 
\begin{equation}
(n^{1/2}(\hat{\theta}-\theta_{0}),(n^{1/2}(\hat{\theta}^{\ast}-\hat{\theta}%
)|D_{n}))\overset{w}{\rightarrow}_{w}\left(\ell(\theta_{0}),(\ell(%
\theta_{0})|M)\right)\text{ }  \label{eq bound ave}
\end{equation}
is that: (i) under $\mathscr{G}{}_{1}$ and $\mathscr{G}{}_{2}$, $%
g(\theta_{0})=g^{\ast}(\theta_{0})$ and $\dot{g}=\dot{g}^{\ast}$; (ii) under 
$\mathscr{G}{}_{3}$, either $g(\theta_{0})=g^{\ast}(\theta_{0})$ and $\dot{g}%
=\dot{g}^{\ast}$, or $g(\theta_{0})>\max\{0,g^{\ast}(\theta_{0})\}$.

Moreover, under (\ref{eq bound ave}) the bootstrap is valid in the sense of (%
\ref{eq unconditional validity}) for any pair of statistics $\tau _{n},\tau
_{n}^{\ast }$ such that, under $\mathsf{H}_{0}$, $\tau _{n}=\phi (n^{1/2}(%
\hat{\theta}-\theta _{0}))+o_{p}(1)$ and $\tau _{n}^{\ast }=\phi (n^{1/2}(%
\hat{\theta}^{\ast }-\hat{\theta}))+o_{p}(1)$ for some continuous real
function $\phi $ such that $\phi (\ell (\theta _{0}))$ is well-defined a.s.
\end{corollary}

\noindent The class of functions $g^{\ast}=g-|g|^{1+\kappa}$ for $\kappa>0$
satisfies both conditions (i) and (ii) of Corollary \ref{corollary bootstrap
with boundary}; hence, the ensuing bootstrap inference is valid under all of 
$\mathscr{G}{}_{1}$-$\mathscr{G}{}_{3}$. In contrast, the standard bootstrap
violates condition (i) and, in general, is asymptotically invalid if $%
g(\theta_{0})=0$. An exception is when the discrepancy between the original
and the bootstrap geometry is offset by the use of a test statistic that
takes into account the geometric position of the null hypothesis in the
original parameter space. Section \ref{sec one sided tests} focuses on this
setup.

\bigskip{}

\noindent \textsc{Remark}. The practical implications of Corollary \ref%
{corollary bootstrap with boundary} depend on the choice of the statistic $%
\tau_{n}$ and the respective function $\phi$, which will typically be a
linear $\phi(l)=l^{\prime}\frac{\partial r}{\partial\theta^{\prime}}%
(\theta_{0})$ arising from the delta method, with $l\in\mathbb{R}^{2},\frac{%
\partial r}{\partial\theta^{\prime}}(\theta_{0})\neq0$. For instance, if $%
g(\theta_{0})=0$ and $\phi(\ell)$ depends on $\ell$ only through $\dot{g}%
^{\prime}\ell=\max\{0,\dot{g}^{\prime}M^{-1/2}\xi\}$, then the cdf of $%
\phi(\ell)$ will not be continuous. Still, the bootstrap will be valid in
the sense of (\ref{eq unconditional validity}), meaning that the largest
open subset of $[0,1]$ on which the bootstrap test is correctly sized as $%
n\rightarrow\infty$ coincides with the analogous set for the asymptotic
test. This set will be smaller than $(0,1)$, however. An example is $%
\tau_{n}=n^{1/2}g(\hat{\theta})$, $\tau_{n}^{\ast}=n^{1/2}(g(\hat{\theta}%
^{\ast})-g(\hat{\theta}))$ with $\phi(\ell)=\dot{g}^{\prime}\ell$,
corresponding to a right-sided test of $\mathsf{H}_{0}:g(\theta_{0})=0$. 
\medskip{}

\noindent \textsc{Remark}. Bootstrap validity extends readily to statistics
where\ $n^{1/2}(\hat{\theta}-\theta _{0})$ is normalized by some $\hat{\Sigma%
}=\Sigma (M_{n})+o_{p}(1)$ for a function $\Sigma :\mathbb{R}^{2\times
2}\rightarrow \mathbb{R}^{2\times 2}$ which is continuous on the set of
positive definite matrices. Specifically, bootstrap validity holds if, under 
$\mathsf{H}_{0}$, $\tau _{n}=\phi (n^{1/2}\hat{\Sigma}(\hat{\theta}-\theta
_{0}))+o_{p}(1)$ and $\tau _{n}^{\ast }=\phi (n^{1/2}\hat{\Sigma}(\hat{\theta%
}^{\ast }-\hat{\theta}))+o_{p}(1)$, where $\phi $ is a continuous real
function such that $\phi (\Sigma (M)\ell (\theta _{0}))$ is a.s.
well-defined.$\hfill \square $

\section{Discussion and extensions}

\label{sec:Discussion}

In this section we address the following three issues: (i) the validity of
one-sided bootstrap tests; (ii) a discussion of the bootstrap schemes from
Corollary \ref{corollary bootstrap with boundary} within the paradigm of
some previous works -- specifically, Fang and Santos (2019) and Hong and Li
(2020); and (iii) uniform bootstrap validity.

\subsection{Validity of one-sided standard bootstrap tests}

\label{sec one sided tests}

Under case $\mathscr{G}{}_{1}$, consider testing $\mathsf{H}_{0}:g(\theta
_{0})=0$ against the alternative $H_{1}:g(\theta _{0})>0$ using a one-sided
test and the standard bootstrap, i.e., with $g^{\ast }=0$. For a test
statistic of the form $\tau _{n}:=n^{1/2}g(\hat{\theta})$, a bootstrap
counterpart is given by $\tau _{n}^{\ast }:=n^{1/2}(g(\hat{\theta}^{\ast
})-g(\hat{\theta}))$ and the associated one-sided bootstrap test rejects for
large values of the bootstrap \emph{p}-value $p_{n}^{\ast }:=P^{\ast }(\tau
_{n}^{\ast }\leq \tau _{n})$; equivalently, for small values of $\tilde{p}%
_{n}^{\ast }:=1-p_{n}^{\ast }$. As for $\hat{\theta}^{\ast }$, also $\tau
_{n}^{\ast }$ is affected in the limit by extra randomness due to $\theta
_{0}$ being on the boundary. From (\ref{eq asy distrib for general
bootstraps}), which reduces to (\ref{eq asy distribution}) and (\ref{eq asy
for standard BS with param on the boundary}), it follows by the delta method
that 
\begin{equation}
(\tau _{n},(\tau _{n}^{\ast }|D_{n}))\overset{w}{\rightarrow }_{w}\left( 
\dot{g}^{\prime }\ell \text{,}\left( \dot{g}^{\prime }\ell ^{\ast }|(M,\ell
)\right) \right) =(\dot{g}^{\prime }\ell ,(\max \{-\dot{g}^{\prime }\ell ,%
\dot{g}^{\prime }\tilde{\ell}^{\ast }\}|(M,\ell ))),
\label{eq convergence after delta method}
\end{equation}%
with $\ell $, $\ell ^{\ast }$ and $\tilde{\ell}^{\ast }$ as previously
defined. For $\tau _{n}^{\ast }$, however, the randomness induced by
conditioning on $\ell $ affects the sample paths of the associated random
cdf on the negative half-line alone, because $\dot{g}^{\prime }\ell \geq 0$,
and is thus irrelevant for bootstrap tests with nominal size in $(0,\frac{1}{%
2})$. Put differently, the bootstrap \emph{p}-values $\tilde{p}_{n}^{\ast }$
are asymptotically uniformly distributed below $\frac{1}{2}$. This follows
rigorously from the next generalization of Theorem 3.1 in Cavaliere and
Georgiev (2020), the proof being analogous, where conditions for bootstrap
validity restricted to a subset of nominal testing levels are formulated.

\begin{theorem}
\label{th2}Let there exist a random variable $\tau $ and a random element $X$%
, both defined on the same probability space, such that the support of $\tau
_{n}$ is contained in a finite or infinite closed interval $T$, and $(\tau
_{n},F_{n}^{\ast })\overset{w}{\rightarrow }(\tau ,F)$ in $\mathbb{R}\times %
\mathscr{D}(T)$ for $F_{n}^{\ast }(u):=P(\tau _{n}^{\ast }\leq u|D_{n})$ and 
$F(u):=P(\tau \leq u|X)$, $u\in T$. If the possibly random cdf $F$ is
sample-path continuous on $T$, then the bootstrap p-value $p_{n}^{\ast
}:=F_{n}^{\ast }(\tau _{n})$\ satisfies 
\begin{equation*}
P(p_{n}^{\ast }\leq q)\rightarrow q
\end{equation*}%
for $q$ such that $q\in F(T)$ a.s.
\end{theorem}

\noindent By Theorem \ref{th2} with $T=[0,\infty)$, which corresponds to the
support of $\tau_{n}$ and $\tau:=\dot{g}^{\prime}\ell$, it follows that the
standard bootstrap applied to the one-sided statistic $\tau_{n}$ is
asymptotically correctly sized for nominal test sizes in $(0,\frac{1}{2})$.

\subsection{Fang and Santos (2019) and Hong and Li (2020)}

\label{subsec:Fang-and-Santos}

In this section we put the geometric considerations of Section \ref{sec par
on the boundary main} in the perspective of Fang and Santos (2019), and of
the numerical bootstrap of Hong and Li (2020). The discussion is often
specialized to the case of an affine constraint.

Consider the constrained OLS estimator $\hat{\theta}$ of Section \ref%
{subsec:Asymptotic-distributions}. Its limit distribution, see (\ref{eq asy
distribution}), is the distribution of $\ell (\theta _{0})=\varphi _{\theta
_{0}}(M^{-1/2}\xi )$ with 
\begin{equation*}
\varphi _{\theta _{0}}(u)=%
\begin{cases}
\begin{array}{c}
u \\ 
\dot{g}_{\perp }(\dot{g}_{\perp }^{\prime }M\dot{g}_{\perp })^{-1}\dot{g}%
_{\perp }^{\prime }Mu+M^{-1}\dot{g}(\dot{g}^{\prime }M^{-1}\dot{g})^{-1}\max
\{0,\dot{g}^{\prime }u\}%
\end{array}
& 
\begin{array}{c}
\text{if}\text{ $g(\theta _{0})>0$} \\ 
\text{if}\text{ $g(\theta _{0})=0$}%
\end{array}%
,%
\end{cases}%
\end{equation*}%
with $u\in \mathbb{R}^{2}$. By a projection identity, the expression in the
second line of the previous display collapses to $u$ whenever $\dot{g}%
^{\prime }u\geq 0$. Note that the distribution of $M^{-1/2}\xi $ conditional
on $M$ can be estimated consistently by the distribution of the
unconstrained bootstrap OLS estimator conditional on the data; that is, 
\begin{equation*}
n^{1/2}\tilde{(\theta ^{\ast }}-\hat{\theta})\overset{w^{\ast }}{\rightarrow 
}_{w}M^{-1/2}\xi |M\text{.}
\end{equation*}%
One can then ask what properties of an estimator $\hat{\varphi}_{n}$ of $%
\varphi _{\theta _{0}}$ are sufficient for $\hat{\varphi}_{n}(n^{1/2}\tilde{%
(\theta ^{\ast }}-\hat{\theta}))\overset{w^{\ast }}{\rightarrow }_{w}\varphi
_{\theta _{0}}(M^{-1/2}\xi )|M$ to hold. Fang and Santos (2019) address this
question in the setup of deterministic transformations of non-random limit
distributions, instead of the random transformation $\varphi _{\theta _{0}}$
of the random distribution $M^{-1/2}\xi |M$. Although not directly
applicable here, Theorem 3.2 of Fang and Santos (2019) provides the key
insight: there should be sufficient uniformity in the convergence of $\hat{%
\varphi}_{n}$ to $\varphi _{\theta _{0}}$. Consider for instance 
\begin{eqnarray}
\hat{\varphi}_{n}(u) &=&\hat{\dot{g}}_{\perp }(\hat{\dot{g}}_{\perp
}^{\prime }M_{n}\hat{\dot{g}}_{\perp })^{-1}\hat{\dot{g}}_{\perp }^{\prime
}M_{n}u  \label{eq:phiest} \\
&&+M_{n}^{-1}\hat{\dot{g}}(\hat{\dot{g}}^{\prime }M_{n}^{-1}\hat{\dot{g}}%
)^{-1}\max \{-n^{1/2}|g(\hat{\theta})|^{1+\kappa },\hat{\dot{g}}^{\prime }u\}%
\text{, }u\overset{}{\in }\mathbb{R}^{2}\text{,}  \notag
\end{eqnarray}%
where $\hat{\dot{g}}=\frac{\partial }{\partial \theta ^{\prime }}g(\hat{%
\theta})$, $M_{n}=n^{-1}\sum_{t=1}^{n}\tilde{x}_{t}\tilde{x}_{t}^{\prime }$
with $\tilde{x}_{t}=(1,x_{n,t-1})^{\prime }$, and $\kappa >0$. Given that $%
M_{n}\overset{w}{\rightarrow }M$, $\hat{\dot{g}}\overset{p}{\rightarrow }%
\dot{g}$ and $n^{1/2}|g(\hat{\theta})|^{1+\kappa }\overset{p}{\rightarrow }%
\infty \mathbb{I}_{\{g(\theta _{0})>0\}}$, it is easily checked that $\tilde{%
\varphi}_{n}\overset{w}{\rightarrow }\varphi _{\theta _{0}}$ on $\mathscr{C}%
_{2}(\mathbb{R}^{2})$, and the convergence of $\hat{\varphi}_{n}$ is joint
with that of $n^{1/2}(\hat{\theta}-\theta _{0})$ and $n^{1/2}\tilde{(\theta
^{\ast }}-\hat{\theta})$, the latter one given the data. These facts are
sufficient to ensure that 
\begin{equation*}
(n^{1/2}(\hat{\theta}-\theta _{0}),(\hat{\varphi}_{n}(n^{1/2}\tilde{(\theta
^{\ast }}-\hat{\theta}))|D_{n}))\overset{w}{\rightarrow }_{w}\left( \ell
(\theta _{0}),(\ell (\theta _{0})|M)\right) \text{ }
\end{equation*}%
on $\mathbb{R}^{4}$, essentially as a consequence of the continuous mapping
theorem (CMT) and the continuity of the evaluation map from $\mathscr{C}_{2}(%
\mathbb{R}^{2})\times \mathbb{R}^{2}$ to $\mathbb{R}^{2}$. As the previous
limit is the same as in Corollary \ref{corollary bootstrap with boundary},
it follows that bootstrap inference based on the distribution of $\hat{%
\varphi}_{n}(n^{1/2}\tilde{(\theta ^{\ast }}-\hat{\theta}))$ conditional on
the data is valid. Moreover, for the valid bootstrap schemes obtained from
Corollary \ref{corollary bootstrap with boundary} with $g^{\ast
}=g-|g|^{1+\kappa }$, $\kappa >0$, the bootstrap estimator $\hat{\theta}%
^{\ast }$ satisfies $n^{1/2}(\hat{\theta}^{\ast }-\hat{\theta})=\hat{\varphi}%
_{n}(n^{1/2}\tilde{(\theta ^{\ast }}-\hat{\theta}))$ for affine functions $g$%
. It can be concluded that $\hat{\varphi}_{n}$ of (\ref{eq:phiest})
implicitly performs the geometric approximation proposed in Section \ref{sec
par on the boundary main}, and so does any other estimator of $\varphi
_{\theta _{0}}$ that converges like $\hat{\varphi}_{n}$.

We now argue that such an estimator of $\varphi_{\theta_0}$ is embedded in
the numerical bootstrap of Hong and Li (2020). This ensures the validity of
the numerical bootstrap for the predictive regression of interest here,
though at the cost of a slower consistency rate of the bootstrap estimator
than in Corollary \ref{corollary bootstrap with boundary}. Let $%
s_{n}\rightarrow \infty $ be a sequence such that $n^{-1/2}s_{n}\rightarrow
0 $. Hong and Li (2020) propose in their eq. (4.9) a bootstrap estimator $%
\hat{\theta}_{nb}^{\ast }$ where the constraint set of our $\ell (\theta
_{0})$ (i.e., $\mathbb{R}^{2}$ if $\theta _{0}\in \mathrm{int}\Theta $ and
the half-plane $\text{$\Lambda $}$ if $\theta _{0}\in \partial \Theta $),
would be estimated by $\Lambda _{nb}^{\ast }=\{\lambda \in \mathbb{R}^{2}:g(%
\hat{\theta}+s_{n}^{-1}\lambda )\geq 0\}$, the implied bootstrap parameter
space being $\Theta _{nb}^{\ast }=\hat{\theta}+s_{n}^{-1}\Lambda _{nb}^{\ast
}=\Theta .$ The bootstrap estimator itself, adapted to our setup, could be
written as 
\begin{equation*}
\hat{\theta}_{nb}^{\ast }=\underset{g(\theta )\geq 0}{\arg \min }%
||s_{n}(\theta -\hat{\theta})-M_{n}^{-1/2}\xi _{n}^{\ast }||_{M_{n}},
\end{equation*}%
where $\xi _{n}^{\ast }$ is a bootstrap variable such that $\xi _{n}^{\ast }%
\overset{w^{\ast }}{\rightarrow }_{p}N(0,I_{2})$; e.g., $\xi _{n}^{\ast
}=n^{1/2}M_{n}^{1/2}(\tilde{\theta}^{\ast }-\hat{\theta})$. In the simple
case of an affine $g$ we find the explicit expression 
\begin{equation*}
s_{n}(\hat{\theta}_{nb}^{\ast }-\hat{\theta})=\bar{\varphi}%
_{n}(M_{n}^{-1/2}\xi _{n}^{\ast })
\end{equation*}%
for $\bar{\varphi}_{n}$ defined similarly to $\hat{\varphi}_n$, with the
only difference that in (\ref{eq:phiest}) the term $n^{1/2}|g(\hat{\theta}%
)|^{1+\kappa}$ is replaced by $s_{n}g(\hat{\theta})$. As $s_{n}g(\hat{\theta}%
)\overset{p}{\rightarrow}\infty\mathbb{I}_{\{g(\theta_0)>0\}}$ similarly to $%
n^{1/2}|g(\hat{\theta})|^{1+\kappa }$, $\kappa >0$, it follows that $\bar{%
\varphi}_{n}$ converges similarly to $\hat{\varphi}_n$. As a result, 
\begin{equation*}
(n^{1/2}(\hat{\theta}-\theta _{0}),(s_{n}(\hat{\theta}_{nb}^{\ast }-\hat{%
\theta})|D_{n}))\text{ }\overset{w}{\rightarrow }_{w}\left( \ell (\theta
_{0}),(\ell (\theta _{0})|M)\right) ,
\end{equation*}%
ensuring the validity of the numerical bootstrap, though the consistency
rate of $\hat{\theta}_{nb}^{\ast }$ is $s_{n}=o(n^{1/2})$ instead of $%
n^{1/2} $. In contrast, the rate of $n^{1/2}$ would be achieved by our
proposed bootstrap estimator, with $n^{1/2}(\hat{\theta}^{\ast }-\hat{\theta}%
)=\bar{\varphi}_{n}(M_{n}^{-1/2}\xi _{n}^{\ast })$, if $\Theta ^{\ast
}=\{\theta \in \mathbb{R}^{2}:g(\theta )\geq g_{n}^{\ast }(\hat{\theta})\}$
with $g_{n}^{\ast }=g-n^{-1/2}s_{n}|g|$ is specified in Assumption 3.

\subsection{Uniformity considerations}

In agreement with Chatterjee and Lahiri (2011), Remark 3, the focus in this
paper is on pointwise bootstrap validity. For situations where uniform
bootstrap validity is of interest, our key takeaways are similar to the
literature on non-random limiting bootstrap measures. First, for the null
hypothesis $\mathscr{G}{}_{1}$ that the true parameter value lies on the
boundary of the parameter space, the pointwise-valid bootstrap schemes
outlined in Corollary \ref{corollary bootstrap with boundary} display
asymptotic rejection probabilities matching the local power of the bootstrap
test whenever the true parameter value varies along a sequence that is local
to the boundary at the $n^{-1/2}$ rate. This fact is associated with
rejection frequencies above the nominal test size along
local-to-the-boundary parameter sequences (cf. Fang and Santos, 2019, Remark
3.6). Second, if conservative bootstrap inference along such parameter
sequences is desired, it can be achieved for hypotheses $\mathscr{G}{}_{1}$--%
$\mathscr{G}{}_{3}$ by adapting the approach of Doko Tchatoka and Wang
(2021), and Cavaliere et al. (2024), at the cost of a potential decrease in
power.

To illustrate these points, consider a sequence of true parameter values $%
\theta _{n}=\theta _{0}+n^{-1/2}\vartheta $ such that $g(\theta _{0})=0$ and 
$\dot{g}^{\prime }\vartheta =c>0$ with $g(\theta _{n})=n^{-1/2}c+o(n^{-1/2})$%
. Moreover, let 
\begin{equation}
\ell (\vartheta ,c):=\vartheta +\underset{\lambda \in \Lambda ^{c}}{\arg
\min }||\lambda -M^{-1/2}\xi ||_{M}\text{, }\Lambda ^{c}=\{\lambda \in 
\mathbb{R}^{2}:\dot{g}^{\prime }\lambda +c\geq 0\},  \label{eq:ltc}
\end{equation}%
and $\ell (0,0)=\ell $ of eq. (\ref{eq asy distribution}). Then, the joint
convergence result 
\begin{equation}
(n^{1/2}(\hat{\theta}-\theta _{0}),(n^{1/2}(\hat{\theta}^{\ast }-\hat{\theta}%
)|D_{n}))\overset{w}{\rightarrow }_{w}(\ell (\vartheta ,c),(\ell (0,0)|M))
\label{eq:lct 2}
\end{equation}%
holds for the bootstrap schemes satisfying conditions (i) and (ii) of
Corollary \ref{corollary bootstrap with boundary}. For a function $r:\mathbb{%
R}^{2}\rightarrow \mathbb{R}$ which is continuously differentiable close to $%
\theta _{0}$, consider the statistics $\tau _{n}=n^{1/2}r(\hat{\theta})$ and 
$\tau _{n}^{\ast }=n^{1/2}(r(\hat{\theta}^{\ast })-r(\hat{\theta}))$, and
distinguish among the extreme possibilities $\dot{r}=\alpha \dot{g}$ with $%
\alpha >0$, and $\dot{r}=\alpha \dot{g}_{\perp }$ with $\alpha \neq 0$,
where $\dot{r}=\frac{\partial r}{\partial \theta ^{\prime }}(\theta _{0})$.
The former possibility arises in testing the null hypothesis that $\theta
_{0}$ lies on the boundary (e.g., with $r=g$), whereas the latter one arises
when the null is orthogonal to the boundary (e.g., with $r(\theta )=\theta
_{1}$, $\mathsf{H}_{0}:\theta _{1}=0$ and $\Omega =\mathbb{R}\times \lbrack
0,\infty )$). If $\dot{r}=\dot{g}$ and, without loss of generality, $\alpha
=1$, the delta method yields 
\begin{equation*}
(\tau _{n},(\tau _{n}^{\ast }|D_{n}))\overset{w}{\rightarrow }_{w}(\max \{0,%
\dot{g}^{\prime }M^{-1/2}\xi +c\},(\max \{0,\dot{g}^{\prime }M^{-1/2}\xi
\}|M)).
\end{equation*}%
With $\gamma _{M}:=(\dot{g}^{\prime }M^{-1}\dot{g})^{-1/2}$, it follows that 
\begin{align*}
P^{\ast }(\tau _{n}^{\ast }\leq \tau _{n})& \overset{w}{\rightarrow }\pi
(c;M,\xi ):=\tfrac{1}{2}\mathbb{I}_{\{\dot{g}^{\prime }M^{-1/2}\xi
+c<0\}}+\Phi (\gamma _{M}(\dot{g}^{\prime }M^{-1/2}\xi +c))\mathbb{I}_{\{%
\dot{g}^{\prime }M^{-1/2}\xi +c\geq 0\}} \\
& >\pi (0;M,\xi ),
\end{align*}%
where $\pi (0;M,\xi )\overset{d}{=}\tfrac{1}{2}\mathbb{I}_{\{U<0.5\}}+U%
\mathbb{I}_{\{U\geq 0.5\}}$, $U\sim U_{[0,1]}$, represents the limit
distribution of the bootstrap \emph{p}-value under the null. The inequality
above implies that bootstrap tests rejecting for large bootstrap \emph{p}%
-values will exhibit rejection frequencies above the nominal test size.

On the other hand, if $\dot{r}=\alpha \dot{g}_{\perp }$, it holds that 
\begin{equation*}
(\tau _{n},(\tau _{n}^{\ast }|D_{n}))\overset{w}{\rightarrow }_{w}(\dot{r}%
^{\prime }\gamma _{M}^{\perp }M{}^{1/2}\xi +\dot{r}^{\prime }\gamma
_{M}^{\perp }M\vartheta ,(\dot{r}^{\prime }\gamma _{M}^{\perp }M{}^{1/2}\xi
|M)),
\end{equation*}%
with $\gamma _{M}^{\perp }:=\dot{g}_{\perp }\dot{g}_{\perp }^{\prime }(\dot{g%
}_{\perp }^{\prime }M\dot{g}_{\perp })^{-1}$, such that the boundary is
asymptotically irrelevant. Bootstrap tests of the null that $r(\theta
_{0})=0 $ could be conservative or liberal according to the sign of $\dot{r}%
^{\prime }\dot{g}_{\perp }\dot{g}_{\perp }^{\prime }M\vartheta $. Similar
considerations apply whenever $\dot{r}^{\prime }\dot{g}_{\perp }\neq 0$.

For situations where liberal tests are not desirable, a possible remedy is
suggested next. It involves a continuum of boundaries for the bootstrap
parameter space and its implementation requires a discretization of that
continuum.

Let $\tilde{\theta}$ be the unrestricted OLS estimator of $\theta $ in
regression (\ref{eq:pr}). For every $s\in I_{n}:=[-|g(\tilde{\theta}%
)|^{1-\mu },g(\hat{\theta})]$, let $\hat{\theta}_{s}^{\ast }$ be the
bootstrap estimator over the parameter space $\Theta _{s}^{\ast }:=\{\theta
\in \mathbb{R}^{2}:g(\theta )\geq s-g(\hat{\theta})^{1+\kappa }\}$, where $%
\mu \in (0,1)$ and $\kappa >0$ are fixed. For a continuously differentiable
function $r$, let $p_{n}^{\ast }(s)$ be the \emph{p}-value of a test based
on $\tau _{n}=n^{1/2}r(\hat{\theta})$ and $\tau _{n}^{\ast }=n^{1/2}(r(\hat{%
\theta}^{\ast })-r(\hat{\theta}))$. Then 
\begin{equation*}
\text{$\underset{n\rightarrow \infty }{\limsup }$}P(\sup_{s\in
I_{n}}p_{n}^{\ast }(s)\leq q)\leq q
\end{equation*}%
for all $q\in \limfunc{int}C$, where $C$ is the set from display (\ref{eq
unconditional validity}) for the benchmark asymptotic test based on the
unfeasible statistic $n^{1/2}(r(\hat{\theta})-r(\theta _{n}))$ and the
simple null hypothesis that $\theta _{n}$ is the true parameter value. This
conservative generalization of the validity property (\ref{eq unconditional
validity}) holds irrespective of the values of the drift parameter $c$.
Specifically, the role of $-|g(\tilde{\theta})|^{1-\mu }$ in the definition
of $I_{n}$ is to guarantee that $g(\hat{\theta})-cn^{-1/2}\in I_{n}$ with
probability approaching one. Conservative size control then follows from the
fact that $\hat{\theta}_{s}^{\ast }$ with $s=g(\hat{\theta})-cn^{-1/2}$
satisfies 
\begin{equation*}
(n^{1/2}(\hat{\theta}-\theta _{n}),(n^{1/2}(\hat{\theta}_{s}^{\ast }-\hat{%
\theta})|D_{n}))\overset{w}{\rightarrow }\left( \ell (0,c),(\ell
(0,c)|M)\right) ;
\end{equation*}%
see eqs. (\ref{eq:ltc})--(\ref{eq:lct 2}).

\section{Numerical results and choice of the tuning parameters}

\label{sec:Numerical}

In this section we analyze the finite sample performance of the proposed
bootstrap methodology by means of numerical simulations. The purpose is
twofold: first, to investigate the practical advantage of our methodology
over standard bootstrap methods; second, to provide some practical guidance
on how to choose the functions $g^{\ast }$ and the tuning parameter $\kappa $
in the definition of the bootstrap parameter space. Simulations are based on
setup $\mathcal{G}_{3}$ of Section \ref{sec par on the boundary setup}, as
it covers the general case of a true parameter value that could, but need
not, lie on the boundary of the parameter space under the null hypothesis.
This section is organized as follows. In Section \ref{sec MC design} we
describe the data generating processes, the null hypotheses and the adopted
bootstrap schemes. In Section \ref{sec MC comment} we discuss the
performance of the tests both under the null and under local alternatives.
Section \ref{sec tuning par} deals with the choice of $g^{\ast }$ and $%
\kappa $. Additional numerical results are provided in the accompanying
supplement, Section \ref{sec:additional MC}.

\subsection{Monte Carlo design}

\label{sec MC design}

We consider the same data generating process (DGP) as in (\ref{eq:prr}),
where $x_{n,t}=n^{-1/2}x_{t}$, $x_{t}:=\sum_{i=1}^{t}\varepsilon_{x,i}$, $%
\varepsilon_{x,t}\sim iid$ $N(0,1)$, with the following specifications of $%
\varepsilon_{t}$:

\begin{enumerate}
\item $\varepsilon_{t}\sim iid$ $N(0,1);$

\item $\varepsilon_{t}=\sigma_{t}\nu_{t}$, where $\sigma_{t}^{2}=0.7+0.3%
\varepsilon_{t-1}^{2}$ and $\nu_{t}\sim iid$ $N(0,1)$;

\item $\varepsilon_{t}=\sqrt{0.5}\varepsilon_{x,t}+\sqrt{0.5}\eta_{t}$,
where $\eta_{t}\sim iid$ $N(0,1)$.
\end{enumerate}

In each case, $\{\varepsilon_{x,t}\}$ is independent of, respectively, $%
\{\varepsilon_{t}\},$ $\{\nu_{t}\}$ and $\{\eta_{t}\}$. In Case 1, the
regression errors are independent and Gaussian, while in Case 2 they exhibit
ARCH-type conditional heteroskedasticity. Case 3 allows for correlation
between $\varepsilon_{t}$ and the regressor's innovation $\varepsilon_{x,t}$.

The parameter space is specified as $\Theta :=\{\theta \in \mathbb{R}%
^{2}:g(\theta )\geq 0\}$ where $g(\theta )=\theta _{2}$. That is, $\Theta :=%
\mathbb{R}\times \left[ 0,\infty \right) $ -- such that its boundary is
given by $\partial \Theta =\mathbb{R}\times \{0\}$. For all parameter
values, we test the null hypothesis $\mathsf{H}_{0}:h(\theta _{0})=0$, with $%
h(\theta )=\theta _{1}+\theta _{2}$, against the two-sided alternative $%
h(\theta _{0})\neq 0$. To do so, we employ the test statistics $\tau
_{n}=\phi (\sqrt{n}h(\hat{\theta}))$ and $\tau _{n}^{\ast }=\phi (\sqrt{n}(h(%
\hat{\theta}^{\ast })-h(\hat{\theta})))$, where $\phi (x)=x^{2}$, while $%
\hat{\theta}$ and $\hat{\theta}^{\ast }$ denote the original and bootstrap
constrained LS estimators, respectively. In order to analyze size control
and power of the proposed tests, we consider both empirical rejection
probabilities {[}ERPs{]} under the null and under local alternatives. For
tests performed under the null, we consider three different choices of the
true value $\theta _{0}$, one located on $\partial \Theta $ and two located
on $\Theta \backslash \partial \Theta $; specifically, $\theta_0 \in
\{(0,0)^{\prime },(-0.75,0.75)^{\prime },(-1.5,1.5)^{\prime }\}$. Under $%
\mathsf{H}_{1}$, we employ a local alternative of the form $\theta _{0}={a}%
_{0}n^{-1/2}$, ${a}_{0}\in \mathbb{R}^{2}$, such that $h(\theta _{0})\neq 0$
unless $a=(0,0)^{\prime }$.

Tests are based on \emph{p}-values obtained using a `standard' -- i.e., with 
$\Theta ^{\ast }=\Theta $ -- fixed-regressor Gaussian wild bootstrap and the
proposed `corrected' bootstrap scheme. For the latter, the bootstrap
parameter space is set to $\Theta ^{\ast }=\mathbb{R}\times \lbrack g^{\ast
}(\hat{\theta}_{2}),\infty )$, where the function $g^{\ast }$ satisfies the
assumptions of Corollary \ref{corollary bootstrap with boundary}, see also
Section \ref{sec tuning par}. In order to assess the impact of the tuning
parameter $\kappa $, we consider a grid of possible values for $\kappa $.
Numerical results are based on $50,000$ Monte Carlo simulations, each
involving $B=999$ bootstrap repetitions. Sample sizes are set to $n\in
\{100,200,400,800,1600\}$.

\subsection{Empirical rejection probabilities}

\label{sec MC comment}

We now discuss the ERPs of the bootstrap tests. Specifically, the Monte
Carlo results in Table 1 and 2 refer to the case in which data are generated
under the null and under local alternatives, respectively. The proposed
modified bootstrap parameter space is based on the function $%
g^{\ast}=g-|g|^{1+\kappa}$ for several values of $\kappa>0$.

Table 1 shows that the `standard' bootstrap scheme typically under-rejects
the true null hypothesis when the parameter lies on the boundary of the
parameter space $\Theta $ whereas, as expected, its ERPs are closer to the
nominal level when $\theta _{0}$ is in the interior of $\Theta $. Our
proposed bootstrap performs similarly to the `standard' bootstrap for very
small values of $\kappa $, with the impact of the correction becoming more
relevant as $\kappa $ increases. If the parameter is on the boundary of the
parameter space ($\theta _{0}\in \partial \Theta $), our proposed bootstrap
scheme gives rise to smaller absolute size distortions than the `standard'
bootstrap, for all the considered DGPs and all values of $\kappa $. When $%
\theta _{0}\in \mathrm{int}\Theta $, we observe very little variability in
the ERPs across the different bootstrap methods, at least for reasonably
small values of $\kappa $.

Table 2 reports the ERPs of the tests when data are generated under local
alternatives $\theta _{0}=a_{0}n^{-1/2}$, $a_{0}\in \{(-3,0)^{\prime
},(3,0)^{\prime },(5,0)^{\prime }\}$, such that the true parameter values
lie on the boundary of the parameter space. Results show that both bootstrap
schemes have power under local alternatives, with the `corrected' bootstrap
generally showing higher ERPs than the `standard' bootstrap, in line with
the results obtained under the null. Finally, we notice that the sign of the
deviations from the null hypothesis matters, with positive deviations
showing higher ERPs. This finding can be explained by the fact that the
limit distribution of $n^{1/2}(h(\hat{\theta})-h(\theta _{0}))$ is
asymmetric when $\theta _{0}$ lie on the boundary of $\Theta $. Results
about local alternatives such that $\theta _{0}$ are $n^{-1/2}$-local to the
boundary are substantially similar and are reported in Section S.2 of the
supplement.

\subsection{Choice of $g^{\ast}$ and $\protect\kappa$}

\label{sec tuning par}

We now consider the practical issue of choosing the function $g^{*}$ and the
tuning parameter $\kappa$ used to construct the modified bootstrap parameter
space $\Theta^{*}$.

Regarding $g^{\ast }$, in Section 4 we discussed the functions $%
g_{(1)}^{\ast }:=g-|g|^{1+\kappa }$, $\kappa >0$, which satisfy the
assumptions of Corollary \ref{corollary bootstrap with boundary} and were
employed in the simulations so far, whereas in Section \ref%
{subsec:Fang-and-Santos} we considered also $g_{(2)}^{\ast }:=g-n^{-\kappa
}|g|$, $\kappa \in (0,1/2)$, corresponding to $s_n=n^{1/2-\kappa}$ in the
concluding paragraph of Section \ref{subsec:Fang-and-Santos}. Numerical
results in Table 1 and 2 and in the accompanying Supplement, Section S.2,
show that both choices of $g^{\ast }$ deliver good test performance, both
under the null and under local alternatives. The most salient difference
between $g_{(1)}^{\ast }$ and $g_{(2)}^{\ast }$ is that tests based on $%
g_{(1)}^{\ast }$ tend to be more robust to the choice of $\kappa $ when $%
g(\theta _{0})\geq 1$.

Concerning the choice of the tuning parameter $\kappa $, we focus on $%
g^{\ast }=g_{(1)}^{\ast }.$ Preliminary considerations point at a possible
trade-off between the cases of a boundary and an interior location of the
true parameter $\theta _{0}$. Thus, for $\theta _{0}\in \partial \Theta $,
larger values of $\kappa $ accelerate the convergence of $g(\hat{\theta}%
)^{1+\kappa }$ to zero, which can be expected to favor bootstrap performance
as the bootstrap true value $\hat{\theta}$ is put at a smaller distance from
the bootstrap boundary. On the other hand, if $\theta _{0}\in \text{int}%
(\Theta )$ and $g(\theta _{0})\in (0,1)$, in small samples large values of $%
\kappa $ may put $\hat{\theta}$ too close to the bootstrap boundary,
yielding inferior bootstrap performance.

Our Monte Carlo study indeed confirms that small values of $\kappa $ are
preferable when $\theta _{0}\in \text{int}(\Theta )$ and $g(\theta _{0})\in
(0,1)$; however, it also shows that the proposed correction quickly provides
satisfactory size control for small values of $\kappa $ even when $\theta
_{0}\in \partial \Theta $. Finally, we notice that when $\theta _{0}\in 
\text{int}(\Theta )$ and $g(\theta _{0})\geq 1$ the choice of $\kappa $ has
negligible impact on the ERPs. Overall, our numerical analysis suggests that
choices of $\kappa $ close to $0.5$ provide quite satisfactory size control
across all the considered scenarios.

\medskip {}

\noindent \textsc{Remark}. The above guideline about the choice of $\kappa $
is based on numerical evidence; it delivers a reasonable simple choice which
can be easily implemented. It is not optimal in any sense, and indeed
alternative methods could be employed to obtain data-driven choices of $%
\kappa $. For instance, the unrestricted parameter estimates could be used
to assess how far the true parameter value $\theta _{0}$ is from the
boundary of the parameter space, and then calibrate the choice of $\kappa $
accordingly. This approach would be in the spirit of Romano, Shaikh and Wolf
(2014), who suggest to improve the power of tests of moment inequalities by
introducing a first step, where a confidence region for the moments is
constructed using their unrestricted estimates. Although this approach may
improve the finite sample properties of our tests, it would require a
preliminary choice of further tuning parameters, such as $\beta $ in Romano
et al. (2014),\ hence introducing an extra layer of complexity.$\hfill
\square $

\section{Conclusions}

\label{sec conclusion}

In this paper we analyzed the problem of bootstrap hypotheses tests on the
parameters $(\alpha,\beta)$ of a predictive regression $y_{t}=\alpha+\beta
x_{t-1}+\varepsilon_{t}$, generalizable to higher dimensions, when the
parameter space is defined by means of a smooth constraint $%
g(\alpha,\beta)\geq0$ and the true parameter vector under the null
hypothesis may lie on the boundary of the parameter space. In the framework
of constrained parameter estimation, implementation of the bootstrap is not
straightforward, as the presence of a parameter on the boundary of the
parameter space makes the bootstrap measure random in the limit.

We discussed possible solutions to this inference problem. Specifically, we
presented some modifications of standard bootstrap schemes where the
bootstrap parameter space is shifted by a data-dependent function, thus
allowing us to regain control over the boundary as a source of limiting
bootstrap randomness. We also proved validity of the associated bootstrap
inference in the cases where the posited predicting variable is I(1).

Our contribution is novel in the framework of predictive regression, in that
the existing literature has not analyzed the bootstrap in contexts combining
non-stationarity of the posited predictor with a priori knowledge about the
possible form of predictability, represented by a restricted parameter
space. The value of our work is to provide valid bootstrap implementations
in this setting.

\section*{Supplementary Material}

Cavaliere, G., Georgiev, I. \& Zanelli, E. (2024). Supplement to:
\textquotedblleft Parameters on the boundary in predictive
regression,\textquotedblright\ \emph{Econometric Theory }Supplementary
Material. To view, please visit [[doi to be inserted here by typesetter]]

\section*{References}

\begin{description}
\item \textsc{Andrews,} D.W.K. (1999): Estimation when a parameter is on a
boundary, \emph{Econometrica} 67, 1341--1383.

\item ------ (2000): Inconsistency of the bootstrap when a parameter is on
the boundary of the parameter space, \emph{Econometrica} 68, 399--405.

\item \textsc{Berti, P., L. Pratelli and P. Rigo} (2006): Almost sure weak
convergence of random probability measures, \emph{Stochastics: an
International Journal of Probability and Stochastic Processes} 78, 91--97.

\item \textsc{Cavaliere, G. and I. Georgiev} (2020): Inference under random
limit bootstrap measures, \emph{Econometrica}\ 88, 2547--2574.

\item \textsc{Cavaliere, G., H.B. Nielsen, R.S. Pedersen and A. Rahbek}
(2022): Bootstrap inference on the boundary of the parameter space, with
application to conditional volatility models, \emph{Journal of Econometrics}%
\ 227, 241--263.

\item \textsc{Cavaliere, G., H.B. Nielsen and A. Rahbek} (2017): On the
consistency of bootstrap testing for a parameter on the boundary of the
parameter space, \emph{Journal of Time Series Analysis} 38, 513--534.

\item \textsc{Cavaliere, G., I. Perera and A. Rahbek} (2024): Specification
tests for GARCH processes with nuisance parameters on the boundary, \emph{%
Journal of Business \& Economic Statistics} 42, 197--214.

\item \textsc{Chatterjee, A., and S. N. Lahiri} (2011): Bootstrapping Lasso
estimators, \emph{Journal of the American Statistical Association }106,
608--25.

\item \textsc{Chen, Q. and Z. Fang }(2019): Inference on functionals under
first order degeneracy, \emph{Journal of Econometrics} 210, 459--481.

\item \textsc{Doko Tchatoka, F., and W. Wang} (2021): Uniform inference
after pretesting for exogeneity with heteroskedastic data, MPRA paper
106408, University Library of Munich, Germany.

\item \textsc{D\"{u}mbgen, L.} (1993): On nondifferentiable functions and
the bootstrap, \emph{Probability Theory and Related Fields }95, 125--140.

\item \textsc{Fang Z. }(2014): Optimal plug-in estimators of directionally
differentiable functionals. Unpublished manuscript.

\item \textsc{Fang Z. and A. Santos} (2019): Inference on directionally
differentiable functions, \emph{Review of Economic Studies} 86, 377--412.

\item \textsc{Georgiev, I., D. Harvey, S. Leybourne and A.M.R. Taylor}
(2019): A bootstrap stationarity test for predictive regression invalidity, 
\emph{Journal of Business \& Economic Statistics} 37, 528--541.

\item \textsc{Geyer, C.J.} (1994): On the asymptotics of constrained
M-estimation, \emph{Annals of Statistics} 22, 1993--2010.

\item \textsc{Hirano, K. and Porter, J.R.} (2012): Impossibility results for
nondifferentiable functionals, \emph{Econometrica} 80, 1769-.1790.

\item \textsc{Hong, H. and J. Li} (2020): The numerical bootstrap, \emph{%
Annals of Statistics} 48, 397--412.

\item \textsc{Kallenberg} O. (1997): \emph{Foundations of Modern Probability}%
, Springer-Verlag: Berlin.

\item \textsc{Kallenberg} O. (2017): \emph{Random measures: theory and
applications}, Springer-Verlag: Berlin.

\item \textsc{M\"{u}ller, U.K. and M. Watson} (2008):\ Testing models of low
frequency variability, \emph{Econometrica} 76, 979--1016.

\item \textsc{Phillips,} P.C.B. (2014): On confidence intervals for
autoregressive roots and predictive regression, \emph{Econometrica} 82,
1177--1195.

\item \textsc{Romano, J.P, A.M. Shaikh and M. Wolf }(2014): A practical
two-step method for testing moment inequalities, \emph{Econometrica} 82,
1979--2002.

\item \textsc{Sweeting T.J.} (1989): On conditional weak convergence, \emph{%
Journal of Theoretical Probability} 2, 461--474.
\end{description}

\appendix

\label{Sec Appendix}

\setcounter{theorem}{0}\setcounter{equation}{0}\setcounter{lemma}{0} 

\numberwithin{theorem}{section} \numberwithin{lemma}{section} %
\numberwithin{remark}{section}

\section{Mathematical Appendix}

\subsection{Proof of Theorem \protect\ref{Lemma bootstrap with boundary}}

\label{sec all proofs}

\noindent Introduce $\tilde{x}_{t}:=(1,x_{n,t-1})^{\prime}$. Let $%
\mu_{n}:=n^{1/2}(\hat{\theta}-\theta_{0})$, $M_{n}:=n^{-1}\sum_{t=1}^{n}%
\tilde{x}_{t}\tilde{x}_{t}^{\prime}$ and $N_{n}^{\ast}:=n^{-1/2}%
\sum_{t=1}^{n}\varepsilon_{t}^{\ast}\tilde{x}_{t}$. Moreover, let the
normalized bootstrap estimator be denoted by $\mu_{n}^{\ast}:=n^{1/2}(\hat{%
\theta}^{\ast}-\hat{\theta})$; similarly, $\tilde{\mu}_{n}^{\ast}:=n^{1/2}(%
\tilde{\theta}^{\ast}-\hat{\theta})$, where $\tilde{\theta}^{\ast}$ is the
unrestricted (OLS) bootstrap estimator. On the event $\{\det(M_{n})>0\}$
with $P(\det(M_{n})>0)\rightarrow1$, the estimator $\tilde{\theta}^{\ast}$
is well-defined and unique. As we are interested in distributional
convergence results, without loss of generality we proceed as if $%
P(\det(M_{n})>0)=1$.

By arguments similar to the proof of Theorem 4.1 in Cavaliere and Georgiev
(2020), it can be concluded that $(\mu_{n},M_{n},N_{n}^{\ast})\overset{%
w^{\ast}}{\rightarrow}_{w}(\ell(\theta_{0}),M,M^{1/2}\xi^{\ast})|(M,\ell(%
\theta_{0}))$ in $\mathbb{R}^{2\times4}$, where $M$ is of full rank with
probability one, $\xi^{\ast}|(M,\ell(\theta_{0}))\sim
N(0,\sigma_{e}^{2}I_{2})$ and $\sigma_{e}^{2}$ denotes the variance of $%
\varepsilon_{t}$ corrected for $\Delta x_{n,t}$. To derive the result (\ref%
{eq asy distrib for general bootstraps}), we analyze the properties of $%
\mu_{n}^{\ast}$ on a special probability space where $(\mu_{n},M_{n},N_{n}^{%
\ast})$ given the data converge weakly a.s. rather than weakly in
distribution. Specifically, by Lemma A.2(a) in Cavaliere and Georgiev (2020)
we can consider a probability space (where $\ell(\theta_{0}),M$ and, for
every $n\in N$, also the original data and the bootstrap sample can be
redefined, maintaining their distribution), such that 
\begin{equation}
\mu_{n}\overset{a.s.}{\rightarrow}\ell(\theta_{0})\text{, }M_{n}\overset{a.s.%
}{\rightarrow}M\text{, }N_{n}^{\ast}\overset{w^{\ast}}{\rightarrow}%
_{a.s.}M^{1/2}\xi^{\ast}|(M,\ell(\theta_{0}))\overset{}{=}%
M^{1/2}\xi^{\ast}|M,  \label{eq weak conv in prob}
\end{equation}
the last equality being an a.s. equality of conditional distributions.

Let $q_{n}^{\ast}(\theta):=n^{-1}\sum_{t=1}^{n}(y_{t}^{\ast}-\theta^{\prime}%
\tilde{x}_{t})^{2}$ with $\tilde{\theta}^{\ast}:=\arg\min_{\theta\in\mathbb{R%
}^{2}}q_{n}^{\ast}(\theta)$ being well-defined and unique for outcomes in
the event $\{\det(M_{n})>0\}$. On the special probability space, the
asymptotic distribution of $\tilde{\mu}_{n}^{\ast}=n^{1/2}(\tilde{\theta}%
^{\ast}-\hat{\theta})=M_{n}^{-1}N_{n}^{\ast}$ follows from (\ref{eq weak
conv in prob}) and a CMT (Theorem 10 of Sweeting, 1989): 
\begin{equation}
\tilde{\mu}_{n}^{\ast}\overset{w^{\ast}}{\rightarrow}_{a.s.}\tilde{\ell}%
^{\ast}|(M,\ell(\theta_{0}))\overset{}{=}\tilde{\ell}^{\ast}|M\text{, }%
\tilde{\ell}^{\ast}:=\sigma_{e}^{2}M^{-1/2}\xi^{\ast}\text{.}
\label{eq separata}
\end{equation}

Let us turn now to the bootstrap estimator $\hat{\theta}^{\ast}$. If $%
g(\theta_{0})>g^{\ast}(\theta_{0})$, then the consistency facts $\hat{\theta}%
\overset{a.s.}{\rightarrow}\theta_{0}$ (from (\ref{eq weak conv in prob}))
and $\tilde{\theta}^{\ast}\overset{w^{\ast}}{\rightarrow}_{a.s.}\theta_{0}$
(from (\ref{eq separata})), jointly with the continuity of $g,g^{\ast}$ at $%
\theta_{0}$, imply that $P^{\ast}(g(\tilde{\theta}^{\ast})\geq g^{\ast}(\hat{%
\theta}))\overset{a.s.}{\rightarrow}1$. Hence, $\tilde{\theta}^{\ast}$
uniquely minimizes $q_{n}^{\ast}$ on $\Theta^{\ast}$ with $P^{\ast}$%
-probability approaching one a.s. This establishes the existence of $\hat{%
\theta}^{\ast}$ with $P^{\ast}$-probability approaching one a.s., as well as
the facts $P^{\ast}(\hat{\theta}^{\ast}=\tilde{\theta}^{\ast})\overset{a.s.}{%
\rightarrow}1$ and $P^{\ast}(\mu_{n}^{\ast}=\tilde{\mu}_{n}^{\ast})\overset{%
a.s.}{\rightarrow}1$. Using also (\ref{eq separata}), it follows that $%
\mu_{n}^{\ast}\overset{w}{\rightarrow}_{a.s.}\tilde{\ell}^{\ast}|M$ on the
special probability space, and since $\mu_{n}\overset{a.s.}{\rightarrow}%
\ell(\theta_{0})$ on this space, it follows further that $%
(\mu_{n},(\mu_{n}^{\ast}|D_{n}))\overset{w}{\rightarrow}_{w}(\ell(%
\theta_{0}),(\tilde{\ell}^{\ast}|M))$ on a general probability space, as
asserted in (\ref{eq semno}).

In the case where $g^{\ast}(\theta_{0})=g(\theta_{0})$, it still holds that $%
\tilde{\theta}^{\ast}$ uniquely minimizes $q_{n}^{\ast}$ on $\Theta^{\ast}$
whenever $g(\tilde{\theta}^{\ast})\geq g^{\ast}(\hat{\theta})$, such that $%
\hat{\theta}^{\ast}$ exists and equals $\tilde{\theta}^{\ast}$ on the event $%
\{g(\tilde{\theta}^{\ast})\geq g^{\ast}(\hat{\theta})\}$. However, the
probability of this event no longer tends to one. Whenever $g(\tilde{\theta}%
^{\ast})<g^{\ast}(\hat{\theta})$, a minimizer of $q_{n}^{\ast}$ on $%
\Theta^{\ast}$ exists if and only if a minimizer, say $\check{\theta}^{\ast}$%
, of $q_{n}^{\ast}$ on $\partial\Theta^{\ast}$ exists and minimizes $%
q_{n}^{\ast}$ over the entire $\Theta^{\ast}$ (this claim is due to the fact
that, for outcomes in the event $\{\det(M_{n})>0\},$ the function $%
q_{n}^{\ast}(\theta)$ is locally minimized uniquely at $\tilde{\theta}%
^{\ast} $). Let $\mathbb{I}_{n}^{\ast}:=\mathbb{I}_{\{b(\tilde{\theta}%
^{\ast})\geq0\}}$ with $b(\theta):=g(\theta)-g^{\ast}(\hat{\theta})$. We
show in Section \ref{se wedge} below that $\check{\theta}^{\ast}(1-\mathbb{I}%
_{n}^{\ast})$, with a measurable $\check{\theta}^{\ast}$, is well-defined
with $P^{\ast}$-probability approaching one a.s. and $(q_{n}^{\ast}(\check{%
\theta}^{\ast})-q_{n}^{\ast}(\theta))(1-\mathbb{I}_{n}^{\ast})\leq0$ for all 
$\theta\in\Theta^{\ast}$, with $P^{\ast}$-probability approaching one a.s.
This establishes the possibility to define the bootstrap estimator $\hat{%
\theta}^{\ast}$ as 
\begin{equation}
\hat{\theta}^{\ast}=\tilde{\theta}^{\ast}\mathbb{I}_{n}^{\ast}+\check{\theta}%
^{\ast}(1-\mathbb{I}_{n}^{\ast})
\label{eq decomposition for the restricted estimator}
\end{equation}
and, therefore, the existence of $\hat{\theta}^{\ast}$ with $P^{\ast}$%
-probability approaching one a.s. The existence result carries over to a
general probability space with $P^{\ast}$-probability approaching one in
probability.

In Section \ref{se wedge} we also show that $\Vert\check{\theta}^{\ast}-\hat{%
\theta}\Vert(1-\mathbb{I}_{n}^{\ast})=O_{p^{\ast}}(n^{-1/2})$ a.s., and as a
result, $\Vert\hat{\theta}^{\ast}-\hat{\theta}\Vert=O_{p^{\ast}}(n^{-1/2})$
a.s., using also (\ref{eq separata}). We do not discuss the uniqueness of $%
\check{\theta}^{\ast}$ but instead we argue next that the measurable
minimizers of $q_{n}^{\ast}$ over the bootstrap boundary are asymptotically
equivalent, as they give rise to the same asymptotic distribution of $\hat{%
\theta}^{\ast}$.

To accomplish this, we use the result of Section \ref{se wedge} that $\check{%
\theta}^{\ast}$ satisfies a first-order condition {[}foc{]} with $P^{\ast}$%
-probability approaching one a.s. Let dots over function names denote
differentiation w.r.t. $\theta$ (e.g., $\dot{q}_{n}^{\ast}(\theta):=(\partial%
\dot{q}_{n}^{\ast}/\partial\theta^{\prime})(\theta)$, a column vector). Then
the foc takes the form 
\begin{equation*}
\{\dot{q}_{n}^{\ast}(\check{\theta}^{\ast})+\check{\delta}_{n}\dot{b}(\check{%
\theta}^{\ast})\}(1-\mathbb{I}_{n}^{\ast})=\{\dot{q}_{n}^{\ast}(\check{\theta%
}^{\ast})+\check{\delta}_{n}\dot{g}(\check{\theta}^{\ast})\}(1-\mathbb{I}%
_{n}^{\ast})=0\text{, }b(\check{\theta}^{\ast})(1-\mathbb{I}_{n}^{\ast})=0,
\end{equation*}
where $\check{\delta}_{n}\in\mathbb{R}$ is a Lagrange multiplier. The foc
implies, by means of a standard argument, the existence of a measurable $%
\bar{\theta}^{\ast}$ between $\check{\theta}^{\ast}$ and $\hat{\theta}$ such
that 
\begin{equation*}
\{n^{1/2}(\check{\theta}^{\ast}-\hat{\theta})-(I_{2}-A_{n}^{\ast}\dot{g}(%
\bar{\theta}^{\ast})^{\prime})\tilde{\mu}_{n}^{\ast}+A_{n}^{\ast}n^{1/2}b(%
\hat{\theta})\}(1-\mathbb{I}_{n}^{\ast})=0,
\end{equation*}
where $A_{n}^{\ast}:=M_{n}^{-1}\dot{g}(\check{\theta}^{\ast})[\dot{g}(\bar{%
\theta}^{\ast})^{\prime}M_{n}^{-1}\dot{g}(\check{\theta}^{\ast})]^{-1}$ is
well-defined with $P^{\ast}$-probability approaching one a.s. As further $%
\Vert\check{\theta}^{\ast}-\hat{\theta}\Vert(1-\mathbb{I}_{n}^{\ast})=O_{p^{%
\ast}}(n^{-1/2})$ a.s., $\Vert\bar{\theta}^{\ast}-\hat{\theta}\Vert(1-%
\mathbb{I}_{n}^{\ast})=O_{p^{\ast}}(n^{-1/2})$ a.s. and $\hat{\theta}%
-\theta_{0}=O(n^{-1/2})$ a.s., using the continuity of $\dot{g}(\theta)$ at $%
\theta_{0}$ it follows that 
\begin{equation*}
\{n^{1/2}(\check{\theta}^{\ast}-\hat{\theta})-[(I_{2}-A^{\ast}\dot{g}%
^{\prime})\tilde{\mu}_{n}^{\ast}-A^{\ast}(\dot{g}-\dot{g}^{\ast})^{%
\prime}n^{1/2}(\hat{\theta}-\theta_{0})]\}(1-\mathbb{I}_{n}^{\ast})=o_{p^{%
\ast}}(1)~~\text{a.s.,}
\end{equation*}
where $A^{\ast}:=M^{-1}\dot{g}[\dot{g}^{\prime}M^{-1}\dot{g}]^{-1}$ and $%
P^{\ast}(|o_{p^{\ast}}(1)|>\eta)\overset{a.s.}{\rightarrow}1$ for all $%
\eta>0 $.

Returning to (\ref{eq decomposition for the restricted estimator}), we
conclude that 
\begin{equation}
n^{1/2}(\hat{\theta}^{\ast}-\hat{\theta})=\tilde{\mu}_{n}^{\ast}\mathbb{I}%
_{n}^{\ast}+\{(I_{2}-A^{\ast}\dot{g}^{\prime})\tilde{\mu}_{n}^{\ast}-A^{%
\ast}(\dot{g}-\dot{g}^{\ast})^{\prime}n^{1/2}(\hat{\theta}-\theta_{0})\}(1-%
\mathbb{I}_{n}^{\ast})+o_{p^{\ast}}(1)~~\text{a.s.}  \label{eq dithe}
\end{equation}
Consider the event indicated by $\mathbb{I}_{n}^{\ast}$. As $\Vert\hat{\theta%
}^{\ast}-\hat{\theta}\Vert=O_{p^{\ast}}(n^{-1/2})$ a.s. and $\hat{\theta}%
-\theta_{0}=O(n^{-1/2})$ a.s., by the mean value theorem and the continuous
differentiability of $g$, $g^{\ast}$ it holds that 
\begin{equation*}
n^{1/2}b(\tilde{\theta}^{\ast})=\dot{g}^{\prime}\tilde{\mu}_{n}^{\ast}+(\dot{%
g}-\dot{g}^{\ast})^{\prime}\mu_{n}+o_{p^{\ast}}(1)~~\text{a.s.}
\end{equation*}
Then $\mathbb{I}_{n}^{\ast}\overset{w^{\ast}}{\rightarrow}_{a.s.}\mathbb{I}%
_{\infty}|(M,\ell(\theta_{0}))$ with $\mathbb{I}_{\infty}:=\mathbb{I}_{\{%
\dot{g}^{\prime}\tilde{\ell}^{\ast}\geq(\dot{g}^{\ast}-\dot{g}%
)^{\prime}\ell(\theta_{0})\}}$, by (\ref{eq weak conv in prob})-(\ref{eq
separata}) and the CMT for weak a.s. convergence (Theorem 10 of Sweeting,
1989), as the probability of the limiting discontinuities is $0$: $P(\dot{g}%
^{\prime}\tilde{\ell}^{\ast}=(\dot{g}^{\ast}-\dot{g})^{\prime}\ell(%
\theta_{0})|(M,\ell(\theta_{0})))=0$ a.s. By exactly the same facts, passage
to the limit directly in (\ref{eq dithe}) yields 
\begin{equation*}
n^{1/2}(\hat{\theta}^{\ast}-\hat{\theta})\overset{w^{\ast}}{\rightarrow}%
_{a.s.}\{\tilde{\ell}^{\ast}\mathbb{I}_{\infty}+\check{\ell}^{\ast}(1-%
\mathbb{I}_{\infty})\}|(M,\ell(\theta_{0}))\text{, }\check{\ell}%
^{\ast}:=(I_{2}-A^{\ast}\dot{g}^{\prime})\tilde{\ell}^{\ast}-A^{\ast}(\dot{g}%
-\dot{g}^{\ast})^{\prime}\ell
\end{equation*}
on the special probability space, where also $\mu_{n}\overset{a.s.}{%
\rightarrow}\ell(\theta_{0})$ by (\ref{eq weak conv in prob}). Therefore, on
a general probability space it holds that 
\begin{equation*}
(\mu_{n},(n^{1/2}(\hat{\theta}^{\ast}-\hat{\theta})|D_{n}))\overset{w^{\ast}}%
{\rightarrow}_{w}(\ell(\theta_{0}),[\{\tilde{\ell}^{\ast}\mathbb{I}_{\infty}+%
\check{\ell}^{\ast}(1-\mathbb{I}_{\infty})\}|(M,\ell(\theta_{0}))]).
\end{equation*}
As $I_{2}-A^{\ast}\dot{g}^{\prime}=\dot{g}_{\perp}(\dot{g}_{\perp}^{\prime}M%
\dot{g}_{\perp})^{-1}\dot{g}_{\perp}^{\prime}M$ and $\tilde{\ell}%
^{\ast}=M^{-1/2}\xi^{\ast}$, it follows that 
\begin{eqnarray*}
\tilde{\ell}^{\ast}\mathbb{I}_{\infty}+\check{\ell}^{\ast}(1-\mathbb{I}%
_{\infty}) & = & \dot{g}_{\perp}(\dot{g}_{\perp}^{\prime}M\dot{g}%
_{\perp})^{-1}\dot{g}_{\perp}^{\prime}M^{1/2}\xi^{\ast} \\
& & +M^{-1}\dot{g}(\dot{g}^{\prime}M^{-1}\dot{g})^{-1}\max\{(\dot{g}^{\ast}-%
\dot{g})^{\prime}\ell,\dot{g}^{\prime}M^{-1/2}\xi^{\ast}\},
\end{eqnarray*}
which is $\arg\min_{\{\dot{g}^{\prime}\lambda\geq(\dot{g}^{\ast}-\dot{g}%
)^{\prime}\ell\}}||\lambda-M^{-1/2}\xi^{\ast}||_{M}$ a.s. as asserted in (%
\ref{eq amans}).$\hfill\square$

For use in the proof of Corollary \ref{corollary bootstrap with boundary},
we notice here a useful consequence of the previous argument. Return to the
special probability space where 
\begin{equation*}
(\mu _{n},(n^{1/2}(\hat{\theta}^{\ast }-\hat{\theta})|D_{n}))\overset{%
w^{\ast }}{\rightarrow }_{a.s.}(\ell (\theta _{0}),[\{\tilde{\ell}^{\ast }%
\mathbb{I}_{\infty }+\check{\ell}^{\ast }(1-\mathbb{I}_{\infty })\}|(M,\ell
(\theta _{0}))]).
\end{equation*}%
Let $\tau _{n}:=\phi (\mu _{n})$, $\tau _{n}^{\ast }:=\phi (n^{1/2}(\hat{%
\theta}^{\ast }-\hat{\theta}))$, $\tau :=\phi (\ell (\theta _{0}))$ and $%
\tau ^{\ast }:=\phi (\tilde{\ell}^{\ast }\mathbb{I}_{\infty }+\check{\ell}%
^{\ast }(1-\mathbb{I}_{\infty }))$ for a continuous $\phi :\mathbb{R}%
\rightarrow \mathbb{R}$. Then 
\begin{equation*}
(\tau _{n},(\tau _{n}^{\ast }|D_{n}))\overset{w^{\ast }}{\rightarrow }%
_{a.s.}(\tau ,\tau ^{\ast }|(M,\ell (\theta _{0})))
\end{equation*}%
by the CMT of Sweeting (1989). Furthermore, the regular conditional
distributions $\tau _{n}^{\ast }|D_{n}$ converge weakly to the regular
conditional distribution $\tau ^{\ast }|(M,\ell (\theta _{0}))$ for almost
all outcomes; see Theorem 2.2 of Berti, Pratelli and Rigo, (2006). For any
fixed outcome such that the previous convergence holds, also $F_{n}^{\ast
-1}(q_{i})\rightarrow F_{M,\ell }^{-1}(q_{i})$, $i=1,2$, hold for the sample
paths of the respective conditional quantile functions, provided that $%
q_{1},q_{2}$ are continuity points of the sample path of $F_{M,\ell }^{-1}$.
If $q_{1},q_{2}$ are continuity points of almost all sample paths of $%
F_{M,\ell }^{-1}$, it follows that $F_{n}^{\ast -1}(q_{i})\rightarrow
_{a.s.}F_{M,\ell }^{-1}(q_{i})$, $i=1,2$. Therefore, on a general
probability space, 
\begin{equation}
(\tau _{n},F_{n}^{\ast -1}(q_{1}),F_{n}^{\ast -1}(q_{2}),(\tau _{n}^{\ast
}|D_{n}))\overset{w^{\ast }}{\rightarrow }_{w}(\tau ,F_{M,\ell
}^{-1}(q_{1}),F_{M,\ell }^{-1}(q_{2}),\tau ^{\ast }|(M,\ell (\theta _{0})))
\label{eq quant}
\end{equation}%
provided that $F_{M,\ell }^{-1}$ is a.s. continuous at $q_{1},q_{2}$.

\subsection{Details of the proof of Theorem \protect\ref{Lemma bootstrap
with boundary}}

\label{se wedge}

Let $g^{\ast}(\theta_{0})=g(\theta_{0})$ throughout this subsection. For
outcomes such that $\tilde{\theta}^{\ast}\notin\Theta^{\ast}$ and $%
\lambda_{\min}(M_{n})>0$, the quadratic function $q_{n}^{\ast}$ is not
minimized over $\Theta^{\ast}$ at any interior point of $\Theta^{\ast}$ (for
otherwise this point would have to be the stationary point $\tilde{\theta}%
^{\ast}$$\notin\Theta^{\ast}$ of $q_{n}^{\ast}$, a contradiction). For such
outcomes, if $q_{n}^{\ast}$ is at all minimized over $\Theta^{\ast}$, then
this has to occur at a boundary point of $\Theta^{\ast}$. Since $%
\partial\Theta^{\ast}\subseteq\{\theta\in\mathbb{R}^{2}:g(\theta)=g^{\ast}(%
\hat{\theta})\}=:\tilde{\partial}\Theta^{\ast}$, we proceed by constructing
a minimizer of $q_{n}^{\ast}$ over the latter set and by showing that this
minimizer is in fact a global one over $\Theta^{\ast}.$ This (and some added
measurability considerations) establishes the well-definition of $\check{%
\theta}^{\ast}$ in (\ref{eq decomposition for the restricted estimator}).
Then we establish the $n^{-1/2}$ consistency rate of $\check{\theta}^{\ast}$
in the sense that $\Vert\check{\theta}^{\ast}-\hat{\theta}\Vert(1-\mathbb{I}%
_{n}^{\ast})=O_{p^{\ast}}(n^{-1/2})$ a.s.\medskip{}

\noindent \textsc{Step 1. Existence of a minimizer of} $q_{n}^{\ast }$ 
\textsc{over a portion of} $\tilde{\partial}\Theta ^{\ast }$ \textsc{close to%
} $\theta _{0}$. The point $(\theta ^{\prime },c)^{\prime }=(\theta
_{0}^{\prime },g(\theta _{0}))^{\prime }\in \mathbb{R}^{3}$ trivially
satisfies the equation $g(\theta )=c$. Since $g$ is continuously
differentiable in a neighborhood of $\theta _{0}$ and $\dot{g}=(\dot{g}%
_{1}(\theta _{0}),\dot{g}_{2}(\theta _{0}))^{\prime }\neq 0$ (say that $\dot{%
g}_{1}(\theta _{0})\neq 0$, with the subscript denoting partial
differentiation), by the implicit function theorem there exist an $r>0$ and
a unique function $\gamma :[\theta _{2,0}-r,\theta _{2,0}+r]\times \lbrack
g(\theta _{0})-r,g(\theta _{0})+r]\rightarrow \lbrack \theta _{1,0}-r,\theta
_{1,0}+r]$ such that $\gamma (\theta _{2,0},g(\theta _{0}))=\theta _{1,0}$, $%
g(\gamma (\theta _{2},c),\theta _{2})=c$;\ moreover, $\gamma $ is
continuously differentiable. For outcomes such that $|g^{\ast }(\hat{\theta}%
)-g(\theta _{0})|\leq r$, the (non-empty) portion of the curve $\tilde{%
\partial}\Theta ^{\ast }=\{\theta \in \mathbb{R}^{2}:g(\theta )=g^{\ast }(%
\hat{\theta})\}$ contained in the square $\Pi :=[\theta _{1,0}-r,\theta
_{1,0}+r]\times \lbrack \theta _{2,0}-r,\theta _{2,0}+r]$ can be
parameterized as $\theta _{1}=\gamma (\theta _{2},g^{\ast }(\hat{\theta}))$, 
$\theta _{2}\in \lbrack \theta _{2,0}-r,\theta _{2,0}+r]$. Define $\check{%
\theta}^{\ast }:=(\gamma (\check{\theta}_{2}^{\ast },g^{\ast }(\hat{\theta}%
^{r})),\check{\theta}_{2}^{\ast })^{\prime }$, where $\check{\theta}%
_{2}^{\ast }$ is a measurable minimizer of the continuous function $%
q_{n}^{\ast }(\gamma (\theta _{2},g^{\ast }(\hat{\theta}^{r})),\theta _{2})$
over $\theta _{2}\in \lbrack \theta _{2,0}-r,\theta _{2,0}+r]$, with $\hat{%
\theta}^{r}:=\hat{\theta}\mathbb{I}_{\{|g^{\ast }(\hat{\theta})-g(\theta
_{0})|\leq r\}}+\theta _{0}\mathbb{I}_{\{|g^{\ast }(\hat{\theta})-g(\theta
_{0})|>r\}}$. Since $\mathbb{I}_{\{|g^{\ast }(\hat{\theta})-g(\theta
_{0})|\leq r\}}\overset{a.s.}{\rightarrow }1$ under $g^{\ast }(\theta
_{0})=g(\theta _{0})$, it follows that $\check{\theta}^{\ast }$ minimizes $%
q_{n}^{\ast }$ over $\tilde{\partial}\Theta ^{\ast }\cap \Pi $ with $P^{\ast
}$-probability approaching one a.s.\medskip {}

\noindent \textsc{Step 2. Minimization of} $q_{n}^{\ast }$ \textsc{over the
entire bootstrap parameter space.} For outcomes in 
\begin{equation*}
\mathcal{A}_{n}:=\{|g^{\ast }(\hat{\theta})-g(\theta _{0})|\leq r\}\cap \{g(%
\tilde{\theta}^{\ast })<g^{\ast }(\hat{\theta})\}\cap \{\Vert \hat{\theta}%
-\theta _{0}\Vert +\Vert \tilde{\theta}^{\ast }-\hat{\theta}\Vert \leq 
\tfrac{r}{2}\},
\end{equation*}%
the minimum of $q_{n}^{\ast }$ over the entire bootstrap parameter space $%
\Theta ^{\ast }$ exists and is attained only in $\Pi $ (e.g., at $\check{%
\theta}^{\ast }$ defined in Step 1), provided that 
\begin{equation*}
\alpha _{n}:=\lambda _{\min }(M_{n})\tfrac{r^{2}}{4}-\lambda _{\max
}(M_{n})\Vert \tilde{\theta}^{\ast }-\hat{\theta}\Vert ^{2}>0.
\end{equation*}%
To see this, consider $\theta ^{c}:=c\hat{\theta}+(1-c)\tilde{\theta}^{\ast
} $ where $c:=\inf \{a\in \lbrack 0,1]:b(a\hat{\theta}+(1-a)\tilde{\theta}%
^{\ast })=0\}$; $\theta ^{c}$ is well-defined whenever $g(\tilde{\theta}%
^{\ast })<g^{\ast }(\hat{\theta})$ because $g(\hat{\theta})\geq g^{\ast }(%
\hat{\theta})$ and $b$ is continuous. Moreover, $\theta ^{c}\in \Pi $ for
outcomes in $\mathcal{A}_{n}$ because $\Vert \theta ^{c}-\theta _{0}\Vert
\leq \Vert \hat{\theta}-\theta _{0}\Vert +\Vert \tilde{\theta}^{\ast }-\hat{%
\theta}\Vert \leq \frac{r}{2}$ and, hence, $q_{n}^{\ast }(\theta ^{c})\geq
q_{n}^{\ast }(\check{\theta}^{\ast })$ for outcomes in $\mathcal{A}_{n}$, by
the minimizing property of $\check{\theta}^{\ast }$ on $\tilde{\partial}%
\Theta ^{\ast }\cap \Pi $ and the fact that $b(\theta ^{c})=0$. For any $%
\theta \not\in \Pi $ and outcomes in $\mathcal{A}_{n}$, we therefore find
that 
\begin{eqnarray*}
q_{n}^{\ast }(\theta )-q_{n}^{\ast }(\check{\theta}^{\ast }) &\geq
&q_{n}^{\ast }(\theta )-q_{n}^{\ast }(\theta ^{c})=q_{n}^{\ast }(\theta
)-q_{n}^{\ast }(\tilde{\theta}^{\ast })+q_{n}^{\ast }(\tilde{\theta}^{\ast
})-q_{n}^{\ast }(\theta ^{c}) \\
&\geq &\lambda _{\min }(M_{n})\Vert \theta -\tilde{\theta}^{\ast }\Vert
^{2}-\lambda _{\max }(M_{n})\Vert \tilde{\theta}^{\ast }-\theta ^{c}\Vert
^{2} \\
&\geq &\lambda _{\min }(M_{n})\{\Vert \theta -\theta _{0}\Vert -\Vert \tilde{%
\theta}^{\ast }-\theta _{0}\Vert \}^{2}-\lambda _{\max }(M_{n})\Vert \tilde{%
\theta}^{\ast }-\hat{\theta}\Vert ^{2} \\
&\geq &\lambda _{\min }(M_{n})\{r-\Vert \tilde{\theta}^{\ast }-\hat{\theta}%
\Vert -\Vert \hat{\theta}-\theta _{0}\Vert \}^{2}-\lambda _{\max
}(M_{n})\Vert \tilde{\theta}^{\ast }-\hat{\theta}\Vert ^{2} \\
&\geq &\lambda _{\min }(M_{n})\tfrac{r^{2}}{4}-\lambda _{\max }(M_{n})\Vert 
\tilde{\theta}^{\ast }-\hat{\theta}\Vert ^{2}=\alpha _{n}.
\end{eqnarray*}%
Thus, for outcomes in $\mathcal{A}_{n}\cap \{\alpha _{n}>0\}$, $q_{n}^{\ast
} $ out of $\Pi $ is larger than $\min_{\theta \in \tilde{\partial}\Theta
^{\ast }\cap \Pi }q_{n}^{\ast }(\theta )$. As $\tilde{\partial}\Theta ^{\ast
}\subseteq \Theta ^{\ast },$ it follows that $\min_{\theta \in \Theta ^{\ast
}\cap \Pi }q_{n}^{\ast }(\theta )$ (which exists) for such outcomes is
actually $\min_{\theta \in \Theta ^{\ast }}q_{n}^{\ast }(\theta )$.
Moreover, 
\begin{equation*}
\text{$\min_{\theta \in \Theta ^{\ast }}q_{n}^{\ast }$($\theta $)=$%
\min_{\theta \in \Theta ^{\ast }\cap \Pi }q_{n}^{\ast }$($\theta $)=$%
\min_{\theta \in \tilde{\partial}\Theta ^{\ast }\cap \Pi }q_{n}^{\ast }$($%
\theta $)=$\min_{\theta \in \partial \Theta ^{\ast }\cap \Pi }q_{n}^{\ast }$(%
$\theta $)},
\end{equation*}%
for if $\min_{\theta \in \Theta ^{\ast }\cap \Pi }q_{n}^{\ast }(\theta )<$$%
\min_{\theta \in \partial \Theta ^{\ast }\cap \Pi }q_{n}^{\ast }(\theta )$,
then $\min_{\theta \in \Theta ^{\ast }\cap \Pi }q_{n}^{\ast }(\theta )$ (and
thus, $\min_{\theta \in \Theta ^{\ast }}q_{n}^{\ast }(\theta )$) is achieved
at an interior point of $\Theta ^{\ast }$, which can only be $\tilde{\theta}%
^{\ast }$, a contradiction with $\tilde{\theta}^{\ast }\notin \Theta ^{\ast
} $ (i.e., with $g(\tilde{\theta}^{\ast })<g^{\ast }(\hat{\theta})$). To
summarize, for outcomes in $\mathcal{A}_{n}\cap \{\alpha _{n}>0\}$, $\check{%
\theta}^{\ast }$ minimizes $q_{n}^{\ast }$ over $\Theta ^{\ast }$ and is at
the boundary of $\Theta ^{\ast }$.

We find the associated probability 
\begin{eqnarray*}
& & P^{\ast}\left((1-\mathbb{I}_{n}^{\ast})q_{n}^{\ast}(\check{\theta}%
^{\ast})<(1-\mathbb{I}_{n}^{\ast})q_{n}(\theta)~~\forall\theta\in\Theta^{%
\ast}\setminus\Pi\right) \\
& & \hspace{0.5in}\overset{}{\geq}P^{\ast}\left(|g^{\ast}(\hat{\theta}%
)-g(\theta_{0})|\leq r,\Vert\hat{\theta}-\theta_{0}\Vert\leq\tfrac{r}{4}%
,\Vert\tilde{\theta}^{\ast}-\hat{\theta}\Vert\leq\frac{r}{4}%
,\alpha_{n}>0\right) \\
& & \hspace{0.5in}\overset{}{=}\mathbb{I}_{\{|g^{\ast}(\hat{\theta}%
)-g(\theta_{0})|\leq r\}\cap\{\Vert\hat{\theta}-\theta_{0}\Vert\leq
r/4\}}P^{\ast}\left(\Vert\tilde{\theta}^{\ast}-\hat{\theta}\Vert\leq\tfrac{r%
}{4},\alpha_{n}>0\right)\overset{a.s.}{\rightarrow}1
\end{eqnarray*}
because $g(\hat{\theta})\overset{a.s.}{\rightarrow}g(\theta_{0})$, $%
\lambda_{\min}(M_{n})\rightarrow\lambda_{\min}(M)>0$ a.s., $%
\lambda_{\max}(M_{n})\rightarrow\lambda_{\max}(M)<\infty$ a.s. and $\Vert%
\tilde{\theta}^{\ast}-\hat{\theta}\Vert\overset{w^{\ast}}{\rightarrow}%
_{a.s.}0$. This establishes the fact that $\hat{\theta}^{\ast}$ of (\ref{eq
decomposition for the restricted estimator}), with $\check{\theta}^{\ast}$
as defined in Step 1, minimizes $q_{n}^{\ast}$ over the bootstrap parameter
space $\Theta^{\ast}$ with $P^{\ast}$-probability approaching one a.s.%
\medskip{}

\noindent \textsc{Step 3. Consistency rate of} $\check{\theta}^{\ast }$.
Similarly to Step 2, for outcomes in $\mathcal{A}_{n}$, 
\begin{equation*}
0\geq q_{n}^{\ast }(\check{\theta}^{\ast })-q_{n}^{\ast }(\theta ^{c})\geq
\lambda _{\min }(M_{n})\Vert \check{\theta}^{\ast }-\tilde{\theta}^{\ast
}\Vert ^{2}-\lambda _{\max }(M_{n})\Vert \tilde{\theta}^{\ast }-\hat{\theta}%
\Vert ^{2}\text{,}
\end{equation*}%
the first inequality by the minimizing property of $\check{\theta}^{\ast }$
over $\tilde{\partial}\Theta ^{\ast }\cap \Pi $. Therefore, 
\begin{eqnarray*}
&&P^{\ast }\left( (1-\mathbb{I}_{n}^{\ast })\Vert \check{\theta}^{\ast }-%
\tilde{\theta}^{\ast }\Vert ^{2}\leq (1-\mathbb{I}_{n}^{\ast })\frac{\lambda
_{\max }(M_{n})}{\lambda _{\min }(M_{n})}\Vert \tilde{\theta}^{\ast }-\hat{%
\theta}\Vert ^{2}\right) \\
&\geq &P^{\ast }\left( |g^{\ast }(\hat{\theta})-g(\theta _{0})|\leq r,\Vert 
\hat{\theta}-\theta _{0}\Vert \leq \tfrac{r}{4},\Vert \tilde{\theta}^{\ast }-%
\hat{\theta}\Vert \leq \tfrac{r}{4}\right) \\
&=&\mathbb{I}_{\{|g^{\ast }(\hat{\theta})-g(\theta _{0})|\leq r\}\cap
\{\Vert \hat{\theta}-\theta _{0}\Vert \leq r/4\}}P^{\ast }\left( \Vert 
\tilde{\theta}^{\ast }-\hat{\theta}\Vert \leq \tfrac{r}{4}\right) \overset{%
a.s.}{\rightarrow }1.
\end{eqnarray*}%
As $\lambda _{\max }(M_{n})/\lambda _{\min }(M_{n})\overset{a.s.}{%
\rightarrow }\lambda _{\max }(M)/\lambda _{\min }(M)$ and $\Vert \tilde{%
\theta}^{\ast }-\hat{\theta}\Vert =O_{p^{\ast }}(n^{-1/2})$ $P$-a.s. (the
latter, by (\ref{eq separata})), it follows that $(1-\mathbb{I}_{n}^{\ast
})\Vert \check{\theta}^{\ast }-\tilde{\theta}^{\ast }\Vert =O_{p^{\ast
}}(n^{-1/2})$ $P$-a.s. and $\Vert \hat{\theta}^{\ast }-\tilde{\theta}^{\ast
}\Vert =O_{p^{\ast }}(n^{-1/2})$ $P$-a.s. for $\hat{\theta}^{\ast }$ of (\ref%
{eq decomposition for the restricted estimator}). Thus, $\hat{\theta}^{\ast
} $ has the same consistency rate as $\tilde{\theta}^{\ast }$. This argument
applies to any $\check{\theta}^{\ast }$ which is measurable and minimizes $%
q_{n}^{\ast }$ over $\tilde{\partial}\Theta ^{\ast }\cap \Pi $ for outcomes
in $\mathcal{A}_{n}$. This completes Step 3.\medskip {}

Finally, consider the first-order condition {[}foc{]} for minimization of $%
q_{n}^{\ast}$ on $\tilde{\partial}\Theta^{\ast}$. As $\Vert\check{\theta}%
^{\ast}-\theta_{0}\Vert(1-\mathbb{I}_{n}^{\ast})\leq\{\Vert\check{\theta}%
^{\ast}-\tilde{\theta}^{\ast}\Vert+\Vert\check{\theta}^{\ast}-\theta_{0}%
\Vert\}(1-\mathbb{I}_{n}^{\ast})\overset{w^{\ast}}{\rightarrow}_{a.s.}0$, it
follows that $\mathbb{I}_{\{\check{\theta}^{\ast}\in\mathrm{\limfunc{int}}%
(\Pi)\}}(1-\mathbb{I}_{n}^{\ast})+\mathbb{I}_{n}^{\ast}\overset{w}{%
\rightarrow}_{a.s.}1$. As additionally $\dot{g}(\theta_{0})\neq0$, by
continuity of $\dot{g}(\theta):=(\partial g/\partial\theta^{\prime})(\theta)$%
, the foc takes the form 
\begin{equation*}
P^{\ast}\left(\{\dot{q}_{n}(\check{\theta}^{\ast})+\check{\delta}_{n}\dot{g}(%
\check{\theta}^{\ast})\}(1-\mathbb{I}_{n}^{\ast})=0\right)\overset{a.s.}{%
\rightarrow}1,
\end{equation*}
where $\check{\delta}_{n}\in\mathbb{R}$ are measurable Lagrange multipliers
that can be determined, for outcomes in the event ${\mathbb{I}_{n}^{\ast}=1}$%
, by involving also the constraint $b(\check{\theta}^{\ast})(1-\mathbb{I}%
_{n}^{\ast})=0$.$\hfill\square$

\subsection{Proof of Corollary \protect\ref{corollary bootstrap with
boundary}}

We only discuss the bootstrap validity part of the corollary, as the
convergence part (\ref{eq bound ave}) was explained in the main text.

Let $\tau _{n}:=\phi (n^{1/2}(\hat{\theta}-\theta _{0}))$, $\tau _{n}^{\ast
}:=\phi (n^{1/2}(\hat{\theta}^{\ast }-\hat{\theta}))$ and $\tau :=\phi (\ell
(\theta _{0}))$. Convergence (\ref{eq bound ave}) and the continuity of $%
\phi $ imply that 
\begin{equation*}
(\tau _{n},(\tau _{n}^{\ast }|D_{n}))\overset{w}{\rightarrow }_{w}(\tau
,(\tau |M)).
\end{equation*}%
If the (random) cdf of $\tau |M$ is sample-path continuous, bootstrap
validity follows from Theorem 3.1 and Lemma A.2(b) of Cavaliere and Georgiev
(2020). We reduce the general case to the globally continuous case by a
local argument for the cdf's $F(\cdot ):=P(\tau \leq \cdot )$ and $%
F_{M}(\cdot ):=P(\tau \leq \cdot |M)$. For concreteness, we focus on the
technically more involved possibility $g(\theta _{0})=0$, such that $\theta
_{0}\in \partial \Theta $ given the assumption $\dot{g}\neq 0$. With 
\begin{equation*}
l(B):=\dot{g}_{\perp }(\dot{g}_{\perp }^{\prime }B\dot{g}_{\perp })^{-1}\dot{%
g}_{\perp }^{\prime }B^{1/2}\xi +B^{-1}\dot{g}(\dot{g}^{\prime }B^{-1}\dot{g}%
)^{-1}\max \{0,\dot{g}^{\prime }B^{-1/2}\xi \}
\end{equation*}%
for positive definite $B\in \mathbb{R}^{2\times 2}$ and with $\ell =l(M)$,
notice the following. If $B$ is a fixed positive definite matrix such that 
\begin{equation}
P\left( \phi \left( l(B)\right) =a\right) >0  \label{eq:proba}
\end{equation}%
for some $a\in \mathbb{R}$, then by equivalence (i.e., mutual absolute
continuity) considerations for non-singular Gaussian distributions, also 
\begin{equation*}
P\left( \phi \left( l(D)\right) =a\right) >0
\end{equation*}%
for any positive definite $D\in \mathbb{R}^{2\times 2}$. In fact, let $\psi :%
\mathbb{R}\rightarrow \mathbb{R}$ be defined as $\psi (\cdot ):=\phi (\dot{g}%
_{\perp }(\cdot ))$ and let $\phi ^{\leftarrow }(\cdot ),\psi ^{\leftarrow
}(\cdot )$ denote inverse images. Then the probability in (\ref{eq:proba})
equals 
\begin{align*}
P\left( l(B)\in \phi ^{\leftarrow }(\{a\})\cap \partial \Lambda \right) 
\overset{}{+}P(l(B)& \in \phi ^{\leftarrow }(\{a\})\cap \mathrm{\limfunc{int}%
}\Lambda ) \\
& \overset{}{=}P(\{\dot{g}^{\prime }B^{-1/2}\xi \leq 0\}\cap \{(\dot{g}%
_{\perp }^{\prime }B\dot{g}_{\perp })^{-1}\dot{g}_{\perp }^{\prime
}B^{1/2}\xi \in \psi ^{\leftarrow }(\{a\})\}) \\
& \overset{}{+}P(\{\dot{g}^{\prime }B^{-1/2}\xi >0\}\cap \{B^{-1/2}\xi \in
\phi ^{\leftarrow }(\{a\})\}) \\
& \overset{}{=}P(\dot{g}^{\prime }B^{-1/2}\xi \leq 0)P((\dot{g}_{\perp
}^{\prime }B\dot{g}_{\perp })^{-1}\dot{g}_{\perp }^{\prime }B^{1/2}\xi \in
\psi ^{\leftarrow }(\{a\})\}) \\
& \overset{}{+}P(B^{-1/2}\xi \in \phi ^{\leftarrow }(\{a\})\cap \mathrm{%
\limfunc{int}}\Lambda ),
\end{align*}%
the equality because $Cov(\dot{g}^{\prime }B^{-1/2}\xi ,(\dot{g}_{\perp
}^{\prime }B\dot{g}_{\perp })^{-1}\dot{g}_{\perp }^{\prime }B^{1/2}\xi )=0$
and $\xi $ is Gaussian. In the previous display, $P(\dot{g}^{\prime
}B^{-1/2}\xi \leq 0)=P(N(0,\dot{g}^{\prime }B^{-1}\dot{g})\leq 0)>0$ for all
positive definite~$B$, 
\begin{equation*}
P((\dot{g}_{\perp }^{\prime }B\dot{g}_{\perp })^{-1}\dot{g}_{\perp }^{\prime
}B^{1/2}\xi \in \psi ^{\leftarrow }(\{a\})\})=P(N(0,(\dot{g}_{\perp
}^{\prime }B\dot{g}_{\perp })^{-1})\in \psi ^{\leftarrow }(\{a\}))
\end{equation*}%
is either $0$ for all positive definite $B$ or positive for all positive
definite $B$, and the same applies to $P(B^{-1/2}\xi \in \phi ^{\leftarrow
}(\{a\})\cap $$\limfunc{int}$$\Lambda )$. Therefore, the sign of the
probability in (\ref{eq:proba}) is the same (zero or positive) for all
positive definite $B$.

The cdf $F_{M}$ is a measurable transformation of $M$ determined a.s.
uniquely by the distribution of ($M,\xi$); it can be identified (up to a set
of measure zero) as 
\begin{align*}
F_{M} &
(\cdot)=\left.P\left(\phi\left(l(B)\right)\leq\cdot\right)\right|_{B=M}
\end{align*}
by virtue of the independence of $M$ and $\xi$. Since $M$ is positive
definite a.s., from the argument in the previous paragraph we can conclude
that every point on the line is either a discontinuity point of almost all
sample paths of $F_{M},$ or a continuity point of almost all sample paths of 
$F_{M}$. By averaging, a point on the line is a discontinuity point of $F$
if and only if it is a discontinuity point of almost all sample paths of $%
F_{M}.$

Let now $q_{0}$ be an interior point of the set 
\begin{equation*}
C=\{q\in (0,1):\lim_{n\rightarrow \infty }P(F(\tau _{n})\leq q)\rightarrow q|%
\mathsf{H}_{0}\}
\end{equation*}%
such that the asymptotic test is correctly sized for $q\in (q_{0}-2\epsilon
,q_{0}+2\epsilon )\subset (0,1)$ for some $\epsilon >0$. As $\tau _{n}%
\overset{w}{\rightarrow }\tau \sim F$, this implies that $F$ and (by the
discussion in previous paragraph) $F_{M}$ skip no values from the interval $%
(q_{0}-2\epsilon ,q_{0}+2\epsilon )$ (for $F_{M}$, a.s.). In particular,
almost all sample paths of $F_{M}$ are continuous on the (random) open
superset $\left( F_{M}^{-1}(q_{0}-\frac{3}{2}\epsilon ),F_{M}^{-1}(q_{0}+%
\frac{3}{2}\epsilon )\right) $ of $I_{\epsilon }:=[F_{M}^{-1}(q_{0}-\epsilon
),F_{M}^{-1}(q_{0}+\epsilon )]$, with 
\begin{equation}
F_{M}^{-1}(q_{0}-\frac{3}{2}\epsilon )<F_{M}^{-1}(q_{0}-\epsilon
)<F_{M}^{-1}(q_{0}+\epsilon )<F_{M}^{-1}(q_{0}+\frac{3}{2}\epsilon )~~\text{%
a.s.}
\end{equation}%
Without loss of generality, $\epsilon $ can be considered such that $%
q_{0}\pm \epsilon $ are continuity points of $F_{M}^{-1}$ a.s. (because $%
F_{M}^{-1}$ is chosen to be c\`{a}dl\`{a}g and its discontinuity points on,
say $[\frac{q_{0}}{2},\frac{q_{0}+1}{2}]$ are countably many). Let $\Psi
^{-}(a,x)$ and $\Psi ^{+}(a,x)$ be generalized inverses of the cdf's of a
standard Gaussian variable conditioned to take values respectively in $%
(-\infty ,a]$ and $[a,\infty )$. On extensions of the probability spaces
where the data and $(\tau ,M)$ are defined, consider a $U_{[0,1]}$ variable $%
\upsilon $. Define $F_{n}^{\ast }(\cdot ):=P^{\ast }(\tau _{n}^{\ast }\leq
\cdot )$, $I_{n,\epsilon }:=[F_{n}^{\ast -1}(q_{0}-\epsilon ),F_{n}^{\ast
-1}(q_{0}+\epsilon )]$ and 
\begin{align*}
\tilde{\tau}_{n}& =\tau _{n}\mathbb{I}_{\{\tau _{n}\in I_{n,\epsilon
}\}}+\Psi ^{-}(F_{n}^{\ast -1}(q_{0}-\epsilon ),\upsilon )\mathbb{\mathbb{I}}%
_{\{\tau _{n}<F_{n}^{\ast -1}(q_{0}-\epsilon )\}} \\
& +\Psi ^{+}(F_{n}^{\ast -1}(q_{0}+\epsilon ),\upsilon )\mathbb{\mathbb{I}}%
_{\{\tau _{n}>F_{n}^{\ast -1}(q_{0}+\epsilon )\}}, \\
\tilde{\tau}_{n}^{\ast }& =\tau _{n}^{\ast }\mathbb{I}_{\{\tau _{n}^{\ast
}\in I_{n,\epsilon }\}}+\Psi ^{-}(F_{n}^{\ast -1}(q_{0}-\epsilon ),\upsilon )%
\mathbb{\mathbb{I}}_{\{\tau _{n}^{\ast }<F_{n}^{\ast -1}(q_{0}-\epsilon )\}}
\\
& +\Psi ^{+}(F_{n}^{\ast -1}(q_{0}+\epsilon ),\upsilon )\mathbb{\mathbb{I}}%
_{\{\tau _{n}^{\ast }>F_{n}^{\ast -1}(q_{0}+\epsilon )\}},\text{{}} \\
\tilde{\tau}& =\tau \mathbb{I}_{\{\tau \in I_{\epsilon }\}}+\Psi
^{-}(F_{M}^{-1}(q_{0}-\epsilon ),\upsilon )\mathbb{\mathbb{I}}_{\{\tau
<F_{M}^{-1}(q_{0}-\epsilon )\}} \\
& +\Psi ^{+}(F_{M}^{-1}(q_{0}+\epsilon ),\upsilon )\mathbb{\mathbb{I}}%
_{\{\tau >F_{M}^{-1}(q_{0}+\epsilon )\}}.
\end{align*}%
Then 
\begin{equation*}
(\tilde{\tau}_{n},(\tilde{\tau}_{n}^{\ast }|D_{n}))\overset{w}{\rightarrow }%
_{w}(\tilde{\tau},(\tilde{\tau}|M))
\end{equation*}%
because 
\begin{equation*}
(f_{1}(\tilde{\tau}_{n}),E\{f_{2}(\tilde{\tau}_{n}^{\ast })|D_{n}\})\overset{%
w}{\rightarrow }(f_{1}(\tilde{\tau}_{n}),E\{f_{2}(\tilde{\tau})|M\})
\end{equation*}%
for any continuous and bounded real functions $f_{1},f_{2},$ as a result of (%
\ref{eq quant}) with $\tau ^{\ast }|(M,\ell (\theta _{0})=\tau |M$ in the
sense of a.s. equality of conditional distributions and the fact that $%
P(\tau =F_{M}^{-1}(q_{0}\pm \epsilon )|M)=0$ a.s. by sample-path continuity
of $F_{M}$ an open superset of $I_{\epsilon }$. As the cdf of $\tilde{\tau}%
|M $ is a.s. sample-path continuous by construction, it follows that $%
P^{\ast }(\tilde{\tau}_{n}^{\ast }\leq \tilde{\tau}_{n})\overset{w}{%
\rightarrow }U_{[0,1]}$, by Theorem 3.1 and Lemma A.2(b) of Cavaliere and
Georgiev (2020).

Let $\tilde{F}_{n}^{\ast }(\cdot ):=P^{\ast }(\tilde{\tau}_{n}^{\ast }\leq
\cdot )$. We now return to the original variables. By considerations of
equalities of events, it holds that 
\begin{equation*}
P(F_{n}^{\ast }(\tau _{n})\leq q_{0})=P(F_{n}^{\ast }(\tilde{\tau}_{n})\leq
q_{0})=P(\tilde{F}_{n}^{\ast }(\tilde{\tau}_{n})\leq q_{0})=P(P^{\ast }(%
\tilde{\tau}_{n}^{\ast }\leq \tilde{\tau}_{n})\leq q_{0})=q_{0}
\end{equation*}%
using the fact that $P^{\ast }(\tilde{\tau}_{n}^{\ast }\leq \tilde{\tau}_{n})%
\overset{w}{\rightarrow }U_{[0,1]}$. This completes the proof.

\newpage {}

\begin{table}[h!]
\begin{footnotesize}
\begin{center}%
\caption*{\textsc{Table 1:} Empirical rejection probabilities (ERPs) of bootstrap tests under the null.}\label{tab:MC}%
\resizebox{1\textwidth}{!}{%
\begin{tabular}
[c]{cc||c|cccc||c|cccc||c|cccccccccc}\hline \hline
\multicolumn{10}{l}{Nominal level: $0.05$} \\\hline \hline
& & \multicolumn{3}{l}{$\theta_0 = (0,0)'$}  & & &  \multicolumn{3}{l}{$\theta_0 = (-0.75,0.75)'$} & & &  \multicolumn{3}{l}{$\theta_0 = (-1.50,1.50)'$}  \\\hline
&  & $b_1$ & $b_2$ &  &  &  &  $b_1$ & $b_2$ &  &  &  &  $b_1$ & $b_2$ \\
  & & &    $\kappa$ & &  & & & $\kappa$ &  & & & & $\kappa$
\\\cline{4-17}
dist.  & $n$ & & 0.25 & 0.50 & 1.0  & 2.0  & & 0.25 & 0.50 & 1.0  & 2.0 & & 0.25 & 0.50 & 1.0  & 2.0
\\\hline
$\xi_1$ &  100 &   4.2 & 4.7 & 5.0 & 5.3 & 5.4 & 6.9 & 7.0 & 7.2 & 7.3 & 7.5 & 6.3 & 6.3 & 6.3 & 6.4 & 6.5 \\ 
& 400 &    3.9 & 4.8 & 5.1 & 5.3 & 5.3 & 5.5 & 5.6 & 5.8 & 6.2 & 6.7 & 5.3 & 5.3 & 5.3 & 5.3 & 5.3 \\
& 800 &    3.7 & 4.8 & 5.1 & 5.2 & 5.2 & 5.2 & 5.3 & 5.4 & 5.6 & 6.2 & 5.2 & 5.2 & 5.2 & 5.2 & 5.2 \\\hline
$\xi_2$ &  100 &   4.2 & 4.7 & 5.0 & 5.3 & 5.5 & 7.1 & 7.3 & 7.4 & 7.6 & 7.8 & 6.2 & 6.3 & 6.3 & 6.4 & 6.5 \\ 
& 400 &    3.8 & 4.7 & 5.0 & 5.1 & 5.2 & 5.7 & 5.9 & 6.1 & 6.4 & 6.9 & 5.3 & 5.3 & 5.3 & 5.3 & 5.3 \\
& 800 &    3.6 & 4.6 & 4.8 & 4.9 & 4.9 & 5.1 & 5.2 & 5.3 & 5.5 & 6.0 & 5.1 & 5.1 & 5.1 & 5.1 & 5.1 \\\hline
$\xi_3$ &  100 &   4.3 & 4.7 & 5.0 & 5.3 & 5.5 & 7.1 & 7.2 & 7.3 & 7.5 & 7.7 & 6.4 & 6.4 & 6.4 & 6.5 & 6.6 \\  
& 400 &    3.7 & 4.6 & 4.9 & 5.1 & 5.1 & 5.5 & 5.7 & 5.9 & 6.2 & 6.7 & 5.2 & 5.2 & 5.2 & 5.2 & 5.2 \\
& 800 &    3.7 & 4.8 & 5.0 & 5.1 & 5.2 & 5.1 & 5.2 & 5.3 & 5.5 & 6.0 & 5.1 & 5.1 & 5.1 & 5.1 & 5.1 \\\hline \hline
\multicolumn{10}{l}{Nominal level: $0.10$} \\\hline \hline
& & \multicolumn{3}{l}{$\theta_0 = (0,0)'$}  & & &  \multicolumn{3}{l}{$\theta_0 = (-0.75,0.75)'$} & & &  \multicolumn{3}{l}{$\theta_0 = (-1.50,1.50)'$}  \\\hline
&  & $b_1$ & $b_2$ &  &  &  &  $b_1$ & $b_2$ &  &  &  &  $b_1$ & $b_2$ \\
  & & &    $\kappa$ & &  & & & $\kappa$ &  & & & & $\kappa$
\\\cline{4-17}
dist.  & $n$ & & 0.25 & 0.50 & 1.0  & 2.0  & & 0.25 & 0.50 & 1.0  & 2.0 & & 0.25 & 0.50 & 1.0  & 2.0
\\\hline
$\xi_1$ &  100 &  8.0 & 9.0 & 9.7 & 10.3 & 10.6 & 13.0 & 13.3 & 13.6 & 14.1 & 14.6 & 11.5 & 11.6 & 11.6 & 11.7 & 11.8 \\   
& 400 &   7.7 & 9.5 & 10.1 & 10.4 & 10.5 & 10.4 & 10.5 & 10.8 & 11.3 & 12.4 & 10.3 & 10.3 & 10.3 & 10.3 & 10.3 \\ 
& 800 &    
7.4 & 9.4 & 9.9 & 10.1 & 10.1 & 10.4 & 10.4 & 10.5 & 10.7 & 11.5 & 10.1 & 10.1 & 10.1 & 10.1 & 10.1 \\ \hline
$\xi_2$ &  100 &   8.1 & 9.0 & 9.7 & 10.3 & 10.5 & 13.2 & 13.5 & 13.8 & 14.3 & 14.7 & 11.3 & 11.3 & 11.4 & 11.5 & 11.6 \\  
& 400 &    7.5 & 9.2 & 9.9 & 10.2 & 10.3 & 10.7 & 10.9 & 11.1 & 11.6 & 12.5 & 10.2 & 10.2 & 10.2 & 10.2 & 10.3 \\
& 800 &    7.2 & 9.2 & 9.8 & 10.0 & 10.0 & 10.2 & 10.3 & 10.3 & 10.5 & 11.3 & 10.3 & 10.3 & 10.3 & 10.3 & 10.3 \\\hline
$\xi_3$ &  100 &   8.3 & 9.2 & 9.9 & 10.5 & 10.8 & 13.3 & 13.7 & 14.0 & 14.5 & 15.0 & 11.7 & 11.7 & 11.8 & 11.9 & 12.0 \\  
& 400 &    7.6 & 9.4 & 10.0 & 10.3 & 10.3 & 10.4 & 10.5 & 10.8 & 11.3 & 12.4 & 10.2 & 10.2 & 10.2 & 10.2 & 10.2 \\
& 800 &    7.4 & 9.3 & 9.9 & 10.1 & 10.1 & 10.1 & 10.1 & 10.2 & 10.4 & 11.2 & 10.0 & 10.0 & 10.0 & 10.0 & 10.0 \\\hline \hline
\end{tabular}}%
\end{center}%

Note: bootstrap tests are based on a standard fixed-regressor wild bootstrap ($b_1$) and on the proposed corrected wild bootstrap method ($b_2$) of Section 4, using $g^*=g-|g|^{1+\kappa}$. ERPs are estimated using 50,000 Monte Carlo replications and 999 bootstrap repetitions. The column “dist." shows the distributions of $\varepsilon_t$: $\xi_1 \sim iid N(0,1)$, $\xi_2 \sim ARCH(1)$ and $\xi_3= \sqrt{0.5} v_t + \sqrt{0.5} \varepsilon_{x,t}$, where  $v_t \sim iidN(0,1)$ and $\varepsilon_{x,t}$ is the error term of the predictive variable $x_{n,t}$.%
\end{footnotesize}%
\end{table}%

\newpage {}

\begin{table}[h!]
\begin{footnotesize}
\begin{center}%
\caption*{\textsc{Table 2:} Empirical rejection probabilities (ERPs) of bootstrap tests under local alternatives.}\label{tab:MC}%
\resizebox{1\textwidth}{!}{%
\begin{tabular}
[c]{cc||c|cccc||c|cccc||c|cccccccccc}\hline \hline
\multicolumn{10}{l}{Nominal level: $0.05$} \\\hline \hline
& & \multicolumn{3}{l}{$a_0 = (-3,0)'$}  & & &  \multicolumn{3}{l}{$a_0 = (3,0)'$} & & &  \multicolumn{3}{l}{$a_0 = (5,0)'$}  \\\hline
&  & $b_1$ & $b_2$ &  &  &  &  $b_1$ & $b_2$ &  &  &  &  $b_1$ & $b_2$ \\
  & & &    $\kappa$ & &  & & & $\kappa$ &  & & & & $\kappa$
\\\cline{4-17}
dist.  & $n$ & & 0.25 & 0.50 & 1.0  & 2.0  & & 0.25 & 0.50 & 1.0  & 2.0 & & 0.25 & 0.50 & 1.0  & 2.0
\\\hline
$\xi_1$ &  100 &   21.0 & 21.0 & 21.1 & 21.2 & 21.3 & 40.6 & 40.9 & 41.0 & 41.0 & 41.0 & 68.0 & 68.0 & 68.0 & 68.0 & 68.0 \\ 
& 400 &    18.9 & 19.1 & 19.3 & 19.4 & 19.5 & 38.5 & 38.8 & 38.8 & 38.8 & 38.8 & 64.9 & 64.9 & 64.9 & 64.9 & 64.9 \\
& 800 &    
18.6 & 18.8 & 19.0 & 19.1 & 19.1 & 37.6 & 37.9 & 37.9 & 37.9 & 38.0 & 64.0 & 64.0 & 64.0 & 64.0 & 64.0 \\\hline
$\xi_2$ &  100 &   21.7 & 21.8 & 21.9 & 22.0 & 22.1 & 41.9 & 42.1 & 42.2 & 42.2 & 42.2 & 68.5 & 68.5 & 68.5 & 68.5 & 68.5 \\
& 400 &    19.2 & 19.4 & 19.5 & 19.7 & 19.8 & 38.3 & 38.7 & 38.7 & 38.7 & 38.7 & 64.7 & 64.8 & 64.8 & 64.8 & 64.8 \\
& 800 &    18.6 & 18.8 & 19.0 & 19.1 & 19.1 & 37.8 & 38.1 & 38.1 & 38.1 & 38.1 & 64.2 & 64.2 & 64.2 & 64.2 & 64.2 \\\hline
$\xi_3$ &  100 &   20.6 & 20.7 & 20.8 & 20.8 & 21.0 & 40.8 & 41.0 & 41.1 & 41.1 & 41.1 & 67.3 & 67.3 & 67.3 & 67.3 & 67.3 \\  
& 400 &    19.0 & 19.1 & 19.3 & 19.4 & 19.4 & 38.1 & 38.4 & 38.5 & 38.5 & 38.5 & 65.0 & 65.0 & 65.0 & 65.0 & 65.0 \\
& 800 &    18.3 & 18.5 & 18.8 & 18.9 & 18.9 & 37.7 & 38.0 & 38.1 & 38.1 & 38.1 & 63.5 & 63.5 & 63.5 & 63.5 & 63.5 \\\hline \hline
\multicolumn{10}{l}{Nominal level: $0.10$} \\\hline \hline
& & \multicolumn{3}{l}{$a_0 = (-3,0)'$}  & & &  \multicolumn{3}{l}{$a_0 = (3,0)'$} & & &  \multicolumn{3}{l}{$a_0 = (5,0)'$}  \\\hline
&  & $b_1$ & $b_2$ &  &  &  &  $b_1$ & $b_2$ &  &  &  &  $b_1$ & $b_2$ \\
  & & &    $\kappa$ & &  & & & $\kappa$ &  & & & & $\kappa$
\\\cline{4-17}
dist.  & $n$ & & 0.25 & 0.50 & 1.0  & 2.0  & & 0.25 & 0.50 & 1.0  & 2.0 & & 0.25 & 0.50 & 1.0  & 2.0
\\\hline
$\xi_1$ &  100 &   29.6 & 29.8 & 29.9 & 30.1 & 30.3 & 54.7 & 55.0 & 55.1 & 55.2 & 55.2 & 81.7 & 81.7 & 81.8 & 81.8 & 81.8 \\  
& 400 &    27.0 & 27.3 & 27.7 & 28.1 & 28.2 & 52.2 & 52.6 & 52.7 & 52.7 & 52.7 & 79.6 & 79.6 & 79.6 & 79.6 & 79.6 \\
& 800 &    26.4 & 26.9 & 27.3 & 27.6 & 27.6 & 51.7 & 52.1 & 52.2 & 52.2 & 52.2 & 78.7 & 78.7 & 78.7 & 78.7 & 78.7 \\\hline
$\xi_2$ &  100 &   30.2 & 30.4 & 30.6 & 30.8 & 31.0 & 55.7 & 55.9 & 56.0 & 56.0 & 56.0 & 82.0 & 82.0 & 82.0 & 82.0 & 82.0 \\  
& 400 &    27.1 & 27.4 & 27.9 & 28.2 & 28.3 & 51.8 & 52.1 & 52.2 & 52.2 & 52.2 & 79.3 & 79.3 & 79.3 & 79.3 & 79.3 \\
& 800 &    26.6 & 27.0 & 27.5 & 27.7 & 27.7 & 51.5 & 51.9 & 51.9 & 51.9 & 51.9 & 78.6 & 78.6 & 78.6 & 78.6 & 78.6 \\\hline
$\xi_3$ &  100 &   29.1 & 29.3 & 29.4 & 29.7 & 29.9 & 54.2 & 54.5 & 54.6 & 54.6 & 54.6 & 80.9 & 80.9 & 80.9 & 80.9 & 80.9 \\  
& 400 &    26.7 & 27.0 & 27.4 & 27.7 & 27.8 & 51.7 & 52.1 & 52.2 & 52.2 & 52.2 & 79.4 & 79.4 & 79.4 & 79.4 & 79.4 \\
& 800 &    26.2 & 26.6 & 27.1 & 27.3 & 27.3 & 51.3 & 51.7 & 51.7 & 51.7 & 51.7 & 78.5 & 78.5 & 78.5 & 78.5 & 78.5 \\\hline \hline
\end{tabular}}%
\end{center}%

Note: bootstrap tests are based on a standard fixed-regressor wild bootstrap ($b_1$) and on the proposed corrected wild bootstrap method ($b_2$) of Section 4, using $g^*=g-|g|^{1+\kappa}$. ERPs are estimated using 50,000 Monte Carlo replications and 999 bootstrap repetitions. The column “dist." shows the distributions of $\varepsilon_t$: $\xi_1 \sim iid N(0,1)$, $\xi_2 \sim ARCH(1)$ and $\xi_3= \sqrt{0.5} v_t + \sqrt{0.5} \varepsilon_{x,t}$, where  $v_t \sim iidN(0,1)$ and $\varepsilon_{x,t}$ is the error term of the predictive variable $x_{n,t}$.%
\end{footnotesize}%
\end{table}%

\newpage {}

\appendix

\setcounter{section}{0}
\def\thesection{S.\arabic{section}}
\def\thesubsection{S.\arabic{section}.\arabic{subsection}}%

\renewcommand{\thelemma}{S.\arabic
{lemma}}
\setcounter{lemma}{0}%

\renewcommand{\theequation}{S.\arabic{equation}}
\setcounter{equation}{0}%

\renewcommand{\thetable}{S.\arabic
{table}}
\setcounter{table}{0}%

\renewcommand{\thefigure}{S.\arabic
{figure}}
\setcounter{figure}{0}%

\setcounter{page}{1}%
\newpage{}

\begin{center}
\bigskip{}
\end{center}

\section*{Supplement to: \textquotedblleft Parameters on the boundary in
predictive regression\textquotedblright {*}}

\begin{center}
\mbox{}

\ \vspace{-0.15cm}

\textsc{Giuseppe Cavaliere}$^{a,b}$\textsc{, Iliyan Georgiev}$^{a}$ \textsc{%
and Edoardo Zanelli}$^{a}$

\textsc{\vspace{-0.15in} }

\vspace{0.45cm}

{\small {}First draft: May 3, 2022; Revised: February 1, 2024 and July 1,
2024; \\This version: September 19, 2024}

\renewcommand{\thefootnote}{}
\footnote{
\hspace{-7.2mm}
$^{*}$ Address correspondence to: Giuseppe Cavaliere, Department of Economics, University of Bologna, 
Piazza Scaravilli 2, 40126 Bologna, Italy; email: giuseppe.cavaliere@unibo.it.
\\
$^{a}$ Department of Economics, University of Bologna, Italy.
\\
$^{b}$ Department of Economics, Exeter Business School, UK.}
\addtocounter{footnote}{-2}
\renewcommand{\thefootnote}{\arabic{footnote}}%
\end{center}

\par\begingroup\leftskip=1 cm
\rightskip=1 cm
\small%

\begin{center}
\textsc{{\small {}Abstract\vspace{-0.15cm} }}{\small {} }
\end{center}

{\small {}This document contains some supplemental material for Cavaliere,
Georgiev and Zanelli (2024), CGZ hereafter. In particular, we consider (i)
generalizations of some of the results in CGZ to the near-I(1) and to the
stationary cases; (ii)\ we report additional Monte Carlo simulations. {}}

{\small {}}%
\par\endgroup\normalsize%

{\small \vspace{2cm} }\setcounter{footnote}{0}

\section{Alternative data generating processes}

{\label{sec more DGPs}} The asymptotic theory in the paper is presented
under the assumption that $x_{n,t}$ is a unit-root non-stationary process.
Here we show that the choice of a bootstrap parameter space is fundamental
for bootstrap validity also under alternative stochastic specifications for $%
x_{n,t}$, e.g., a near-unit root and a stationary specification. More
importantly, a common definition of the bootstrap parameter space could be
appropriate for all the considered specifications of $x_{n,t}$. Still, the
functional forms of the limit distributions are not identical across the
specifications of $x_{n,t}$ and, in the stationary case, we perform OLS
estimation under the additional constraint $\hat{\delta}=0$ in (\ref{eq:pr}%
). The implications for bootstrap inference are discussed below.

\subsection{ Near-unit root regressor}

Consider a modification of Assumption 1 where in part (c) the limit process
becomes%
\begin{equation*}
(X,Z)^{\prime }=\left( \int^{\cdot }e^{c(s-\cdot )}dW(s),Z\right) ^{\prime
},\quad c>0,
\end{equation*}%
for a Brownian motion $(W,Z)^{\prime }\sim BM(0,\Omega )$. Thus, $X$ is an
Ornstein-Uhlenbeck process originating from a near-UR posited predicting
variable $x_{n,t}$. The asymptotic distribution of $\hat{\theta}$ has a more
complex structure than in the unit root case. Now $n^{1/2}(\hat{\theta}%
-\theta _{0})\overset{w}{\rightarrow }M^{-1/2}\xi +v_{c}$ with $%
v_{c}:=(0,c\omega _{xz}\omega _{xx}^{-1})^{\prime }$ if $\theta _{0}\in 
\func{int}\Theta $. On the other hand,%
\begin{equation}
n^{1/2}(\hat{\theta}-\theta _{0})\overset{w}{\rightarrow }\underset{\lambda
\in \Lambda }{\arg \min }||\lambda -M^{-1/2}\xi -v_{c}||_{M},\quad \Lambda
:=\{\lambda \in \mathbb{R}^{2}:\dot{g}^{\prime }\lambda \geq 0\}
\label{eq:asy_distribution-1}
\end{equation}%
if $g(\theta _{0})=0$. The limiting shift by $v_{c}$ is due to the fact that 
$n^{1/2}\Delta x_{n,t}$ in the near-unit root case is not a sufficiently
good proxy for the innovations driving $x_{n,t}$. Eqs. (\ref{eq random limit
depending on M only})--(\ref{eq asy for standard BS with param on the
boundary}) for the standard bootstrap hold in the near-unit root case if $X$
in the definition of $M$ is understood as an Ornstein-Uhlenbeck process.
Therefore, the possibility that $\theta _{0}\in \partial \Theta $ induces
the same kind of limiting bootstrap randomness as in the exact unit-root
case. Additionally, the bootstrap limit distribution does not replicate the
shift in the limit distribution of $n^{1/2}(\hat{\theta}-\theta _{0})$
induced by the vector $v_{c}$, as a consequence of the conditional
independence of the bootstrap innovations and the regressor $x_{n,t-1}$.
This fact is not related to the position of $\theta _{0}$ relative to $%
\Theta $ and requires separate treatment. Consider now the bootstrap
estimator of Corollary \ref{corollary bootstrap with boundary} with the
choice $g^{\ast }=g-|g|^{1+\kappa }$ for $\kappa >0$. In the case where $%
x_{n,t}$ is near-unit root non-stationary, instead of (\ref{eq bound ave})
it holds that%
\begin{equation*}
(n^{1/2}(\hat{\theta}-\theta _{0}),(n^{1/2}(\hat{\theta}^{\ast }-\hat{\theta}%
)|D_{n}))\overset{w}{\rightarrow }_{w}\left( M^{-1/2}\xi +v_{c},(M^{-1/2}\xi
|M)\right)
\end{equation*}%
if $g(\theta _{0})>0$, and%
\begin{align*}
(n^{1/2}(\hat{\theta}-\theta _{0}),(n^{1/2}(\hat{\theta}^{\ast }-\hat{\theta}%
)|D_{n}))& \overset{w}{\rightarrow }_{w}\left( \underset{\lambda \in \Lambda 
}{\arg \min }||\lambda -M^{-1/2}\xi -v_{c}||_{M},\right. \\
& \hspace{2cm}\left. 
\Big(%
\underset{\lambda \in \Lambda }{\arg \min }||\lambda -M^{-1/2}\xi ||_{M}%
\Big|%
M%
\Big)%
\right)
\end{align*}%
if $g(\theta _{0})=0$, where $X$ in the definition of $M$ should again be
read as an Ornstein-Uhlenbeck process. This means that $g^{\ast }$ still
does the job it is designed for (remove the random shift from the half-plane
in the limiting bootstrap distribution). Nevertheless, bootstrap invalidity
due to the limiting shift by $v_{c}$, not related to the position of $\theta
_{0}$ in $\Theta $, remains to be tackled.

\subsection{Stationary regressor}

If $x_{n,t}=x_{t}$ is stationary, then the inclusion of $\Delta
x_{n,t}=\Delta x_{t}$ among the regressors of (\ref{eq:pr}) will, in
general, compromise the consistency of $\hat{\theta}$ for the true value $%
\theta _{0}$ in the predictive regression (\ref{eq:prr}). Assume, however,
that $n^{-1}\sum_{t=1}^{n}\tilde{x}_{t}\tilde{x}_{t}^{\prime }\overset{p}{%
\rightarrow }M$ for $\tilde{x}_{t}:=(1,x_{n,t-1})^{\prime }$ and a
non-random positive definite matrix $M$, and that the unconstrained OLS
estimator of $\theta $ from the predictive regression (\ref{eq:prr}) is
consistent at the $n^{-1/2}$ rate and has asymptotic $N(0,\omega
_{zz}M^{-1}) $ distribution. Then the constrained OLS estimator $\hat{\theta}
$ of (\ref{eq:prr}) subject to $g(\hat{\theta})\geq 0$ (equivalently, the
constrained OLS estimator of (\ref{eq:pr}) subject to $g(\hat{\theta})\geq 0$%
, $\hat{\delta}=0$) satisfies $n^{1/2}(\hat{\theta}-\theta _{0})\overset{w}{%
\rightarrow }\ell _{st}(\theta _{0})=\tilde{\ell}_{st}:=M^{-1/2}\zeta $ with 
$\zeta \sim N(0,\omega _{zz}I_{2})$ in the case where $\theta _{0}\in \func{%
int}\Theta $, and%
\begin{equation*}
n^{1/2}(\hat{\theta}-\theta _{0})\overset{w}{\rightarrow }\ell _{st}(\theta
_{0})=\ell _{st}:\underset{\lambda \in \Lambda }{=\arg \min }||\lambda
-M^{-1/2}\zeta ||_{M},\quad \Lambda :=\{\lambda \in \mathbb{R}^{2}:\dot{g}%
^{\prime }\lambda \geq 0\}
\end{equation*}%
in the case where $g(\theta _{0})=0$. In the stationary case with a
non-random limiting $M$, the limiting behavior of the standard bootstrap is
entirely analogous to the introductory location model example, as the
possibility that $\theta _{0}\in \partial \Theta $ is the only source of
bootstrap randomness in the limit. For $\hat{\theta}$ defined in the
previous paragraph, it holds that $n^{1/2}(\hat{\theta}^{\ast }-\hat{\theta})%
\overset{w^{\ast }}{\rightarrow }_{p}M^{-1/2}\zeta ^{\ast }$ with $\zeta
^{\ast }\sim N(0,\omega _{zz}I_{2})$ in the case where $\theta _{0}\in \func{%
int}\Theta $, such that the limit bootstrap distribution is non-random in
this case, and%
\begin{equation*}
n^{1/2}(\hat{\theta}^{\ast }-\hat{\theta})\overset{w^{\ast }}{\rightarrow }%
_{w}%
\Big(%
\underset{\lambda \in \Lambda _{\ell }^{\ast }}{\arg \min }||\lambda
-M^{-1/2}\zeta ^{\ast }||_{M}%
\Big)%
\Big|%
\ell ,\quad \Lambda _{\ell }^{\ast }:=\{\lambda \in \mathbb{R}^{2}:\dot{g}%
^{\prime }\lambda \geq -\dot{g}^{\prime }\ell \},
\end{equation*}%
with $\zeta ^{\ast }|\ell \sim N(0,\omega _{zz}I_{2})$ in the case where $%
g(\theta _{0})=0$. We conclude that the same discrepancy between $\Lambda $
and $\Lambda _{\ell }^{\ast }$ emerges in the case $g(\theta _{0})=0$
irrespective of the stochastic properties of the regressor. Consider now the
bootstrap estimator of Corollary \ref{corollary bootstrap with boundary}
with the choice $g^{\ast }=g-|g|^{1+\kappa }$ for $\kappa >0$. For a
stationary $x_{n,t}$ and a non-random $M$, the original and the bootstrap
estimators satisfy%
\begin{equation*}
(n^{1/2}(\hat{\theta}-\theta _{0}),(n^{1/2}(\hat{\theta}^{\ast }-\hat{\theta}%
)|D_{n}))\overset{w}{\rightarrow }_{p}\left( \ell _{st}(\theta _{0}),\ell
_{st}(\theta _{0})\right)
\end{equation*}%
and bootstrap validity is restored as in Corollary \ref{corollary bootstrap
with boundary}, in particular because the random shift from the half-plane
in the limiting bootstrap distribution is again removed.

\subsection{Concluding remarks}

An inferential framework that would be asymptotically valid in the unit
root, near-unit root, and stationary cases, allowing the researcher to
remain agnostic to the stochastic properties of the regressor, could be
based on two main ingredients. First, the definition of the bootstrap
parameter space in a way such that it approximates sufficiently well the
geometry of the original parameter space; e.g., by setting $g^{\ast
}=g-|g|^{1+\kappa }$ in the definition of $\Theta ^{\ast }$ for some $\kappa
>0$, see above. This definition is independent of the stochastic properties
of the regressor. Second, the use of an estimator (different from our choice
of OLS) that gives rise to limit distributions that (a) in the near-unit
root case depend on $c$ only through the process $X$ (and thus, the matrix $%
M $), but are free from shifts in the direction of $v_{c}$, and (b) allow
for a common treatment of the contemporaneous correlation between the
innovations of the predictive regression and the shocks driving $x_{n,t}$
(vs. the inclusion or omission of $\Delta x_{n.t}$ in the estimated eq. (\ref%
{eq:pr})). We conjecture that constrained versions of both the IVX (extended
instrumental variables) estimator and the associated bootstrap schemes as
discussed in Demetrescu et al. (2023) would give rise to asymptotically
valid bootstrap inference. A detailed discussion is beyond the scope of this
appendix due to our focus on issues attributable to the boundary of the
parameter space.

\section{Additional Monte Carlo simulations}

{\label{sec:additional MC}}

In this section, we present additional numerical results in support of the
theoretical arguments provided in CGZ. In particular, Tables S.1 and S.2
refer to the same testing procedure considered in Tables 1 and 2 in CGZ,
respectively, but focus on the case $g^{\ast} = g_{2}^{\ast} := g -
n^{-\kappa} |g|$. Furthermore, in Tables S.3 and S.4 we present the
simulated ERPs of bootstrap tests under local alternatives such that $%
\theta_{0} \in \func{int}(\Theta)$, using $g^{\ast} = g_{1}^{\ast}$ and $%
g^{\ast} = g_{2}^{\ast}$, respectively.

\section*{References}

\begin{description}
\item \textsc{Cavaliere, G., I. Georgiev and E. Zanelli} (2024): Parameter
on the boundary in predictive regression, \emph{Econometric Theory},
forthcoming.

\item \textsc{Demetrescu, M., I. Georgiev, A.M.R. Taylor and P.M.M. Rodrigues%
} (2023): Extensions to IVX methods of inference for return predictability, 
\emph{Journal of Econometrics} 237 (Issue 2, Part C).
\end{description}

\bigskip

\bigskip

\newpage {}

\begin{table}[tp]
\begin{footnotesize}
\begin{center}%
\caption*{\textsc{Table S1:} Empirical rejection probabilities (ERPs) of bootstrap tests under the null.}\label{tab:MC}%
\resizebox{1\textwidth}{!}{%
\begin{tabular}
[c]{cc||c|cccc||c|cccc||c|cccccccccc}\hline \hline
\multicolumn{10}{l}{Nominal level: $0.05$} \\\hline \hline
& & \multicolumn{3}{l}{$\theta_0 = (0,0)'$}  & & &  \multicolumn{3}{l}{$\theta_0 = (-0.75,0.75)'$} & & &  \multicolumn{3}{l}{$\theta_0 = (-1.50,1.50)'$}  \\\hline
&  & $b_1$ & $b_2$ &  &  &  &  $b_1$ & $b_2$ &  &  &  &  $b_1$ & $b_2$ \\
  & & &    $\kappa$ & &  & & & $\kappa$ &  & & & & $\kappa$
\\\cline{4-17}
dist.  & $n$ & & 0.05 & 0.10 & 0.20  & 0.40  & & 0.05 & 0.10 & 0.20  & 0.40 & & 0.05 & 0.10 & 0.20  & 0.40
\\\hline
$\xi_1$ &  100 &   4.2 & 4.9 & 5.3 & 5.5 & 5.6 & 6.9 & 7.0 & 7.3 & 8.3 & 9.6 & 6.3 & 6.4 & 6.6 & 7.3 & 9.6 \\ 
& 400 &   3.9 & 4.8 & 5.1 & 5.3 & 5.3 & 5.5 & 5.7 & 6.0 & 7.1 & 9.2 & 5.3 & 5.3 & 5.3 & 5.7 & 8.6 \\
& 800 &    3.7 & 4.7 & 5.0 & 5.2 & 5.2 & 5.2 & 5.3 & 5.6 & 6.7 & 9.4 & 5.2 & 5.2 & 5.2 & 5.3 & 8.4 \\\hline
$\xi_2$ &  100 &   4.2 & 4.9 & 5.3 & 5.6 & 5.7 & 7.1 & 7.3 & 7.5 & 8.4 & 9.9 & 6.2 & 6.4 & 6.6 & 7.2 & 9.5 \\  
& 400 &    3.8 & 4.6 & 5.0 & 5.1 & 5.2 & 5.7 & 6.0 & 6.3 & 7.3 & 9.4 & 5.3 & 5.3 & 5.3 & 5.7 & 8.7 \\
& 800 &    3.6 & 4.5 & 4.8 & 4.9 & 4.9 & 5.1 & 5.2 & 5.5 & 6.7 & 9.3 & 5.1 & 5.1 & 5.1 & 5.3 & 8.6 \\\hline
$\xi_3$ &  100 &   4.3 & 4.9 & 5.3 & 5.6 & 5.7 & 7.1 & 7.2 & 7.4 & 8.5 & 9.9 & 6.4 & 6.5 & 6.7 & 7.4 & 9.8 \\
& 400 &    3.7 & 4.6 & 4.9 & 5.1 & 5.1 & 5.5 & 5.8 & 6.1 & 7.2 & 9.3 & 5.2 & 5.2 & 5.2 & 5.6 & 8.6 \\
& 800 &    3.7 & 4.6 & 5.0 & 5.1 & 5.2 & 5.1 & 5.2 & 5.4 & 6.5 & 9.1 & 5.1 & 5.1 & 5.1 & 5.3 & 8.4 \\\hline \hline
\multicolumn{10}{l}{Nominal level: $0.10$} \\\hline \hline
& & \multicolumn{3}{l}{$\theta_0 = (0,0)'$}  & & &  \multicolumn{3}{l}{$\theta_0 = (-0.75,0.75)'$} & & &  \multicolumn{3}{l}{$\theta_0 = (-1.50,1.50)'$}  \\\hline
&  & $b_1$ & $b_2$ &  &  &  &  $b_1$ & $b_2$ &  &  &  &  $b_1$ & $b_2$ \\
  & & &    $\kappa$ & &  & & & $\kappa$ &  & & & & $\kappa$
\\\cline{4-17}
dist.  & $n$ & & 0.05 & 0.10 & 0.20  & 0.40  & & 0.05 & 0.10 & 0.20  & 0.40 & & 0.05 & 0.10 & 0.20  & 0.40
\\\hline
$\xi_1$ &  100 &   8.0 & 9.1 & 9.9 & 10.5 & 10.7 & 13.0 & 13.3 & 13.7 & 15.4 & 18.6 & 11.5 & 11.7 & 12.0 & 12.9 & 17.2 \\
& 400 &    7.7 & 9.2 & 9.9 & 10.3 & 10.5 & 10.4 & 10.6 & 11.1 & 12.9 & 17.6 & 10.3 & 10.3 & 10.3 & 10.7 & 15.9 \\
& 800 &    7.4 & 9.0 & 9.7 & 10.0 & 10.1 & 10.4 & 10.4 & 10.7 & 12.2 & 18.1 & 10.1 & 10.1 & 10.1 & 10.2 & 15.5 \\\hline
$\xi_2$ &  100 &   8.1 & 9.2 & 9.9 & 10.5 & 10.7 & 13.2 & 13.5 & 13.9 & 15.6 & 18.7 & 11.3 & 11.5 & 11.8 & 12.7 & 16.9 \\  
& 400 &    7.5 & 9.0 & 9.7 & 10.2 & 10.3 & 10.7 & 11.0 & 11.4 & 13.2 & 18.0 & 10.2 & 10.3 & 10.3 & 10.7 & 15.9 \\
& 800 &    
7.2 & 8.9 & 9.5 & 9.9 & 10.0 & 10.2 & 10.3 & 10.5 & 12.0 & 17.7 & 10.3 & 10.3 & 10.3 & 10.4 & 15.7 \\\hline
$\xi_3$ &  100 &   8.3 & 9.4 & 10.2 & 10.8 & 11.0 & 13.3 & 13.7 & 14.1 & 15.8 & 19.0 & 11.7 & 11.9 & 12.2 & 13.2 & 17.5 \\
& 400 &    7.6 & 9.1 & 9.8 & 10.2 & 10.3 & 10.4 & 10.6 & 11.1 & 13.1 & 17.7 & 10.2 & 10.2 & 10.2 & 10.6 & 15.9 \\
& 800 &    7.4 & 9.0 & 9.6 & 10.0 & 10.1 & 10.1 & 10.1 & 10.4 & 11.9 & 17.6 & 10.0 & 10.0 & 10.0 & 10.1 & 15.5 \\\hline \hline
\end{tabular}}%
\end{center}%

Note: bootstrap tests are based on a standard fixed-regressor wild bootstrap ($b_1$) and on the proposed corrected wild bootstrap method ($b_2$) of Section 4, using $g^*=g-n^{-\kappa}|g|$. ERPs are estimated using 50,000 Monte Carlo replications and 999 bootstrap repetitions. The column “dist." shows the distributions of $\varepsilon_t$: $\xi_1 \sim iid N(0,1)$, $\xi_2 \sim ARCH(1)$ and $\xi_3= \sqrt{0.5} v_t + \sqrt{0.5} \varepsilon_{x,t}$, where  $v_t \sim iidN(0,1)$ and $\varepsilon_{x,t}$ is the error term of the predictive variable $x_{n,t}$.%
\end{footnotesize}%
\end{table}%

\begin{table}[tp]
\begin{footnotesize}
\begin{center}%
\caption*{\textsc{Table S2:} Empirical rejection probabilities (ERPs) of bootstrap tests under local alternatives.}\label{tab:MC}%
\resizebox{1\textwidth}{!}{%
\begin{tabular}
[c]{cc||c|cccc||c|cccc||c|cccccccccc}\hline \hline
\multicolumn{10}{l}{Nominal level: $0.05$} \\\hline \hline
& & \multicolumn{3}{l}{$a_0 = (-3,0)'$}  & & &  \multicolumn{3}{l}{$a_0 = (3,0)'$} & & &  \multicolumn{3}{l}{$a_0 = (5,0)'$}  \\\hline
&  & $b_1$ & $b_2$ &  &  &  &  $b_1$ & $b_2$ &  &  &  &  $b_1$ & $b_2$ \\
  & & &    $\kappa$ & &  & & & $\kappa$ &  & & & & $\kappa$
\\\cline{4-17}
dist.  & $n$ & & 0.05 & 0.10 & 0.20  & 0.40  & & 0.05 & 0.10 & 0.20  & 0.40 & & 0.05 & 0.10 & 0.20  & 0.40
\\\hline
$\xi_1$ &  100 &   21.0 & 21.1 & 21.3 & 21.5 & 21.5 & 40.6 & 40.8 & 40.9 & 41.0 & 41.0 & 68.0 & 68.0 & 68.0 & 68.0 & 68.0 \\
& 400 &    18.9 & 19.1 & 19.3 & 19.5 & 19.5 & 38.5 & 38.7 & 38.8 & 38.8 & 38.8 & 64.9 & 64.9 & 64.9 & 64.9 & 64.9 \\
& 800 &    18.6 & 18.8 & 19.0 & 19.1 & 19.1 & 37.6 & 37.8 & 37.9 & 37.9 & 37.9 & 64.0 & 64.0 & 64.0 & 64.0 & 64.0 \\\hline
$\xi_2$ &  100 &   21.7 & 21.9 & 22.0 & 22.2 & 22.3 & 41.9 & 42.1 & 42.2 & 42.2 & 42.3 & 68.5 & 68.5 & 68.5 & 68.5 & 68.5 \\  
& 400 &    19.2 & 19.4 & 19.6 & 19.7 & 19.8 & 38.3 & 38.6 & 38.7 & 38.7 & 38.7 & 64.7 & 64.8 & 64.8 & 64.8 & 64.8 \\
& 800 &    18.6 & 18.8 & 19.0 & 19.1 & 19.1 & 37.8 & 38.0 & 38.1 & 38.1 & 38.1 & 64.2 & 64.2 & 64.2 & 64.2 & 64.2 \\\hline
$\xi_3$ &  100 &   20.6 & 20.7 & 20.9 & 21.2 & 21.3 & 40.8 & 41.0 & 41.1 & 41.1 & 41.1 & 67.3 & 67.3 & 67.3 & 67.3 & 67.3 \\
& 400 &    19.0 & 19.1 & 19.3 & 19.4 & 19.4 & 38.1 & 38.3 & 38.4 & 38.5 & 38.5 & 65.0 & 65.0 & 65.0 & 65.0 & 65.0 \\
& 800 &    18.3 & 18.5 & 18.7 & 18.8 & 18.9 & 37.7 & 38.0 & 38.0 & 38.1 & 38.1 & 63.5 & 63.5 & 63.5 & 63.5 & 63.5 \\\hline \hline
\multicolumn{10}{l}{Nominal level: $0.10$} \\\hline \hline
& & \multicolumn{3}{l}{$a_0 = (-3,0)'$}  & & &  \multicolumn{3}{l}{$a_0 = (3,0)'$} & & &  \multicolumn{3}{l}{$a_0 = (5,0)'$}  \\\hline
&  & $b_1$ & $b_2$ &  &  &  &  $b_1$ & $b_2$ &  &  &  &  $b_1$ & $b_2$ \\
  & & &    $\kappa$ & &  & & & $\kappa$ &  & & & & $\kappa$
\\\cline{4-17}
dist.  & $n$ & & 0.05 & 0.10 & 0.20  & 0.40  & & 0.05 & 0.10 & 0.20  & 0.40 & & 0.05 & 0.10 & 0.20  & 0.40
\\\hline
$\xi_1$ &  100 &   29.6 & 29.8 & 30.1 & 30.5 & 30.7 & 54.7 & 55.0 & 55.1 & 55.2 & 55.2 & 81.7 & 81.7 & 81.7 & 81.8 & 81.8 \\ 
& 400 &    27.0 & 27.3 & 27.8 & 28.1 & 28.2 & 52.2 & 52.5 & 52.6 & 52.7 & 52.7 & 79.6 & 79.6 & 79.6 & 79.6 & 79.6 \\
& 800 &    26.4 & 26.8 & 27.2 & 27.5 & 27.6 & 51.7 & 52.1 & 52.1 & 52.2 & 52.2 & 78.7 & 78.7 & 78.7 & 78.7 & 78.7 \\\hline
$\xi_2$ &  100 &   30.2 & 30.4 & 30.7 & 31.2 & 31.4 & 55.7 & 55.9 & 55.9 & 56.0 & 56.1 & 82.0 & 82.0 & 82.0 & 82.0 & 82.0 \\  
& 400 &   27.1 & 27.4 & 27.9 & 28.2 & 28.3 & 51.8 & 52.0 & 52.1 & 52.2 & 52.2 & 79.3 & 79.3 & 79.3 & 79.3 & 79.3 \\
& 800 &    26.6 & 26.9 & 27.4 & 27.7 & 27.7 & 51.5 & 51.8 & 51.9 & 51.9 & 51.9 & 78.6 & 78.6 & 78.6 & 78.6 & 78.6 \\\hline
$\xi_3$ &  100 &   29.1 & 29.3 & 29.6 & 30.1 & 30.3 & 54.2 & 54.4 & 54.5 & 54.6 & 54.6 & 80.9 & 80.9 & 80.9 & 80.9 & 80.9 \\
& 400 &    26.7 & 27.0 & 27.4 & 27.8 & 27.8 & 51.7 & 52.0 & 52.1 & 52.2 & 52.2 & 79.4 & 79.4 & 79.4 & 79.4 & 79.4 \\
& 800 &    26.2 & 26.5 & 27.0 & 27.3 & 27.3 & 51.3 & 51.6 & 51.7 & 51.7 & 51.8 & 78.5 & 78.5 & 78.5 & 78.5 & 78.5 \\\hline \hline
\end{tabular}}%
\end{center}%

Note: bootstrap tests are based on a standard fixed-regressor wild bootstrap ($b_1$) and on the proposed corrected wild bootstrap method ($b_2$) of Section 4, using $g^*=g-n^{-\kappa}|g|$. ERPs are estimated using 50,000 Monte Carlo replications and 999 bootstrap repetitions. The column “dist." shows the distributions of $\varepsilon_t$: $\xi_1 \sim iid N(0,1)$, $\xi_2 \sim ARCH(1)$ and $\xi_3= \sqrt{0.5} v_t + \sqrt{0.5} \varepsilon_{x,t}$, where  $v_t \sim iidN(0,1)$ and $\varepsilon_{x,t}$ is the error term of the predictive variable $x_{n,t}$.%
\end{footnotesize}%
\end{table}%

\begin{table}[tp]
\begin{footnotesize}
\begin{center}%
\caption*{\textsc{Table S3:} Empirical rejection probabilities (ERPs) of bootstrap tests under local alternatives.}\label{tab:MC}%
\resizebox{1\textwidth}{!}{%
\begin{tabular}
[c]{cc||c|cccc||c|cccc||c|cccccccccc}\hline \hline
\multicolumn{10}{l}{Nominal level: $0.05$} \\\hline \hline
& & \multicolumn{3}{l}{$a_0 = (-3,1)'$}  & & &  \multicolumn{3}{l}{$a_0 = (2,2)'$} & & &  \multicolumn{3}{l}{$a_0 = (3,4)'$}  \\\hline
&  & $b_1$ & $b_2$ &  &  &  &  $b_1$ & $b_2$ &  &  &  &  $b_1$ & $b_2$ \\
  & & &    $\kappa$ & &  & & & $\kappa$ &  & & & & $\kappa$
\\\cline{4-17}
dist.  & $n$ & & 0.25 & 0.50 & 1.0  & 2.0  & & 0.25 & 0.50 & 1.0  & 2.0 & & 0.25 & 0.50 & 1.0  & 2.0
\\\hline
$\xi_1$ &  100 &   12.8 & 12.9 & 13.0 & 13.2 & 13.4 & 48.4 & 49.6 & 50.1 & 50.3 & 50.4 & 73.0 & 73.9 & 74.4 & 74.7 & 74.7 \\ 
& 400 &    11.4 & 11.6 & 11.9 & 12.2 & 12.3 & 45.4 & 47.2 & 47.5 & 47.6 & 47.6 & 70.0 & 71.6 & 72.0 & 72.0 & 72.0 \\ 
& 800 &    10.9 & 11.2 & 11.6 & 11.7 & 11.8 & 44.8 & 46.9 & 47.1 & 47.1 & 47.2 & 69.3 & 71.1 & 71.4 & 71.4 & 71.4 \\\hline
$\xi_2$ &  100 &   13.1 & 13.2 & 13.3 & 13.5 & 13.6 & 49.6 & 50.8 & 51.3 & 51.6 & 51.6 & 73.2 & 74.1 & 74.7 & 75.0 & 75.0 \\  
& 400 &    11.4 & 11.6 & 11.8 & 12.1 & 12.2 & 46.1 & 48.0 & 48.3 & 48.3 & 48.3 & 70.2 & 71.8 & 72.2 & 72.3 & 72.3 \\
& 800 &    11.0 & 11.3 & 11.7 & 11.9 & 11.9 & 45.2 & 47.2 & 47.4 & 47.4 & 47.4 & 69.6 & 71.5 & 71.7 & 71.7 & 71.7 \\\hline
$\xi_3$ &  100 &   12.3 & 12.4 & 12.5 & 12.7 & 12.9 & 48.1 & 49.3 & 49.9 & 50.1 & 50.1 & 72.4 & 73.2 & 73.8 & 74.1 & 74.1 \\  
& 400 &    11.4 & 11.6 & 11.9 & 12.2 & 12.3 & 46.0 & 47.8 & 48.2 & 48.2 & 48.3 & 69.9 & 71.5 & 72.0 & 72.0 & 72.0 \\
& 800 &    11.1 & 11.4 & 11.8 & 12.0 & 12.1 & 45.0 & 46.9 & 47.1 & 47.1 & 47.1 & 69.4 & 71.3 & 71.6 & 71.6 & 71.6 \\\hline \hline
\multicolumn{10}{l}{Nominal level: $0.10$} \\\hline \hline
& & \multicolumn{3}{l}{$a_0 = (-3,1)'$}  & & &  \multicolumn{3}{l}{$a_0 = (2,2)'$} & & &  \multicolumn{3}{l}{$a_0 = (3,4)'$}  \\\hline
&  & $b_1$ & $b_2$ &  &  &  &  $b_1$ & $b_2$ &  &  &  &  $b_1$ & $b_2$ \\
  & & &    $\kappa$ & &  & & & $\kappa$ &  & & & & $\kappa$
\\\cline{4-17}
dist.  & $n$ & & 0.25 & 0.50 & 1.0  & 2.0  & & 0.25 & 0.50 & 1.0  & 2.0 & & 0.25 & 0.50 & 1.0  & 2.0
\\\hline
$\xi_1$ &  100 &   21.2 & 21.5 & 21.6 & 22.0 & 22.4 & 58.8 & 60.4 & 61.1 & 61.5 & 61.5 & 80.7 & 81.6 & 82.2 & 82.5 & 82.5 \\ 
& 400 &    19.2 & 19.6 & 20.2 & 21.0 & 21.2 & 56.0 & 58.2 & 58.6 & 58.7 & 58.7 & 78.2 & 79.9 & 80.3 & 80.4 & 80.4 \\
& 800 &    18.3 & 18.9 & 19.7 & 20.2 & 20.2 & 55.8 & 58.1 & 58.5 & 58.5 & 58.5 & 77.8 & 79.8 & 80.1 & 80.1 & 80.2 \\\hline
$\xi_2$ &  100 &   21.8 & 22.0 & 22.1 & 22.5 & 23.0 & 59.6 & 61.1 & 61.8 & 62.1 & 62.2 & 81.0 & 81.9 & 82.5 & 82.9 & 82.9 \\
& 400 &    19.1 & 19.5 & 20.1 & 20.7 & 21.0 & 56.8 & 59.0 & 59.5 & 59.6 & 59.6 & 78.6 & 80.4 & 80.8 & 80.8 & 80.9 \\
& 800 &    18.9 & 19.5 & 20.2 & 20.7 & 20.8 & 56.0 & 58.4 & 58.7 & 58.8 & 58.8 & 78.0 & 79.9 & 80.2 & 80.3 & 80.3 \\\hline
$\xi_3$ &  100 &   20.6 & 20.8 & 20.9 & 21.3 & 21.8 & 58.5 & 60.1 & 60.8 & 61.1 & 61.2 & 80.2 & 81.2 & 81.7 & 82.0 & 82.1 \\
& 400 &    19.1 & 19.5 & 20.1 & 20.8 & 21.0 & 56.6 & 58.7 & 59.2 & 59.3 & 59.3 & 78.3 & 80.1 & 80.5 & 80.6 & 80.6 \\
& 800 &    18.7 & 19.2 & 20.0 & 20.5 & 20.6 & 55.7 & 58.2 & 58.5 & 58.6 & 58.6 & 77.8 & 79.5 & 79.9 & 79.9 & 79.9 \\\hline \hline
\end{tabular}}%
\end{center}%

Note: bootstrap tests are based on a standard fixed-regressor wild bootstrap ($b_1$) and on the proposed corrected wild bootstrap method ($b_2$) of Section 4, using $g^*=g-|g|^{1+\kappa}$. ERPs are estimated using 50,000 Monte Carlo replications and 999 bootstrap repetitions. The column “dist." shows the distributions of $\varepsilon_t$: $\xi_1 \sim iid N(0,1)$, $\xi_2 \sim ARCH(1)$ and $\xi_3= \sqrt{0.5} v_t + \sqrt{0.5} \varepsilon_{x,t}$, where  $v_t \sim iidN(0,1)$ and $\varepsilon_{x,t}$ is the error term of the predictive variable $x_{n,t}$.%
\end{footnotesize}%
\end{table}%

\begin{table}[tp]
\begin{footnotesize}
\begin{center}%
\caption*{\textsc{Table S4:} Empirical rejection probabilities (ERPs) of bootstrap tests under local alternatives.}\label{tab:MC}%
\resizebox{1\textwidth}{!}{%
\begin{tabular}
[c]{cc||c|cccc||c|cccc||c|cccccccccc}\hline \hline
\multicolumn{10}{l}{Nominal level: $0.05$} \\\hline \hline
& & \multicolumn{3}{l}{$a_0 = (-3,1)'$}  & & &  \multicolumn{3}{l}{$a_0 = (2,2)'$} & & &  \multicolumn{3}{l}{$a_0 = (3,4)'$}  \\\hline
&  & $b_1$ & $b_2$ &  &  &  &  $b_1$ & $b_2$ &  &  &  &  $b_1$ & $b_2$ \\
  & & &    $\kappa$ & &  & & & $\kappa$ &  & & & & $\kappa$
\\\cline{4-17}
dist.  & $n$ & & 0.05 & 0.10 & 0.20  & 0.40  & & 0.05 & 0.10 & 0.20  & 0.40 & & 0.05 & 0.10 & 0.20  & 0.40
\\\hline
$\xi_1$ &  100 &   12.8 & 13.0 & 13.2 & 13.6 & 13.7 & 48.4 & 49.6 & 50.1 & 50.4 & 50.4 & 73.0 & 74.0 & 74.5 & 74.7 & 74.7 \\  
& 400 &    11.4 & 11.6 & 12.0 & 12.2 & 12.3 & 45.4 & 47.0 & 47.4 & 47.6 & 47.6 & 70.0 & 71.4 & 71.9 & 72.0 & 72.0 \\
& 800 &    10.9 & 11.1 & 11.5 & 11.7 & 11.8 & 44.8 & 46.5 & 47.0 & 47.1 & 47.2 & 69.3 & 70.8 & 71.3 & 71.4 & 71.4 \\\hline
$\xi_2$ &  100 &   13.1 & 13.3 & 13.5 & 13.9 & 14.0 & 49.6 & 50.8 & 51.3 & 51.6 & 51.6 & 73.2 & 74.2 & 74.7 & 75.0 & 75.0 \\  
& 400 &    11.4 & 11.6 & 11.9 & 12.1 & 12.2 & 46.1 & 47.7 & 48.1 & 48.3 & 48.3 & 70.2 & 71.6 & 72.1 & 72.3 & 72.3 \\
& 800 &    11.0 & 11.3 & 11.7 & 11.8 & 11.9 & 45.2 & 46.9 & 47.3 & 47.4 & 47.4 & 69.6 & 71.2 & 71.6 & 71.7 & 71.7 \\\hline
$\xi_3$ &  100 & 12.3 & 12.4 & 12.8 & 13.2 & 13.3 & 48.1 & 49.3 & 49.9 & 50.1 & 50.2 & 72.4 & 73.4 & 73.9 & 74.1 & 74.2 \\  
& 400 &    11.4 & 11.7 & 12.0 & 12.2 & 12.3 & 46.0 & 47.6 & 48.0 & 48.2 & 48.3 & 69.9 & 71.4 & 71.8 & 72.0 & 72.0 \\
& 800 &    11.1 & 11.4 & 11.8 & 12.0 & 12.1 & 45.0 & 46.5 & 47.0 & 47.1 & 47.2 & 69.4 & 71.0 & 71.5 & 71.6 & 71.6 \\\hline \hline
\multicolumn{10}{l}{Nominal level: $0.10$} \\\hline \hline
& & \multicolumn{3}{l}{$a_0 = (-3,1)'$}  & & &  \multicolumn{3}{l}{$a_0 = (2,2)'$} & & &  \multicolumn{3}{l}{$a_0 = (3,4)'$}  \\\hline
&  & $b_1$ & $b_2$ &  &  &  &  $b_1$ & $b_2$ &  &  &  &  $b_1$ & $b_2$ \\
  & & &    $\kappa$ & &  & & & $\kappa$ &  & & & & $\kappa$
\\\cline{4-17}
dist.  & $n$ & & 0.05 & 0.10 & 0.20  & 0.40  & & 0.05 & 0.10 & 0.20  & 0.40 & & 0.05 & 0.10 & 0.20  & 0.40
\\\hline
$\xi_1$ &  100 &   21.2 & 21.5 & 21.9 & 22.7 & 23.0 & 58.8 & 60.2 & 60.9 & 61.4 & 61.5 & 80.7 & 81.6 & 82.1 & 82.4 & 82.5 \\
& 400 &    19.2 & 19.6 & 20.3 & 21.0 & 21.2 & 56.0 & 57.7 & 58.3 & 58.6 & 58.7 & 78.2 & 79.6 & 80.1 & 80.4 & 80.4 \\
& 800 &   18.3 & 18.8 & 19.6 & 20.1 & 20.2 & 55.8 & 57.7 & 58.2 & 58.5 & 58.5 & 77.8 & 79.4 & 79.9 & 80.1 & 80.1 \\\hline
$\xi_2$ &  100 &   21.8 & 22.0 & 22.5 & 23.3 & 23.7 & 59.6 & 61.0 & 61.6 & 62.1 & 62.2 & 81.0 & 81.9 & 82.5 & 82.9 & 82.9 \\  
& 400 &    19.1 & 19.5 & 20.1 & 20.8 & 21.0 & 56.8 & 58.5 & 59.2 & 59.5 & 59.6 & 78.6 & 80.0 & 80.6 & 80.8 & 80.8 \\
& 800 & 18.9 & 19.4 & 20.1 & 20.7 & 20.8 & 56.0 & 57.9 & 58.5 & 58.7 & 58.8 & 78.0 & 79.5 & 80.1 & 80.3 & 80.3 \\\hline
$\xi_3$ &  100 &   20.6 & 20.8 & 21.3 & 22.2 & 22.6 & 58.5 & 59.9 & 60.6 & 61.1 & 61.2 & 80.2 & 81.1 & 81.7 & 82.0 & 82.1 \\  
& 400 & 19.1 & 19.5 & 20.2 & 20.8 & 21.0 & 56.6 & 58.3 & 58.9 & 59.2 & 59.3 & 78.3 & 79.7 & 80.3 & 80.5 & 80.6 \\ 
& 800 & 18.7 & 19.1 & 19.9 & 20.5 & 20.6 & 55.7 & 57.7 & 58.3 & 58.6 & 58.6 & 77.8 & 79.2 & 79.7 & 79.9 & 79.9 \\\hline \hline
\end{tabular}}%
\end{center}%

Note: bootstrap tests are based on a standard fixed-regressor wild bootstrap ($b_1$) and on the proposed corrected wild bootstrap method ($b_2$) of Section 4, using $g^*=g-n^{-\kappa}|g|$. ERPs are estimated using 50,000 Monte Carlo replications and 999 bootstrap repetitions. The column “dist." shows the distributions of $\varepsilon_t$: $\xi_1 \sim iid N(0,1)$, $\xi_2 \sim ARCH(1)$ and $\xi_3= \sqrt{0.5} v_t + \sqrt{0.5} \varepsilon_{x,t}$, where  $v_t \sim iidN(0,1)$ and $\varepsilon_{x,t}$ is the error term of the predictive variable $x_{n,t}$.%
\end{footnotesize}%
\end{table}%

\end{document}